\documentclass[twocolumn,superscriptaddress,showpacs,prd,aps,amsmath,amssymb,nofootinbib]{revtex4-1}
\usepackage{graphicx,color}
\usepackage{amsmath,amssymb}
\usepackage{verbatim}
\usepackage{float}
\usepackage{wasysym}
\usepackage{amssymb,graphicx}
\usepackage{epsfig}
\usepackage{psfrag}
\usepackage{dsfont}
\usepackage{amsfonts}
\usepackage{mathrsfs}
\usepackage{multirow}
\usepackage{bm}
\usepackage{hyperref}
\usepackage{pifont}
\usepackage{tabularx}
\usepackage{aas_macros}
\usepackage{graphicx}
\newcommand{\cmark}{\ding{51}}%
\newcommand{\xmark}{\ding{55}}%

\usepackage{soul,xcolor}
\hypersetup{
  colorlinks=true,        % false: boxed links; true: colored links
  linkcolor=blue,         % color of internal links
  citecolor=cyan,         % color of links to bibliography
}

\begin{document}
\title{Jet Launching from Binary Neutron Star Mergers: Incorporating Neutrino Transport and Magnetic Fields}
\author{Lunan Sun}
 \email{lsun11@illinois.edu}
\affiliation{Department of Physics, University of Illinois at
  Urbana-Champaign, Urbana, IL 61801, USA}
\author{Milton Ruiz}
\affiliation{Department of Physics, University of Illinois at
  Urbana-Champaign, Urbana, IL 61801, USA}
  \author{Stuart L. Shapiro}
\affiliation{Department of Physics, University of Illinois at
  Urbana-Champaign, Urbana, IL 61801, USA}
\affiliation{Department of Astronomy \& NCSA, University of Illinois at
  Urbana-Champaign, Urbana, IL 61801, USA}
\author{Antonios Tsokaros}
\affiliation{Department of Physics, University of Illinois at
  Urbana-Champaign, Urbana, IL 61801, USA}
%
%%%%%%%%%%%%%%%%
%%% Abstract %%%
%%%%%%%%%%%%%%%%
\begin{abstract}
  We perform general relativistic, magnetohydrodynamic (GRMHD) simulations  of merging binary neutron stars
  incorporating neutrino transport and magnetic fields. Our new radiative transport module for
  neutrinos adopts a general relativistic, truncated-moment (M1) formalism. The binaries consist of
  two identical, irrotational stars modeled by the SLy nuclear equation of state
  (EOS). They are initially in quasicircular orbit and threaded with a poloidal magnetic field that extends from the
  stellar interior into the exterior, as in typical pulsars. We insert neutrino processes shortly after the merger
  and focus on the role of neutrinos in launching a jet following the collapse of the hypermassive neutron
  star (HMNS) remnant to a spinning black hole (BH). We treat two microphysical versions: one (a ``warm-up")
  evolving a single neutrino species and considering only charged-current processes, and the other evolving
  three species $(\nu_e,  \bar{\nu}_e, \nu_{\rm x})$ and related processes. We trace the evolution
  until the system reaches a quasiequilibrium state after BH formation. We find that the BH + disk remnant
  eventually launches an incipient jet. The electromagnetic Poynting luminosity is $\sim
  10^{53} \rm \, erg\, s^{-1}$, consistent with that of typical short gamma-ray bursts (sGRBs). The effect of
  neutrino cooling shortens the lifetime of the HMNS, and lowers the amplitude of the major peak of
  the gravitational wave (GW) power spectrum somewhat. After BH formation, neutrinos help clear out the matter
  near the BH poles, resulting in lower baryon-loaded surrounding debris. The neutrino luminosity resides
  in the range $\sim 10^{52-53} \rm \,erg\,s^{-1}$ once quasiequilibrium is achieved. Comparing with the
  neutrino-free models, we observe that the inclusion of neutrinos yields similar ejecta masses and is
  inefficient in carrying off additional angular momentum.
\end{abstract}

\maketitle

%%%%%%%%%%%%%%%%%%%%%%
%%%  Introduction  %%%
%%%%%%%%%%%%%%%%%%%%%%
\section{Introduction}
\label{sec:intro}
The coincident detection of GWs and electromagnetic (EM) signals from the first established neutron star
binary (NSNS) merger GW170817~\cite{TheLIGOScientific:2017qsa, Savchenko17, Cowperthwaite17,Soares-Santos17,
  Smartt17, Kasliwal17,Chornock17} provides a unique opportunity to study systematically the properties of
compact objects, nuclear physics, and electromagnetism in strong gravity. The simultaneous detection of GW
and EM signals from the NSNS mergers is the prime target of multimessenger astronomy (MA) and can provide
us with important information about high-energy astrophysics phenomena in strong gravity, the synthesis of
heavy nuclei, the properties of dense, nuclear matter, etc. In particular, the observation of GW170817 coincident
with the sGRB GRB170817A~\cite{Monitor:2017mdv} demonstrated that NSNS remnants can power sGRBs
\cite{TheLIGOScientific:2017qsa,FERMI2017GCN,Savchenko:2017ffs}. In addition, this GW observation
and its association with kilonova AT 2017gfo/DLT17ck~\cite{Valenti:2017ngx} indicate that, in contrast to
BH binary mergers, compact binary mergers where at least one of the companions is a NS are likely to be followed
by various processes involving  electromagnetic and neutrino emission. To systematically interpret multimessenger
signals and their relation to the properties of the binary system, it is necessary to perform full general
relativistic, magnetohydrodynamic (GRMHD) simulations incorporating detailed microphysical processes. 

It has been long established that magnetic fields play a critical role in the fate of NSNS merger remnants.
The Kelvin-Helmholtz instability (KHI) at the contact surface once the NSs touch for the first time triggers
the exponential amplification of the magnetic field strength~\cite{Obergaulinger10,Zrake_2013,Kiuchi:2015sga,
  Palenzuela:2021gdo}. Subsequently, material is ejected due to orbital angular momentum advection and to
torques~\cite{Siegel_2014, Kiuchi_2012}. There have been a series of systematic numerical studies of the effect
of magnetic fields on NSNS mergers~(see~e.g.~\cite{Price719,dlsss06b,AHLLMNPT08,rgbgka11,grb11,PhysRevD.88.044020,
  PhysRevD.89.104029, PhysRevD.92.044045, Ruiz:2016rai,Ciolfi:2019fie, Ruiz:2019, Ciolfi:2020mnras,
  Aguilera-Miret:2021fre,Palenzuela:2021gdo} and references therein). In particular, we have shown
that NSNS remnants consisting of a BH + disk can launch a collimated, mildly relativistic outflow--an incipient jet-- with duration and luminosity consistent
with typical sGRB central engine lifetimes and magnitudes, as well as with the Blandford–Znajek mechanism~\cite{Blandford02} (hereafter BZ) for launching
jets and their associated Poynting luminosities \cite{Ruiz:2016rai,Ruiz:2018wah}.
{We also reported that  a jet is launched following the delayed collapse of a hypermassive neutron
star (HMNS) remnant if the {\it initial} pulsar-like magnetic field (i.e. magnetic filed that
extends from the neutron star interior into its exterior) in the binary companions has a
sufficiently large-scale poloidal component aligned to the orbital angular momentum of the system.
The simulations in~\cite{Ruiz20} suggest that this initial poloidal component is needed to
amplify the magnetic energy to $\gtrsim 10^{50}\,\rm erg$ in the BH + disk remnant, a threshold
value that is required for jet launching. It is worth emphasizing that the emergence of a jet does
not require rely on the development of a large-scale poloidal magnetic
field component in the HMNS, but only initially, as in typical pulsars, which is what 
we reported in~\cite{Ruiz:2016rai}.}
%We also reported that  a jet is launched following the delayed collapse of a hypermassive neutron
%star (HMNS) remnant if the {\it initial} magnetic field in the binary companions has a sufficiently
%large-scale poloidal component aligned to
%the orbital angular momentum of the system~\cite{Ruiz20}.
In addition, our GRMHD simulations of merging black hole-neutron star (BHNS) binaries, where the
NS is endowed with a magnetic field that extends from the stellar
interior into the exterior, show that a jet may be launched from the
highly spinning BH + disk remnant if the initial BH spin satisfies $a/M_{\rm BH} \gtrsim 0.4$ and an approximate alignment of the
magnetic moment with the total angular momentum of the system is present~\cite{prs15, Ruiz:2018wah}. 

Neutrino processes (e.g., cooling and emission) may also have significant effects on the fate of NSNS merger remnants.
It has been suggested that neutrino pair annihilation could carry a large amount of energy from the inner region of the
disk~\cite{Popham_1999, DPN02, Lei_2013, jojb15}. The thinning of the disk may result in a favorable geometry for jet launching
\cite{Popham_1999, Ruiz:2019}, though the duration and energy of neutrino emission in NSNS mergers are  likely
to be insufficient for the outflows to break out from the ejecta shell and form relativistic jets~\cite{jojb15}.
It is also believed that neutrino-driven winds, in which neutrinos absorbed in the disk can lift matter out of its
gravitational potential, is a crucial mechanism of mass loss for NSNS mergers~\cite{Rosswog03,
  Dessart09, Perego14, Murguia_Berthier_2021}. Moreover, it was recently suggested that the jet launching mechanism could be a joint
process of two widely accepted GRB central engine models: a BH model, where a stellar-mass BH is surrounded by a neutrino cooling-dominated
accretion flow first proposed in~\cite{Popham_1999}, and the BZ mechanism~\cite{Blandford02}, where
the BH rotational energy is extracted by the magnetic field. According to~\cite{Lei_2017}, an initially non-spinning BH may first
launch a thermal ``fireball” via neutrino annihilation, followed by a more powerful, Poynting-flux dominated jet via the BZ process once the BH is spun up by accretion. Neutrinos may also have important effects on the magnetorotational instability (MRI), which is one of the main mechanisms
to amplify the magnetic field to the strength required for jet launching (typically $\gtrsim 10^{15}\rm\,G$; see e.g.~\cite{Guilet15}). As the joint
contribution and interaction between neutrino transfer and magnetic fields may produce copious interesting effects in compact binary
mergers, numerical simulations of mergers with full global and microphysical ingredients are crucial in determining the real physics
of the jet-launching mechanism underlying recent observations, such as GRB 170817A and GRB 160625B~\cite{Dirirsa:2017OJ,Monitor:2017mdv}.

Neutrino transport schemes in compact binary simulations have been developed and improved somewhat slowly. This is mainly because of
long-standing complications with numerically solving Boltzmann’s equation of radiation transport, whose requirement of evolving a system in
6-dimensional phase space plus time for each species of neutrinos makes its application computationally expensive and impractical.
Moreover, the interaction timescale between matter and radiation is shorter than the dynamical timescale of MHD matter in optically
thick regions, making the numerical simulation more challenging to perform with sufficient resolution. To achieve neutrino transport
in compact binary merger systems, simplifications of the exact transport equations have been applied. These include the simplest
leakage schemes, which are based on the assumption that the neutrino diffusion timescale is much longer than the weak interaction
timescale~\cite{Sekiguchi_2010, Deaton_2013}. With a leakage scheme, one can trace the approximate local energy variation of neutrinos
and neutrino cooling but cannot treat non-local neutrino-matter interactions and neutrino heating. More sophisticated are truncated moment
formalisms, in particular, the two-moment (``M1") scheme with analytic closure~\cite{ShibataM1, Wanajo_2014, Foucart_M1_2015, Fujibayashi_2020},
and a mixed leakage-one moment scheme~\cite{Radice_M0_16} for evolving binary merger systems. Recently, an improved M1 scheme was proposed
in~\cite{Foucart_2016} that evolves the neutrino number density and yields a local estimate of the average
neutrino energy. It shows that the composition of the polar ejecta in a NSNS merger is less neutron rich than in their previous simulations, where the neutrino spectrum was assumed to be energy-independent in optically thin regions. More recently, a general relativistic Monte-Carlo (MC) scheme has been implemented for the simulation of an unequal mass NSNS merger~\cite{Foucart_2020, Foucart_2021}. This has
shown that MC and M1 have similar and consistent outcomes of ejected material and neutrino outflow, except for a factor of $\sim 2$
difference in the luminosity of heavy-lepton neutrinos. However, the studies above are purely hydrodynamic simulations and do not study
the joint effect of neutrino radiation and magnetic fields. 

A recent GRMHD study of the NSNS postmerger phase that includes a neutrino leakage scheme, in which the remnant is seeded with  a poloidal
magnetic field 17~ms after merger, reports that the strong toroidal field ($\sim 10^{16} \, \rm G$) around a HMNS remnant is able to
launch a magnetically driven outflow~\cite{M_sta_2020}. Neutrino cooling may help to reduce the baryon-load in the polar region above the HMNS.
However, the maximum attainable Lorentz factor in these simulations ($\Gamma_L\lesssim 5$) is too low to explain
most sGRBs~\cite{Zou10}. It is also claimed that, with a better neutrino evolution scheme that includes pair annihilation
and radiative processes, the Lorentz factor could be boosted to higher values. Therefore, a numerical GRMHD study of NSNS merger
with a neutrino transport scheme that includes key interactions would provide new information regarding the merger remnant, radiation signals,
jet-launching mechanism, and many other important features. More recently, a GRMHD simulation of a BHNS merger with neutrino transfer using a
combined leakage-moment method has been performed in~\cite{Hayashi2021}. It finds that during the merger, dynamical mass ejection and
accretion disk formation occur shortly after tidal disruption. The magnetic field is strengthened due to magnetic winding and MRI. Moreover,
post-merger mass ejection due to magnetically-induced turbulent viscosity takes place $\sim$ 300--500~ms after the tidal disruption, with
the neutrino luminosity dropping quickly below $\sim 10^{51} \, \rm erg\,s^{-1}$. It also claims that a high-intensity Poynting flux is generated a few hundred
ms after merger and remains strong for 1-2 s, which agrees with the duration of typical sGRBs.

In this work, we perform GRMHD simulations of NSNS mergers  modeled using a piecewise
polytropic representation of the nuclear SLy nuclear EOS~\cite{Douchin01} and initially endowed with a
poloidal magnetic field extending from the stellar interior to the exterior, as in pulsars. Shortly after merger,
we insert neutrino transfer using an M1 closure scheme. We performed two versions: i) a simplified ``warm-up" version
involving only one neutrino species $(\bar{\nu}_e)$ and considering only charged-current interactions (called ``Rad-Simp''); and ii) a full, more realistic
version that evolves three neutrino species and considers additional microphysical processes (``Rad-Full''). We compare our
results to simulations that include neither magnetic fields nor neutrino transport
(``Unmag") as well as those that include only magnetic fields (``Mag") and those that include only neutrinos transport (``Unmag + Rad").

We find that neutrino processes enhance the angular momentum transport accelerating the  collapse of the HMNS.% remnant by inducing an effective viscosity.
As a consequence, in neutrino transport cases, the GW waveform is shorter in duration, though
its strain amplitude remains above the sensitivity curve of next generation GW observatories, such as
the Einstein Telescope (ET), between $1$~kHz and $5$~kHz, assuming a source distance of~$50 \, \rm Mpc$.
We note that the most prominent peak in the strain may be detectable by aLIGO and $A^+$ as well. We also find that neutrinos
do not have significant impact on the growth of the magnetic field, but they have the effect of clearing out
the polar region above the BH poles, inducing a lower baryon-load in surrounding debris. A magnetically-driven jet is
launched  after $\gtrsim 10~\rm ms$ following the collapse of the HMNS. However, the delay  time between
the peak GW (i.e. the binary merger) and the emergence of the jet is significantly shorter
in  neutrino radiation cases. The outgoing EM Poynting luminosity [$L_{\rm EM} \sim {10^{53}} \rm erg \, s^{-1}$]
is roughly consistent with sGRB models~\cite{Beniamini_2020, Bhat16, Lien16} and the luminosity associated with
the BZ mechanism~\cite{Blandford02}. 

A key motivation of our project has been to implement a neutrino transport scheme into our {\tt Illinois GRMHD} code, which has been used
successfully to treat compact binary mergers~(see e.g.~\cite{eflskb08, Etienne:2012te, gpesp14, Tsokaros:2019anx, Tsokaros:2019lnx, Ruiz20b},
including those involving magnetized neutron stars~\cite{lset08, Ruiz20}). While the M1 scheme we adopt is not the most advanced and the results we report here
are not likely the final answers, they are sufficient to generate a preliminary sketch of the combined influence of magnetic fields and neutrino
transport on compact binary mergers and other astrophysical scenarios in strong gravitational fields. 

The structure of the paper is summarized as follows. In Sec.~\ref{sec:method}, we introduce our implementation of M1 radiation transport and
the microphysical interactions responsible for thermal neutrino emission. A more detailed description and a strong-field test of our M1 scheme
are presented in Appendices~\ref{App:A} to~\ref{App:OS}. We next describe our adopted numerical setup and initial conditions in
Sec.~\ref{sec:Numsetup}. We discuss the effects of the magnetic field and thermal neutrino emission on the nature of the merger, properties
of the remnant and ejecta, and the strength and detectability of GW and EM signals in Sec.~\ref{sec:Evolution}. Finally, we summarize our findings
and identify some limitations of our simulations in Sec.\ref{sec:Conclusion}. Throughout the paper, we adopt geometrized units
$(G = c = 1)$ except where stated otherwise. Greek indices denote all four spacetime dimensions, while Latin indices imply spatial parts only.

%%%%%%%%%%%%%%%%%
%%%  Methods  %%%
%%%%%%%%%%%%%%%%%
\section{Formalism and Method}
\label{sec:method}

\subsection{\label{sec:2-0} Numerical Methods}
The simulations in this work are performed using the original {\tt Illinois GRMHD} code~\cite{els10}, coupled with our newly developed
neutrino radiation moment formalism and transport module.
The code is embedded in the \url{Cactus} infrastructure~\cite{web:Cactus} using \url{Carpet} as the moving-mesh
refinement scheme~\cite{web:Carpet}. The code evolves the Baumgarte–Shapiro–Shibata–Nakamura (BSSN) gravitational field equations~\cite{sn95,bs98b}
coupled to the moving puncture gauge conditions~\cite{Goddard1,Campanelli:2005dd}, using fourth-order centered spatial
differencing, except on shift advection terms, where a fourth-order upwind differencing is used. In all our evolution cases,
we set the damping coefficient $\eta_{\tiny{\beta}}$ appearing in the shift condition to $2.7/M$, where $M$ is the ADM mass
of the system. The BSSN evolution equations are embedded with fifth-order Kreiss-Oliger dissipation~\cite{Goddard2}.
We also adopt the generalized Lorenz gauge~\cite{epls2012, fgpes12} to evolve Maxwell’s equations, and employ a damping parameter
$\zeta = 4.1/M$ to avoid the rise of spurious magnetic fields due to interpolations across refinement levels (see~\cite{epls2012}
for details). Time integration is performed using the method of lines with a fourth-order Runge-Kutta
integration scheme with a Courant-Friedrichs-Lewy (CFL) factor of 0.5. In the absence of neutrinos, this code has been thoroughly tested and used
in the past in various scenarios involving compact binaries~(e.g. \cite{eflskb08, lset08, Etienne:2012te, gpesp14, Tsokaros:2019anx,Tsokaros:2019lnx,
  Ruiz20, Ruiz20b}). The incorporation of radiation in the optically thick limit was previously used and tested in~\cite{grrmhd}. For implementation details,
see~\cite{grrmhd, els10, elps2012, fgpes12}.

%%%%%%%%%%%%%%%%%%%%%%%%%%%
%%% Evolution Radiation %%%
%%%%%%%%%%%%%%%%%%%%%%%%%%%
\subsection{ Evolution of Radiation Fields}
\label{sec:2-1}

%%%%%%%%%%%%%%%%%%%%
%%% MI transport %%%
%%%%%%%%%%%%%%%%%%%%
\subsubsection{Radiation fields: M1 Transport}
We adopt and generalize the radiation transport approach described in~\cite{grrmhd}, where a basic radiation transport
scheme applicable for optically thick gases was embedded in the {\tt Illinois GRMHD} code.
The local properties of the radiation can be described by the specific intensity $I_\nu = I (x^\mu ; N^i ; \nu)$,
where $x^\mu$ is the coordinate position, $N^i$ is the direction of radiation, and $\nu$ is the frequency with $N^\mu = p^\mu / (h\nu)$,
with $p^\mu$ the four-momentum of the photon, and $h$ the Plank's constant. The quantities above are measured in the local Lorentz frame
of a fiducial observer. The evolution of $I_\nu$ can be described by the Boltzmann equation for radiation transport, which is a $6+1$-dimensional
integro-differential equation system whose numerical integration requires extremely large computational resources. To realize the numerical
evolution of radiation fields, several approximations need to be imposed. First, we adopt the moment formalism (see, e.g.,~\cite{Thorne81})
and consider the three lowest moments. We use the energy-(frequency-)integrated equations and the energy-averaged emissivities and opacities.
For an observer comoving with the fluid, the energy moments of interest are the radiation energy density,
\begin{equation}
    E = \int d\nu\,d\Omega\,I_\nu\,,
\end{equation}
the radiation flux,
\begin{equation}
    F^\alpha = h^\alpha_{~\gamma}\,\int d\nu\,d\Omega\,I_\nu\,N^\gamma\,,
\end{equation}
and the radiation stress tensor,
\begin{equation}
    \mathcal{P}^{\alpha\beta} = h^\alpha_{~\gamma}\,h^\beta_{~\delta}\,\int\,d\nu\,d\Omega\,I_\nu\,N^\gamma\,N^\delta\,,
\end{equation}
where $d\Omega$ is the solid angle measured in the local Lorentz frame, and $h^{\alpha\beta} = g^{\alpha\beta} +
u^\alpha u^\beta$ is the projection tensor onto the orthogonal slices of the fluid four-velocity $u^\alpha$.
We note that $F^\alpha$ is orthogonal to the fluid four-velocity.  With the
above expressions, one can decompose the radiation tensor in terms of an observer comoving with the fluid as
\begin{equation}\label{eq:Rmunu}
    R^{\alpha\beta} = Eu^\alpha u^\beta + F^\alpha u^\beta + F^\beta u^\alpha + \mathcal{P}^{\alpha\beta}.
\end{equation}
In the numerical evolution, the highest order radiation moments, $E$ and $F^i$, are used as the primitive variables for the radiation
(neutrino) fields. We then obtain $F^0$ using $F^\alpha u_\alpha = 0$, where $u^\alpha$ satisfies
$u^\alpha u_\alpha = -1$ {with $u^\mu = (u^0,0)$}. The radiation
stress tensor $\mathcal{P}^{\alpha\beta}$ is computed as a function of $E$ and $F^\alpha$ according to the Minerbo closure scheme~\cite{MINERBO1978541},
which is an interpolation between the analytical expressions in optically thick and thin limits (see Appendix~\ref{App:A}). At each timestep, the primitive
variables are computed from the conserved variables, which are related to quantities measured by normal observers (see below) using a 4D Newton-Rasphson
solver described in Appendix~\ref{App:B}. 
 
One can also decompose the radiation stress energy tensor according to the normal observer
\begin{equation}\label{eq:Rmunu2}
  R^{\alpha\beta} = \bar{E}n^\alpha n^\beta + \bar{F}^\alpha n^\beta +
  \bar{F}^\beta n^\alpha + \bar{\mathcal{P}}^{\alpha\beta}\,,
\end{equation}
where $n^\alpha$ is the unit vector normal to the constant $t$ slice. The barred quantities,
\begin{equation}\label{eq:Ebar}
    \bar{E} = n_\alpha n_\beta R^{\alpha \beta} = \alpha^2 R^{00}\,,
\end{equation}
\begin{equation}\label{eq:Fbar}
    \bar{F}^\alpha = -\gamma^\alpha_{~\rho} n_\beta R^{\beta \rho} = \alpha(R^{\alpha 0} + \beta^\alpha R^{00})\,,
\end{equation}
\begin{equation}\label{eq:Pbar}
\begin{aligned}
  \bar{\mathcal{P}}^{\alpha \beta} & = \gamma^\alpha_{~\rho} \gamma^\beta_{~\sigma} R^{\rho \sigma} = R^{\alpha\beta} \\ &
  - \alpha (n^\alpha R^{\beta 0} + n^\beta R^{\alpha 0}) + \alpha^2 n^\alpha n^\beta R^{00}\,,
\end{aligned}
\end{equation}
are the radiation energy density, flux, and stress tensor with respect to a normal observer, respectively.
Here $\gamma^{\alpha\beta} = g^{\alpha\beta} + n^\alpha n^\beta$ is the projection operator onto slices of constant $t$,
and $\alpha$ and $\beta^\mu$ are the lapse and shift functions, respectively. Note that Eqs.~(\ref{eq:Ebar})-(\ref{eq:Pbar})
can also be regarded as the purely normal, mixed normal-spatial, and purely spatial components of $R^{\alpha\beta}$,
respectively~\cite{bs20}. They are directly related to the conserved radiation variables for the radiation dynamical
equations (see Sec.~\ref{sec:rad_evo}). The conversion between the fluid-frame variables and the normal-frame variables
is given by~\cite{Cardall2013, bs20}:
\begin{equation}
    \bar{E} = W^2 E + 2 W v_\alpha  \bar F^\alpha + v_\alpha v_\beta \bar{\mathcal{P}}^{\alpha\beta}\,,
\end{equation}
\begin{equation}
\begin{aligned}
  \bar{F}_\alpha =& W^2 v_\alpha E + W (g_{\alpha\beta} - n_\alpha n_\beta) F^\beta \\
  +& W v_\alpha v_\beta F^\beta + W(g_{\alpha \beta} - n_\alpha n_\beta) v_{\gamma}
  \mathcal{P}^{\beta \gamma}\,,
\end{aligned}
\end{equation}
\begin{equation}
\begin{aligned}
  \bar{\mathcal{P}}_{\alpha\beta} =& W^2 v_\alpha v_\beta E +  W (g_{\alpha\gamma} -
  n_\alpha n_\gamma) v_{\beta}F^\gamma \\ +& W (g_{\gamma\beta} - n_\gamma n_\beta) v_\alpha F^\gamma\\
  +& (g_{\alpha\gamma} - n_\alpha n_\gamma)(g_{\beta\delta} - n_\beta n_\delta) \mathcal{P}^{\gamma\delta}\,,
\end{aligned}
\end{equation}
where the 4-velocity is decomposed into its orthogonal and tangent parts %to the spacelike slice
\begin{equation}
    u^\alpha = W (n^\alpha + v^\alpha)\,,
\end{equation}
with $v^\alpha = (0, u^i/W + \beta^i/\alpha)$, and $W \equiv - n_\alpha u^\alpha = \alpha u^0$ is the Lorentz-factor between
normal and fluid observers. Note that by construction, $v^\alpha$ is purely spatial, $n_\alpha v^\alpha = 0$.

%%%%%%%%%%%%%%%%%%%%%%%%%%%
%%% Evolution Radiation %%%
%%%%%%%%%%%%%%%%%%%%%%%%%%%
\subsubsection{Radiation Evolution}
\label{sec:rad_evo}

Following~\cite{grrmhd}, the dynamics of the radiation field can be expressed as
\begin{equation}\label{eq:delRmunu}
    {R^{\alpha\nu}}_{;\nu} = - G^\alpha\,,
\end{equation}
where $G^\alpha = \int d\nu\,d\Omega\,(\kappa_\nu I_\nu - \eta_\nu) N^\alpha$
is the the radiation four-force density which describes the interaction between the radiation field and matter.
In the energy-integrated moment formalism, the four-force terms can be written in the covariant form as
\begin{equation} \label{eq:Gmu2}
    G^\alpha = \rho_0 \left[ (\kappa^a E - \eta)u^\alpha + (\kappa^a +\kappa^s) F^\alpha  \right]\,,
\end{equation}
where $\rho_0$ is the rest mass density, $\eta$, $\kappa^a$, and $\kappa^s$ represent the energy-averaged emissivity, absorption
opacity, and scattering opacity, respectively. They are computed as functions of temperature, density, and the chemical potential
via the major microphysical interactions listed in~\ref{sec:2-4}. Their analytic expressions, derived from~\cite{RF1996}
and~\cite{Burrows2000}, are bounded by the opposite limits of local thermal equilibrium (LTE) and free streaming. In our
implementation, we interpolate the limiting values following the method of interpolating $\mathcal{P}^{\alpha\beta}$ to
accommodate regions with different optical depths (see Appendix~\ref{App:source_term} for a detailed derivation). Note that,
in LTE, $\eta$ can be written by invoking Kirchhoff's law, using the frequency-integrated neutrino thermal equilibrium intensity
$\bar{B}$: $\eta = \kappa^a
4\pi \bar{B} =\kappa^a\,a_R\,T^4$~\cite{grrmhd, bs20}, where $T$ is the temperature of the fluid, and $a_R$ is a constant
chosen based on the type of radiation. For photons, $a_R$ is the radiation constant $a$, and for neutrinos $a_R = 7\,a\,
\mathcal{N}_{\rm \nu}/16$, where $\mathcal{N}_\nu$ is the number of left-handed neutrino and right-handed anti-neutrino species contributing to
thermal processes. 

One can perform a normal projection of Eq.~(\ref{eq:delRmunu}) and obtain the energy equation
\begin{equation}\label{eq:taubar}
    \partial_t \bar{\tau} + \partial_i (\alpha^2 \sqrt{\gamma} R^{0i}) = \bar{s} - (\alpha^2 \sqrt{\gamma}) \, G^0\,,
\end{equation}
where $\bar{\tau} = (\alpha^2 \sqrt{\gamma}) R^{00} = \sqrt{\gamma} \bar{E}$ is the radiation energy density, and
$\bar{s}$ is given by
\begin{eqnarray}
  \bar{s} &= & \alpha\sqrt{\gamma} \left[ (R^{00}\beta^i\beta^j + R^{0i}\beta^j + R^{ij})K_{ij}  \right.\nonumber\\
    && \left. - (R^{00}\beta^i + R^{0i}) \partial_i \alpha \right]\nonumber \\
    &=& \sqrt{\gamma} \left(\alpha \bar{P}^{ij}K_{ij} - \bar{F}^i \partial_i \alpha \right)\,.
\end{eqnarray}
Similarly, the spatial projection of Eq.~(\ref{eq:delRmunu}) yields the momentum equations
(see~\cite{grrmhd,bs20}~for detailed derivations)
\begin{equation}
  \label{eq:Sbar}
  \partial_t \bar{S}_i + \partial_i (\alpha \sqrt{\gamma} R^i_{~j}) =  \alpha \sqrt{\gamma}\left(\frac{1}{2}
  R^{\alpha\beta}g_{\alpha\beta,i} - G_i\right)\,,
\end{equation}
where $\bar{S}_i = \alpha\sqrt{\gamma}R^0_{~i}$ are the radiation
energy density and momentum density, respectively. They form the conserved variables of the dynamical equations.
We note that, although described in different notations, Eqs.~(\ref{eq:taubar}) and~(\ref{eq:Sbar}) are equivalent
to the evolution equations in previous works (e.g. Eqs.~(3.39) and (3.40) in~\cite{ShibataM1}
and Eqs.~(16) and (17) in \cite{Foucart_M1_2015}) in conservative form. 

We use the Harten-Lax-van Leer (HLL) approximate Riemann solver to evolve Eqs.~(\ref{eq:taubar})-(\ref{eq:Sbar}),
in which we adopt the characteristic speed of radiation field following~\cite{Foucart_M1_2015} for the HLL fluxes
at cell interfaces. Note that Eqs.~(\ref{eq:taubar})-(\ref{eq:Sbar}) require the computation of $\mathcal{P}^{\alpha
  \beta}(E, F^\alpha)$ and source terms $G^\alpha$ for closure. We adopt the prescription presented in~\cite{MINERBO1978541}
regarding the radiation pressure $\mathcal{P}^{\alpha\beta} (E, F^\alpha)$, which is an interpolation of its optically
thin and thick limits. Since the closure equation for $\mathcal{P}^{\alpha\beta}$ is a nonlinear function of $E$
and $F^\alpha$, the recovery of primitive variables requires a 4-dimensional solver. A detailed description of the
closure scheme and the recovery of the primitive variables are summarized in Appendices~\ref{App:A} and \ref{App:B},
respectively. For the source terms, we adopted two versions: i) the ``warm-up'' Simplified Scheme (Rad-Simp),
which assumes a purely neutron-rich environment, where the net electron fraction $Y_e$ is zero, and accounts for only the
charged-current interaction of electron anti-neutrinos; and ii) the Full Scheme (Rad-Full), which includes three species of
neutrinos: ${\nu}_e$, $\bar{\nu}_e$, and ${\nu}_{\rm x}$, where ${\nu}_{\rm x}$ is the sum of muon- and tau- neutrinos
and their antiparticles, called the ``heavy-lepton neutrinos" (Rad-Full). This scheme includes all
the important interactions whose reaction rates are based on the evolved electron fraction $Y_e$ using Eq.~(\ref{Eqn:Ye_evolve}),
and the temperature. A description of the source terms and the two schemes is summarized in Sec.~\ref{sec:2-4} (see
also Table~\ref{Tab:Scheme}). 

%%%%%%%%%%%%%%%%%%%%
%%% MHD evoltion %%%
%%%%%%%%%%%%%%%%%%%%
\subsection{Evolution of Metric and MHD Fields}
\label{sec:2-2}
As in our previous magnetized NSNS studies, the {\tt Illinois GRMHD} code solves the equations of ideal GRMHD
in a conservative scheme via high-resolution shock capturing methods for the evolution of matter and magnetic field.
The equations for metric, matter, and magnetic field, including radiation source terms, are given in~\cite{grrmhd}.
Additionally, to compute $Y_e$ used in the radiation source terms, we now evolve the quantity $\rho_* Y_e$
following~\cite{ShibataM1, Foucart_M1_2015} 
\begin{eqnarray}
  \label{Eqn:Ye_evolve}
  \partial_t (\rho_* Y_e) & + & \partial_j \left[(\rho_*Y_e) v^j \right]  =\\ &
  -&{\rm  sgn}(\nu_i) \, \alpha\sqrt{\gamma} \, m_N \, \left(\frac{\eta - \kappa_a E}{\langle \epsilon_\nu \rangle}\right)\,,
  \nonumber
\end{eqnarray}
where
\begin{equation*}
  {\rm  sgn}(\nu_i)=\left\{
  \begin{array}{rl}
    1 & \rm{for}\,\nu_e,\\
   -1 & \rm{for}\,\bar{\nu}_e,\\
    0 & \rm{for}\,{\nu}_x,
  \end{array}
  \right.
\end{equation*}
$m_N$ is the nucleon rest mass, $\langle \epsilon_\nu \rangle$ is the local mean energy of neutrinos,
which is approximated as $F_5(\eta_\nu)/F_4(\eta_\nu)\,T$, with $F_k (\eta_\nu)$ the Fermi integral $F_k
(\eta_\nu) = \int^\infty_0 x^k/(1 + \rm exp (x - \eta_\nu)) \, dx$, and $\eta_\nu = \mu_\nu/(k_B T)$,
with $\mu_\nu$ the chemical potential evaluated using the analytic expressions detailed in
Appendix~\ref{subapp:ACP}. We note that the Fermi integrals can be efficiently evaluated using the polynomial
approximations in~\cite{FermiIntegral78}.

We evolve the MHD equations using the the standard HLL, approximate Riemann solver with a simpler, dissipative
dispersion relation~(see Eq.~50 in~\cite{dlss05a}). Required by the dispersion relation, the sound speed
$c^2_{\rm s} = (d P/ d \rho_0)_{\rm s}/h$ consists of a cold and a thermal component of pressure (see Sec.~\ref{sec:2-3}),
where $h$ is the specific enthalpy and $\rm s$ is the specific entropy. The thermal part is based on an analytic expression for a semi-degenerate
neutron gas of non-zero temperature. The analytic derivation and our implementation of the sound speed is described
in detail in Appendix~\ref{subapp:soundspeed}. Notice that, as in standard hydrodynamic and MHD simulations, we integrate
the ideal GRMHD equations  on the whole numerical grid by imposing a tenuous constant–density atmosphere
$\rho_{0,\rm atm} = 10^{-10}\,\rho_{0,\rm max}(0)$, where $\rho_{0,\rm max}(0)$ is the initial maximum value
of the rest-mass of the system.

%%%%%%%%%%%%%%%%%%%%%%
%%% Sources terms  %%%
%%%%%%%%%%%%%%%%%%%%%%%
\subsection{EOS, Neutrino Processes and Source Terms}
\label{sec:2-3}
\subsubsection{Finite Temperature Treatment}
Our NSNS binaries are  modeled by a piecewise polytropic representation of the SLy nuclear EOS~\cite{Douchin01}.
A cold EOS is adequate to model the NS prior to merger. However, during merger,
considerable shock heating increases the internal energy. To account for this, we adopt an EOS that has both a
thermal and cold contribution to the total energy density, $\mathcal{E}=\mathcal{E}_{\text{cold}} +  \mathcal{E}_{\text{th}}$,
and total pressure, $P=P_{\text{cold}} +  P_{\text{th}}$. Specifically,
$P_{\rm cold}=\kappa_{i}\,\rho_0^{\Gamma_i}$, with $\kappa_i$ and $\Gamma_i$
the corresponding polytropic constant and the polytropic exponent in the rest-mass density range
$\rho_{0,i-1}\leq\rho_0 \leq \rho_{0,i}$, respectively, fitted for SLy~\cite{Read:2008iy}. Note that the cold energy density is given by $\mathcal{E}_{\text{cold}} =
{P}_{\text{cold}}/(\Gamma_i-1)$. The thermal terms are the sum of nucleon and radiation (other than neutrino)
contributions: 1) $\mathcal{E}_{\text{th}} = \mathcal{E}_{\text{rad}}+\mathcal{E}_{\text{nucl}}$; and 2)
$P_{\text{th}} = P_{\text{rad}}+P_{\text{nucl}}$. The nucleon parts are based on a semi-degenerate neutron gas
expression that asymptotes to a Maxwell-Boltzmann gas in the non-degenerate limit.  The radiation part includes
photons and relativistic electron and positron pairs, which follow a Stefan–Boltzmann relation.
Details of treatments of thermal EOS are summarized in Appendix~\ref{subapp:TEP}.

To implement the temperature-dependent thermal energy density and pressure in our simulations, we set our evolved
value of $\mathcal{E}$ to $\mathcal{E}_{\rm cold} + \mathcal{E}_{\rm th}$ using Eq.~(\ref{eq:E_th}) and solve for
the temperature $T$. After obtaining
$T$, we calculate the thermal pressure $P_{\rm th}$ base on Eq.~(\ref{eq:P_th}) and then obtain the total pressure
${P} = P_{\rm th} + P_{\rm cold}$ for the next iteration. This method provides a physically reasonable estimate of
temperature. Note that such formalism requires modifications in the MHD primitive solver for the computation of
the Jacobian. The details of finding the temperature, as well as the modification of the primitives solver, are
summarized in Appendix~\ref{subapp:NIPSM}. See also Appendix~\ref{subapp:ACP} for a detailed description of
this treatment of thermal energy density and pressure.

%%%%%%%%%%%%%%%%%%%%%%%%%%%%%%%%%
%%%  Opacity and Emissivity   %%%
%%%%%%%%%%%%%%%%%%%%%%%%%%%%%%%%%
\subsubsection{Opacity and Emissivity}
\label{sec:2-4}
As shown in Eq.~(\ref{eq:Gmu2}), the source terms $G^\alpha$ are determined by the radiation energy density $E$,
energy flux $F^\mu$, absorption opacity $\kappa^a$, scattering opacity $\kappa^s$, and emissivity $\eta$. The relevant
interactions are summarized in Table~\ref{Tab:Scheme}. For neutrino emission they include the charged-current interactions

\begin{eqnarray}
  e^- &+& p \rightarrow n + \nu_e\,,
    \label{eq:charged_e-}\\
  e^+ &+& n \rightarrow p + \bar{\nu}_e \,,
    \label{eq:charged_e+}
\end{eqnarray}
and three pair process interactions: electron-positron pair annihilation
\begin{equation}\label{eq:pair_annihilation}
    e^+ + e^- \rightarrow \nu_i + \bar{\nu}_i\,,
\end{equation}
plasmon decay,
\begin{equation}\label{eq:plasmon_decay}
    \gamma + \gamma \rightarrow \nu_i + \bar{\nu}_i\,,
\end{equation}
and nucleon-nucleon bremsstrahlung,
\begin{equation}\label{eq:Bremsstrahlung}
    N + N \rightarrow  N + N + \nu_i + \bar{\nu}_i\,.
\end{equation}
Here the subscript ``$i$" represents all neutrino species, and $N$ stands for nucleons. We can safely omit neutron decay
($ n \rightarrow p+ e^- + \bar{\nu}_e $) from consideration here. The reason is that even when the leptons in some regions
become sufficiently hot to lift their degeneracy and partially remove the blocking factor that inhibits the decay at $T=0$,
the decay time ($\gtrsim 14$~min) greatly exceeds the time it takes the HMNS remnant to collapse to a BH, as well
as the time for which we follow the evolution in our simulation.

The inverse reaction of the interactions above are responsible for neutrino absorption. Therefore, we need to compute both $\kappa^a$
and $\eta$ for Eqs.~(\ref{eq:charged_e-})-(\ref{eq:Bremsstrahlung}). Lastly, we also consider the neutral-current
neutrino scattering off nucleons,
\begin{equation}
  \label{eq:scattering}
  \nu_i + N \rightarrow \nu_i + N\,.
\end{equation}
Note that we neglect electron-neutrino/antineutrino scattering for simplicity, noting that its contribution to the opacity is smaller
than that for nucleon scattering by a typical factor of $\sim 100$ in the HMNS remnant (see also~\cite{Fischer2016}). We
do note for future work that electron scattering is inelastic and contributes to the thermalization of neutrinos
\cite{Arnett1977,Tubbs1980}, while nucleon scattering is conservative. We also note that for neutrino energies of $\sim$10 MeV,
cross sections of the charged–current absorptions of electron-type neutrinos are $\sim 100$ times greater than that of the
electron-neutrino scattering~\cite{Burrows2000}. 

Following~\cite{Foucart_M1_2015}, at every iteration we first compute $\kappa^a$ for the inverse of Eqs.~(\ref{eq:charged_e-})
and~(\ref{eq:charged_e+}) (charged-current interactions), $\kappa^s$ for Eq.~(\ref{eq:scattering}), and $\eta$ for
Eqs.~(\ref{eq:pair_annihilation}) and~(\ref{eq:plasmon_decay}) (thermal emission) based on the formalism in~\cite{RF1996}. We then
compute $\eta$ for Eqs.~(\ref{eq:Bremsstrahlung}) using Eq. (49) in~\cite{Burrows2000}. These expressions require
the knowledge of the fluid temperature and the chemical potential, which are computed analytically. 

Next, using either $\kappa_a$ or $\eta$ computed above, we calculate $\eta$ for the two charged-current interactions (Eqs.~(\ref{eq:charged_e-}) and (\ref{eq:charged_e+}) ) and $\kappa_a$ for the three thermal pair processes. In the optically thick region, we impose the LTE condition
and use the energy-integrated version of Kirchhoff’s law
\begin{equation}
\begin{aligned}
    &\eta_{\tiny\rm Kir} = 4\,\pi\,\bar{B}\,\kappa^a\,, \hspace{3mm} (\mbox{charged-current})\,,\\
    &\kappa^a = \eta/4\pi \bar{B}\,, \hspace{3mm} (\rm pair~processes),
\end{aligned}
\end{equation}
where the energy-integrated equilibrium intensity $\bar{B}(T)$ is defined in Sec.~\ref{sec:rad_evo}.
This method maintains the thermal equilibrium of neutrinos with the fluid in the optically thick region.
In the optically thin region, as the neutrinos can be out of thermal equilibrium, $\eta_{\rm Kir}$ may be
inaccurate in estimating the emission of electron-type neutrinos. Therefore, we adopt an expression for the
total emissivity by interpolating between $\eta_{\rm Kir}$ and $\eta_{\rm fs}$, where $\eta_{\rm fs}$ is the expression for the
free-streaming emission rate for charged-current interaction in~\cite{RF1996}. Here, we adopt an interpolation
method similar to the M1 formalism (see Eq.~(\ref{eq:eta_tot})), which estimates the emissivity
based on optical thickness. Note that this is different from the interpolation methods in~\cite{Foucart_M1_2015},
which adopts a step-function in terms of the optical depth.

%%%%%%%%%%%%%%%%%%%%%%%%%%%
%%% Table microphysics  %%%
%%%%%%%%%%%%%%%%%%%%%%%%%%%
%
\begin{table}
\centering
 \begin{tabular}{c| c| c c} 
  \hline
 \hline
    \textbf{\textsf{Microphysics Scheme}}    & \textbf{\textsf{Simplified}} & \textbf{\textsf{Full}} & \\ [0.5ex] 
 \hline
 \hline 
 $Y_e$ & 0 & Evolved &  \\ 
 \hline
 $Y_p$   & 0 & $Y_e$ &  \\
 \hline
 $Y_n$   & 1 & $1-Y_e$ &  \\
 \hline
  \textbf{Emission:} &&& \\
  $e^+ + n \rightarrow p + \bar{\nu}_e$ & \cmark & \cmark & \\ 
  $e^- + p \rightarrow n + \nu_e$ & \xmark & \cmark & \\ 
 $e^- + e^+ \rightarrow \nu_i + \bar{\nu}_i$ & \xmark & \cmark & \\
 $\gamma \rightarrow \nu_i + \bar{\nu}_i$ & \xmark & \cmark & \\
   $N + N \rightarrow N + N + \nu_i + \bar{\nu}_i$ & \xmark & \cmark & \\
   \hline
   \textbf{Absorption:} &&& \\
   $p + \bar{\nu}_e \rightarrow e^+ + n$& \cmark & \cmark & \\
   $n + \nu_e \rightarrow e^- + p$ & \xmark & \cmark & \\
   $\nu_i + \bar{\nu}_i  \rightarrow e^- + e^+ $ & \xmark & \cmark & \\
    $\nu_i + \bar{\nu}_i \rightarrow \gamma $ & \xmark & \cmark & \\
	$N + N + \nu_i + \bar{\nu}_i \rightarrow N + N $ & \xmark & \cmark & \\
 \hline
   \textbf{Scattering:} &&& \\
   $\bar{\nu}_e + n \rightarrow \bar{\nu}_e + n$& \cmark & \cmark & \\
	$\{\nu_e, \nu_i\} + n \rightarrow \{\nu_e, \nu_i\} + n $ & \xmark & \cmark & \\   
	$\{\nu_e, \bar{\nu}_e,\nu_i\} + p \rightarrow \{\nu_e, \bar{\nu}_e,\nu_i\}  + p$& \xmark & \cmark & \\
\hline
\hline
\end{tabular}
\caption{\label{Tab:Scheme} Comparison of the two microphysics versions employed here.}
\end{table}
%
%%%%%%%%%%%%%%%%%%%%%%%%%%%
%%%  Numerical Setup    %%%
%%%%%%%%%%%%%%%%%%%%%%%%%%%

\section{Numerical Setup}
\label{sec:Numsetup}
\subsection{Initial Condition}
\label{sec:3-2} 
We consider initial NSNS configurations  in a quasiequilibrium circular orbit computed using our
Compact Object CALculator~({\tt COCAL}) code~\cite{Tsokaros:2015fea,Tsokaros:2018dqs}.
The binaries consist of two identical, irrotational NSs, modeled by the SLy nuclear EOS~\cite{Douchin01}, as
in~\cite{Read:2008iy}. Specifically, we use the SLyM2.7 (magnetized and
unmagnetized, hereafter denoted as ``Mag" and ``Unmag", respectively) configurations treated  previously
in~\cite{Ruiz2021}, for which the ADM mass of the system is $M_{\rm ADM} = M = 2.7M_\odot$, and the companions have an initial
coordinate separation of $45\,\rm km$ (see Table I in~\cite{Ruiz2021}).
We define $M_{0}=3.02M_\odot$ as the total rest-mass of the system.
We note that the SLy EOS
satisfies current observational constraints on NSs. The maximum mass configuration
  of an  isolated star predicted by SLy is $M_{\rm sph}^{\rm max}= 2.06M_\odot$ consistent with:
  i) $M_{\rm sph}^{\rm max}>2.072^{+0.067}_{-0.066}M_\odot$ from the NICER and XMM
  analysis of PSR J0740+6620~\cite{Riley:2021pdl};
  ii) $M_{\rm sph}^{\rm max}>2.01^{+0.017}_{-0.017}M_\odot$ from  the NANOGrav analysis of PSR J1614-2230
  \cite{Fonseca:2016tux};
  iii) $M_{\rm sph}^{\rm max}>2.01^{+0.14}_{-0.14}M_\odot$ from the pulsar timing analysis of PSR J0348+0432
  \cite{Antoniadis:2013pzd}; and $M_{\rm sph}^{\rm max}>2.14^{+0.20}_{-0.18}M_\odot$ from the NANOGrav and
  the Green Bank Telescope~\cite{NANOGrav:2019jur}.
  In addition, SLy predicts that a star with a mass of $1.4M_\odot$ has a radius of $R= 11.46\,\rm km$,
  and a tidal deformability of $\Lambda_{1.4}=306.4$. These values are consistent with:
  i) $R = 11.94^{+0.76}_{-0.87}\rm\, km$ obtained by a combined analysis of X-ray and
  GW measurements of PSR J0740+6620~\cite{Pang:2021jta};
  ii) $R= 11.9^{+1.4}_{-1.4}\,\rm km$  for a NS with mass in the range  $1.16-1.6M_\odot$ at the  $90\%$
  credible level from the combined analysis of the LIGO/Virgo scientific collaboration (LVSC) of the progenitors
  of GW170817 with the radio-timing observations of the pulsar J0348+0432
  \cite{LIGOScientific:2018cki,Antoniadis:2013pzd}; and 
  iii) $\Lambda_{1.4}= 190^{+390}_{-120}$  for a NS with mass of $1.4\,M_\odot$ at the $90\%$ credible level from
  the LVSC analysis of GW170817~\cite{LIGOScientific:2018cki}.

In the magnetized cases, the NSs are initially endowed with a dipole-like magnetic field generated by the vector potential~\cite{Ruiz:2017inq}
\begin{equation}
    A_{\phi} = \frac{\pi r^2_0 I_0 \varpi^2}{(r^2_0 + r^2)^{3/2}} \left[1 + \frac{15 r^2_0 (r^2_0 + \varpi^2)}{8 (r^2_0 + r^2)^2}\right],
\end{equation}
where $r_0$ and $I_0$ are the loop radius and current, which determine the geometry and strength of the magnetic field.
$r^2 = \varpi^2 + z^2$, $\varpi_{\rm cm}^2 = (x - x_{\rm cm})^2 + (y - y_{\rm cm})^2$, where $(x_{\rm cm}, y_{\rm cm}, 0)$
is the center of mass the NS, determined by the coordinate of the maximum value of the rest mass density. We set $I_0$ and $r_0$
such that the maximum value of the magnetic-to-gas-pressure ratio in the NS interior is $\beta_{\rm ratio}^{-1} \equiv P_{\rm mag}/P_{\rm gas}
= 0.003125$. The resulting initial magnetic field strength at the NS pole measured by a normal observer is $B_{\rm pole}\simeq 10^{15.3}\rm G$.
As in~\cite{Ruiz:2018wah}, we initially impose a low, variable atmospheric density $\rho_{0,\rm atm}$ that satisfies $\beta^{-1}_{\rm ratio}
= 100$ (see~Eq.~(4)~in~\cite{Ruiz:2018wah}).
This setup mimics the magnetically dominant environment in a pulsar-like magnetosphere and validates the evolution of the exterior magnetic field.
While the magnetic field we imposed is astrophysically large, we choose it so that following merger, the rms value of the field
strength in the HMNS remnant is close to the values reported in very-high-resolution (with a local resolution of $\Delta x = 17.5 \,\rm m$)
simulations~\cite{Kiuchi:2015sga}, which showed that the KHI during merger can boost the rms B-field to $10^{15.5} \rm G$, with local values
reaching $10^{17} \rm G$. Moreover, recent NSNS simulations~\cite{Aguilera-Miret20} with a local numerical resolution of $\Delta x = 37
\,\rm m$ have also shown that, a pure poloidal magnetic field with a strength of $\sim 10^{11} \, \rm G$ in the interior of NSs can be amplified
to rms values of $\sim 10^{16} \, \rm G$ within the first $\sim 5 \, \rm ms$ after merger.

%%%%%%%%%%%%%%%%%%%
%%%  Neutrinos  %%%
%%%%%%%%%%%%%%%%%%%
\subsection{Neutrino Insertion}
\label{sec:3-3}
To probe the effects of neutrino emission in our binary remnant, in all our cases we insert the neutrino radiation at
$t_{\rm insert}\sim 1350M \approx 18~\rm ms$ (or $t-t_{\rm mer}\sim 226M\sim 3\,\rm ms$ after merger). The initial
neutrino energy is based on the local fluid temperature, which is related by an analytic expression to the thermal
energy density of fluid matter (see Sec.~\ref{sec:2-3}). In the low-density region where $\rho_0 = \rho_{\rm atm}$,
we set $T = T_{\rm atm} = 10^3 \,K$, which is $\sim 7$ orders of magnitudes lower than the temperature in
the central region of HMNS, and $\sim 6$ orders of magnitude lower than the treatment in~\cite{Foucart_M1_2015}. 

In a hot NS in LTE, the energy density of each chirality state of ultra-relativistic neutrinos is
\begin{equation}
    \mathcal{E}_{\nu, \rm state} = \frac{7}{16}\,a\,T^4\,,
\end{equation}
where $a$ is the usual radiation constant. Let $\mathcal{N}_{\nu}$ be the number of neutrino species
involved. Then $\mathcal{N}_{\nu}$ has the maximum value of 6 when all the possible species of the Standard Model are
considered (i.e.~$\nu_{e}$, $\bar{\nu}_{e}$, $\nu_{\mu}$ and their antiparticles). If we assume that only the left-handed
neutrino states couple to the weak interactions, then the energy density of each neutrino (and anti-neutrino) species $i$
becomes
\begin{equation}
    \mathcal{E}_{\nu, i} = \frac{7}{16} \mathcal{N}_{\nu}\,a\,T^4\,.
\end{equation}
In Rad-Simp, we set $\mathcal{N}_{\nu} = 1$ for $\bar{\nu}_e$, while in Rad-Full we set $\mathcal{N}_{\nu} = 1$
for both $\nu_e$ and $\bar{\nu}_e$, and $\mathcal{N}_{\nu} = 4$ for $\nu_{\rm x}$. Note that the neutrino helicity may be
flipped if the magnetic moment of the neutrinos is considered~\cite{Fujikawa1980}. It is believed that the left-handed
neutrino rotates to the right-handed one when the magnetic fields are transverse to neutrino propagation~\cite{Lim1988}.
Other effects such as nonequilibrium quantum kinematics and coherent flavor evolution of neutrinos are also not captured
in a thermal description~\cite{NQK14}. Here we ignore these effects.  

At $t_{\rm insert}$, we assign the radiation primitive variables $E = \mathcal{E}_{\nu, i}$ for each species
of neutrinos evolved and set the flux $F^\alpha = 0$. This method can well describe the interior of the HMNS system where the density
and temperatures are sufficiently high. In the outer regions of the merger remnant, matter may not have attained thermal
equilibrium. However, as $\mathcal{E}_{\nu, i}$ scales as $T^4$, its value in the atmosphere becomes negligibly
small compared with the interior value. After setting the primitive variables, we compute the conserved variables $\bar{\tau}$
and $\bar{S}_i$, and the source terms $G^\alpha$ for radiation evolution. 

%%%%%%%%%%%%%%%%%%%%%%%%%%%%
%%%  Neutrinos evolution %%%
%%%%%%%%%%%%%%%%%%%%%%%%%%%%
\subsection{Neutrino Evolution}
\label{sec:3-4} 
After initializing $E$ and $F^\alpha$, we compute the conserved variables $(\bar{\tau}, \bar{S}_i)$ and reconstruct the primitive
variables to the left and right of the grid cell surfaces using the monotonized central (MC) scheme~\cite{vL77}. Then we compute
the conserved quantities at the cell faces, together with the left- and right-going wave speed using the method described in
\cite{Foucart_M1_2015} to obtain the flux quantities $\alpha^2\,\sqrt{\gamma}\,R^{0i}$ and $\alpha\,\sqrt{\gamma} R^{i}_{\,j}$. Next,
we compute the spatial derivative terms in Eqs.~(\ref{eq:taubar}) and~(\ref{eq:Sbar}). After this, we compute the source terms
$\bar{s}$ in Eq.~(\ref{eq:taubar}), and the term $R^{\alpha\beta}g_{\alpha\beta,i}$ in Eq.~(\ref{eq:Sbar}). With these values, we
compute the opacities and emissivisies using the method described in Appendix~\ref{App:source_term}, which depends on the matter
density and temperature. To avoid overly stiff source terms, we set the maximum value of frequency-integrated opacity at $10 \,
\rm km^{-1}$, which could yield a sufficiently high optical depth ($\gtrsim 20$) to model the optically thick environment inside the high-density
region. With this prescription, we can evolve the radiation field using explicit time-stepping at a modest resolution.
Finally, we compute the four-force terms which are required to evolve the radiation and MHD variables. 

With the conserved quantities $(\bar{\tau}, \bar{S}^i)$ computed at a given time step, we now need to recover the primitive variables
$(E, F^i)$. Due to the non-linearity in the closure scheme, a numerical root finder is needed. After the fluid velocity $v^i$ is found
by the MHD primitive solver, we adopt a 4D Newton-Raphson solver that computes the radiation primitive variables from the radiation
evolution variables. The solver requires an $4 \times 4$ Jacobian matrix which governs the coefficient of the first order terms. Line
searches and backtracking methods are adopted for faster convergence~\cite{1992nrcabookP}. The elements of the Jacobian can be
evaluated analytically. Detailed descriptions, including the relevant equations, the expression for the Jacobian matrix, and other
numerical settings are given in Appendix~\ref{App:B}. 
%
%%%%%%%%%%%%%%%%%
%%%  Table II %%%
%%%%%%%%%%%%%%%%%
%
\begin{table*}[]
\begin{minipage}{\textwidth}
\centering
\caption{\label{table:Tab2} Key simulation parameters. Here, $\tau_{\rm HMNS} [\rm ms]$ is the lifetime of the HMNS,
  $t_{\rm BH} [\rm ms]$ is the BH formation time. The mass and the dimensionless spin parameter of the BH remnant are given by
  $M_{\rm BH}\,[M_\odot]$ and $\tilde{a}=a_{\rm BH}/M_{\rm BH}$, respectively.
  $\Delta \bar{E}_{\rm GW} \equiv E_{\rm GW}/M_{\rm ADM}$ and $\Delta \bar{J}_{\rm GW} \equiv J_{\rm GW}/J_{\rm ADM}$ denote the
  fraction of energy and angular momentum carried off by gravitational radiation.  $\dot{M}\,[M_\odot/\rm s]$ is the rest-mass
  accretion rate, $M_{\rm disk}\,[M_\odot]$ denotes the rest mass of the accretion disk
  once it begins settle into a steady state ($t-t_{\rm BH}\sim 450M\sim 6\rm\,ms$).
  $M_{\rm esc}/M_0$ denotes the rest-mass fraction of escaping matter following the peak amplitude of GWs. $B_{\rm rms}\,[G]$ is the rms
  value of the magnetic field at the HMNS pole just before collapse. $L_{\rm EM}\,[\rm erg\, s^{-1}]$ is the Poynting luminosity
  driven by the jet, which is averaged over the last $\sim 5\, \rm ms$ before the termination of our simulations. $\alpha_{\rm SS}$
  is the effective Shakura--Sunyaev magnetic viscosity parameter. $L_{\rm knova}\,[\rm erg\, s^{-1}]$ and $\tau_{\rm peak}\,[\rm days]$ are the
  peak luminosity and the rise time of the kilonova projected from our cases, respectively. $L_{\nu}\,[\rm erg \, s^{-1}]$ is the
  total neutrino luminosity and $\Delta \bar{J}_{\nu} \equiv J_{\nu} / J_{\rm ADM}$ is the fraction of angular momentum carried-off by neutrinos (at late times).
  A dash symbol denotes ``not applicable''.
}
 \scalebox{0.88}{
\begin{tabular}{ c|cccccccccccccccccc}
 \hline
 \hline
 Model & $\tau_{\rm HMNS}$ & $t_{\rm BH}$ & $M_{\rm BH}$ & $\tilde{a}$ & $\Delta \bar{E}_{GW}$& $\Delta \bar{J}_{GW}$ & $\dot{M}$& $M_{\rm disk}/M_0^{(\dag)}$ &
 $M_{\rm esc}/M_0$& $B_{\rm rms}$& $L_{\rm EM}$ & $\alpha_{\rm SS}$ & $L_{\rm knova}$ & $\tau_{\rm peak}$ & $L_{\nu}$ & $ \Delta \bar{J}_{\nu}$  \\
 &&&&&$\times 10^{-2}$&$\times 10^{-1}$&&$\times 10^{-2}$&$\times 10^{-2}$&&&&&&&$\times 10^{-4}$ &\\
 \hline

Unmag$^{(\ddag)}$&  8.43  & 23.1 & 2.55  & 0.70 & $3.23$ & 3.31 & 4.07 & $2.24$ & 0.13 & -    &  -         & -           & $10^{41.5}$ & 0.5  & -           &-\\

Unmag +Rad-Simp & 7.55   & 22.0 & 2.59  & 0.71 & $3.09$ & 3.24 & 4.00 & $1.89$ & 0.11 & -    &  -         & -           & $10^{41.1}$ & 0.14        & $10^{52.0}$ & 0.7\\

  \hline
  Mag$^{(\ddag)}$&9.36    & 24.5 & 2.45  & 0.62 & 2.63   & 2.95 &  2.91 & $6.16$ & 0.89 & $10^{16.3}$ & $10^{52.8}$ & 0.02-0.07   & $10^{41.5}$     & 0.47         &  - &-\\

Mag+Rad-Simp      &4.11    & 19.2 & 2.55  & 0.69 & $2.40$ & 2.75 &  1.81 & $3.33$ & $0.58$& $10^{16.1}$ & $10^{52.9}$ &  0.01-0.08  &$10^{41.5}$      &0.41          &$10^{52.6}$ &5.5\\

Mag+Rad-Full      &5.24    & 20.7 & 2.51  & 0.68  &$1.55$ & 2.88 &  2.15 & $4.58$ & $0.95$ & $10^{16.0}$& $10^{52.8}$ & 0.02-0.09   &$10^{41.6}$      &0.46          &$10^{53.1}$ &44.0\\
  \hline
  \hline
\end{tabular}
}
\end{minipage}
\begin{flushleft}
    $^{(\dag)}$ $M_0$ denotes the initial total rest mass of the system.\\
  {$^{(\ddag)}$ Cases we treated previously in}~\cite{Ruiz2021}.
\end{flushleft}
\end{table*}

%%%%%%%%%%%%%%%
%%%  Grid   %%%
%%%%%%%%%%%%%%%
\subsection{Grid Setup}
\label{sec:3-5}

In all our simulations, we use the same grid setup as in~\cite{Ruiz2021}, which consists
of two sets of nested boxes with nine refinement levels centered on each star, with adjacent levels
differing in size and resolution by factors of two. When two boxes overlap, they are replaced by a common
box centered on the center of mass of the system. The half length of each grid level is $(2835.26/ 2^{n-1})~\rm km $,
where $n$ is the level number. The innermost refinement level has a grid spacing of $\sim 0.027M\sim 111~\rm m$.
The number of grid points across the initial NS equatorial radius is $N_{\rm NS} = 82$. In all our cases we use
reflection (equatorial) symmetry over the orbital plane. As reported in~\cite{Ruiz20}, orbital plane symmetry does
not impact the final outcome of the evolutions. The detailed grid hierarchy is summarized in Table II in~\cite{Ruiz2021}.

%%%%%%%%%%%%%%%%%%%%%%
%%%  Diagnostics   %%%
%%%%%%%%%%%%%%%%%%%%%%
\subsection{Diagnostics}
\label{sec:3-6}
To verify the reliability of our numerical results, we monitored the $L_2$ normalized constraints computed from
Eqs.~(40)-(41) in~\cite{eflskb08}. In all our simulations, we find that during the inspiral the Hamiltonian constraint violation remains below
$0.2\%$ and oscillates between $0.2\%$ and $0.4\%$ during the HMNS phase. Then it peaks at $0.8\%$ at BH formation, and then
gradually approaches to $\lesssim 0.1\%$. The normalized momentum constraint violation remains smaller than $0.4\%$ during
the inspiral and peaks at $\sim 5.5\%$ at BH formation and gradually relaxes to $\sim 0.3\%$ following the onset of steady state.
{Similar values were reported in our long-term, simulations of spinning NSNS modeled by piecewise EOSs~\cite{Tsokaros:2019anx}.}

After BH formation, we locate and track the position of the BH apparent horizon using the \url{AHFinderDirect} thorn~\cite{Thornburg_2003}.
The BH mass $M_{\rm BH}$, and its dimensionless spin parameter $\tilde{a}=a/M_{\rm BH}$ are calculated using the isolated horizon
formalism~\cite{isolatedhorizon}. We adopt a modified version of the \url{Psikadelia} thorn to compute the Weyl scalar $\Psi_4$ decomposed into
$s = -2$ spin-weighted spherical harmonics at different radii between $50M\sim 200\rm\,km$ and $300M=1200\,\rm km$~\cite{Ruiz08}. We compute
the rest mass accretion rate $\dot{M}$
following Eq.~(A11) in~\cite{farris09}.  The mass of escaping matter (ejecta) is computed as $M_{\rm esc} = -\int_{r > r_0}\sqrt{\gamma}
\rho_0\,n_\mu\,u^\mu\,d^3x\,$, where the integral is restricted by the two conditions: i) $-1-u_0 > 0$, where $u^0$ is the time component of
4-velocity, and ii) having positive radial velocity $v^r > 0$. $M_{\rm esc}$ is computed at various coordinate radii $r_0$ from $30M\sim 120\,\rm km$
to $100M\sim 400\rm\,km$ in order to verify that it is $r_0$-independent. We compute the rate of escaping mass $\dot{M}_{\rm esc} =
-\int v^r \sqrt{\gamma}\rho_0\,n_\mu \,u^\mu\,d\mathcal{S}$~\cite{farris09} flowing across spherical surfaces at coordinate radii
between $r_{\rm ext}=50M\sim 200\,\rm km$ and $350M\sim 1400\rm\,km$.

%%%%%%%%%%%%%%%%%%%%%%%%%%%%%%%%
%%%   fig 1: Manetized cases %%%
%%%%%%%%%%%%%%%%%%%%%%%%%%%%%%%%
\begin{figure*}
    \centering
    \includegraphics[scale=0.102]{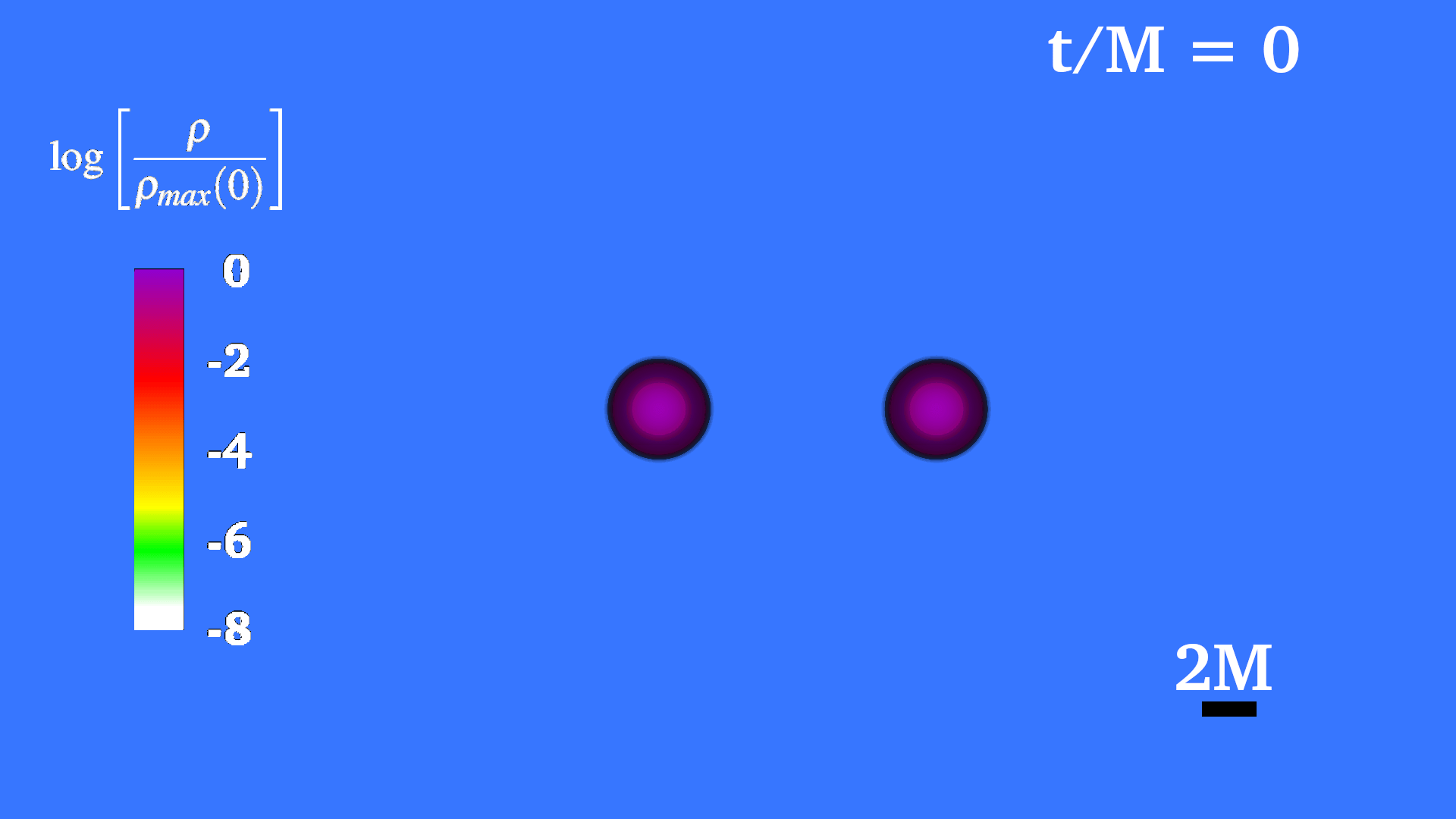}
    \includegraphics[scale=0.102]{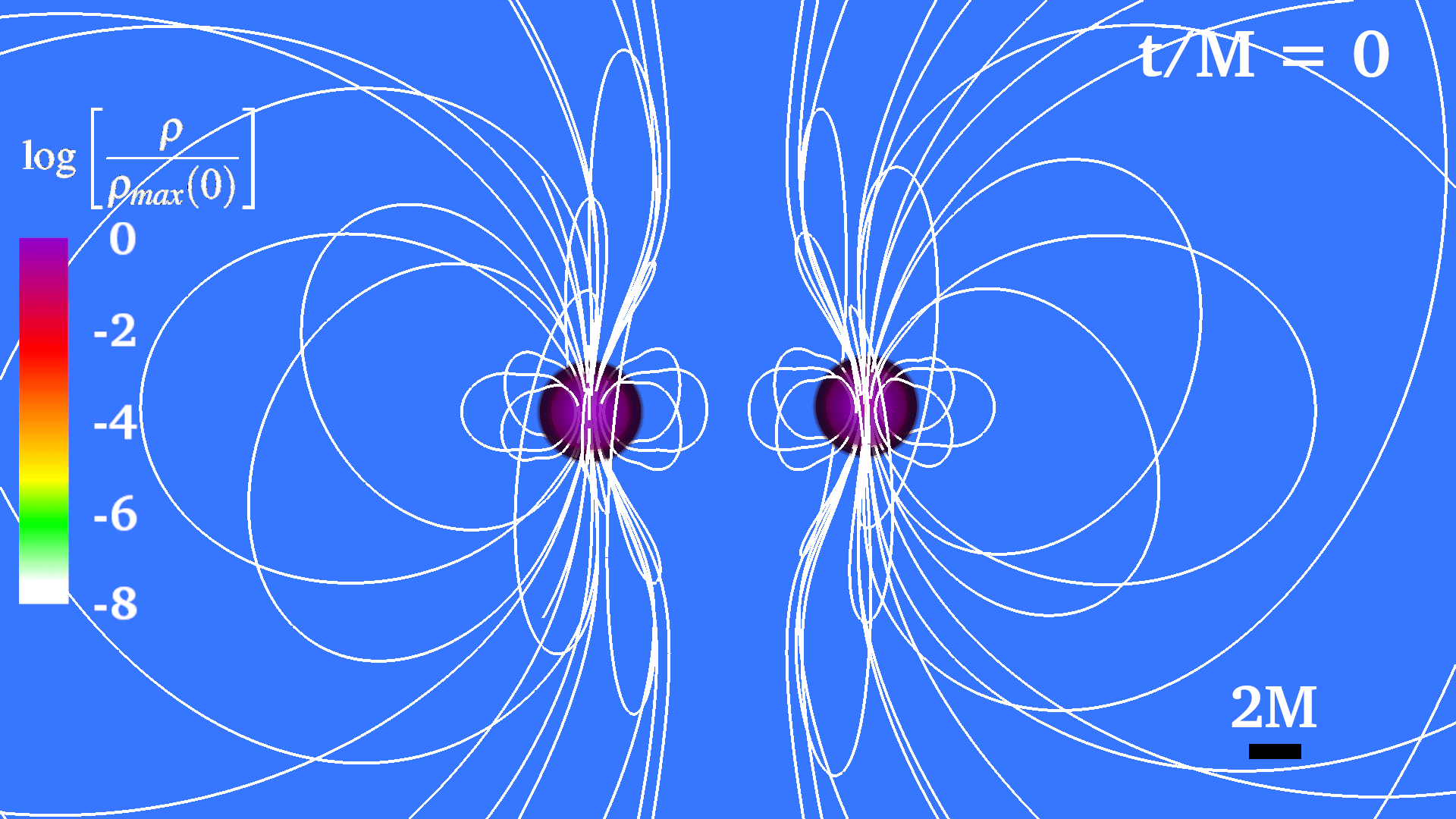}
    \includegraphics[scale=0.102]{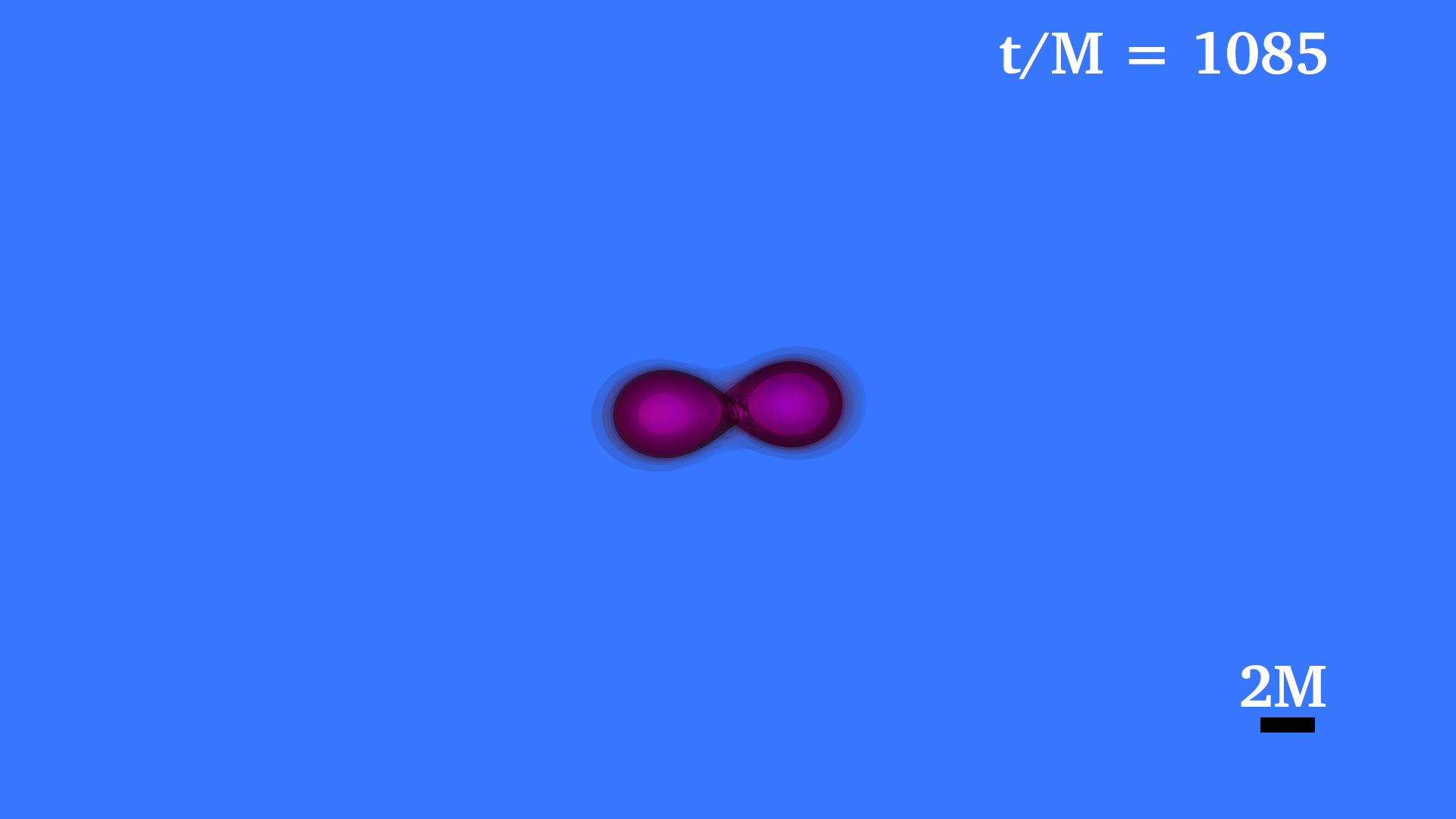}
    \includegraphics[scale=0.102]{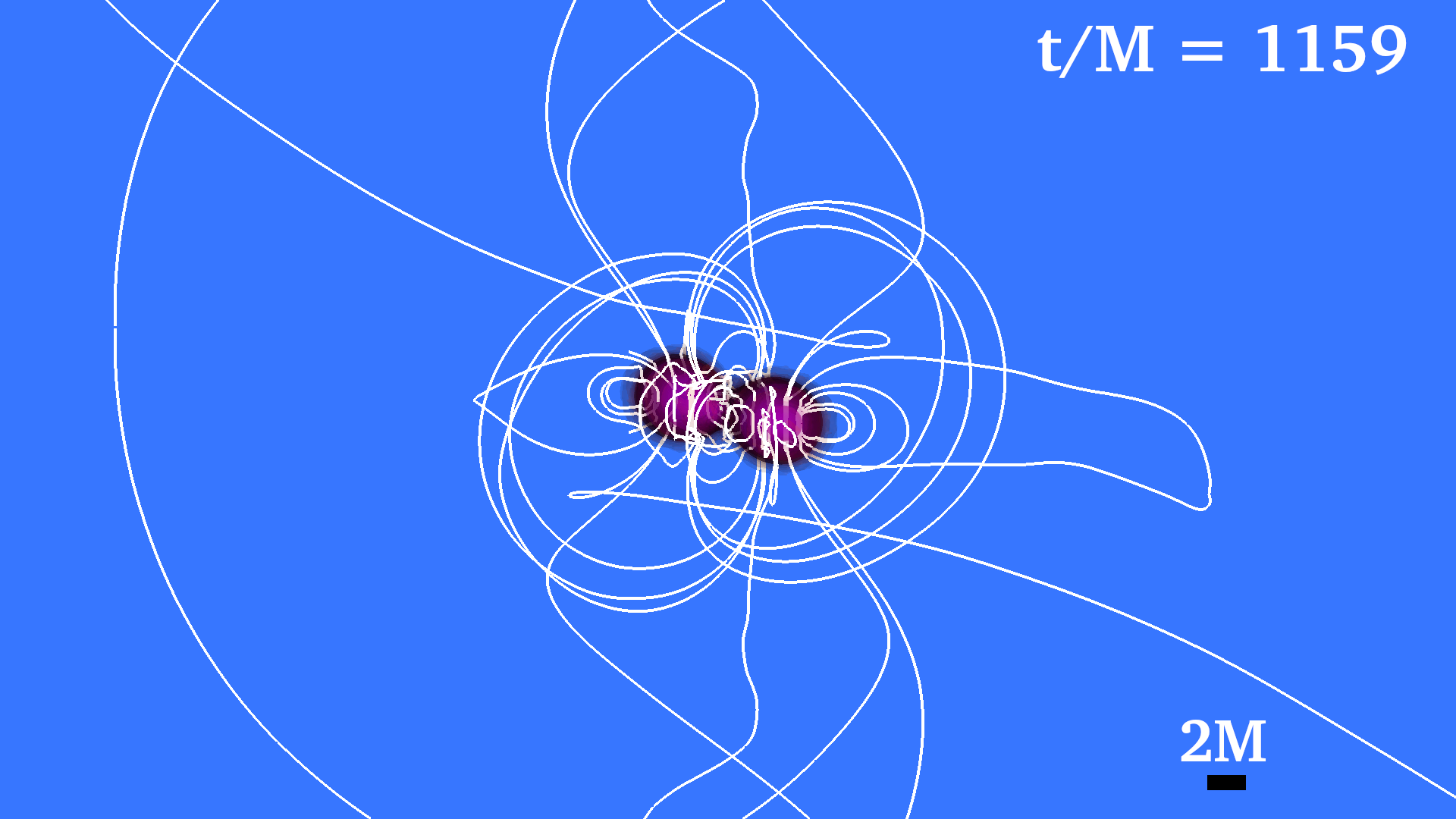}
    \includegraphics[scale=0.102]{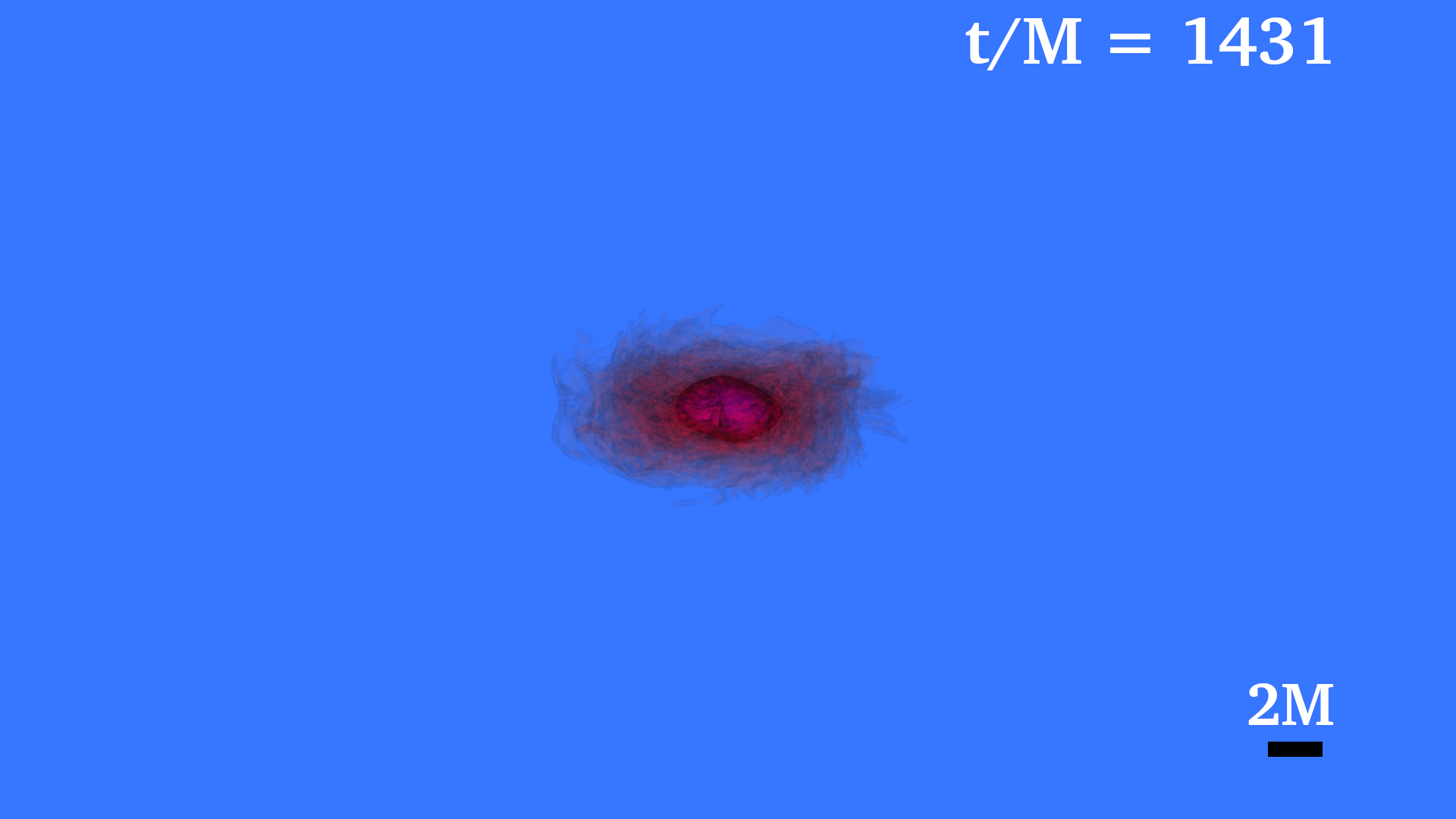}
    \includegraphics[scale=0.102]{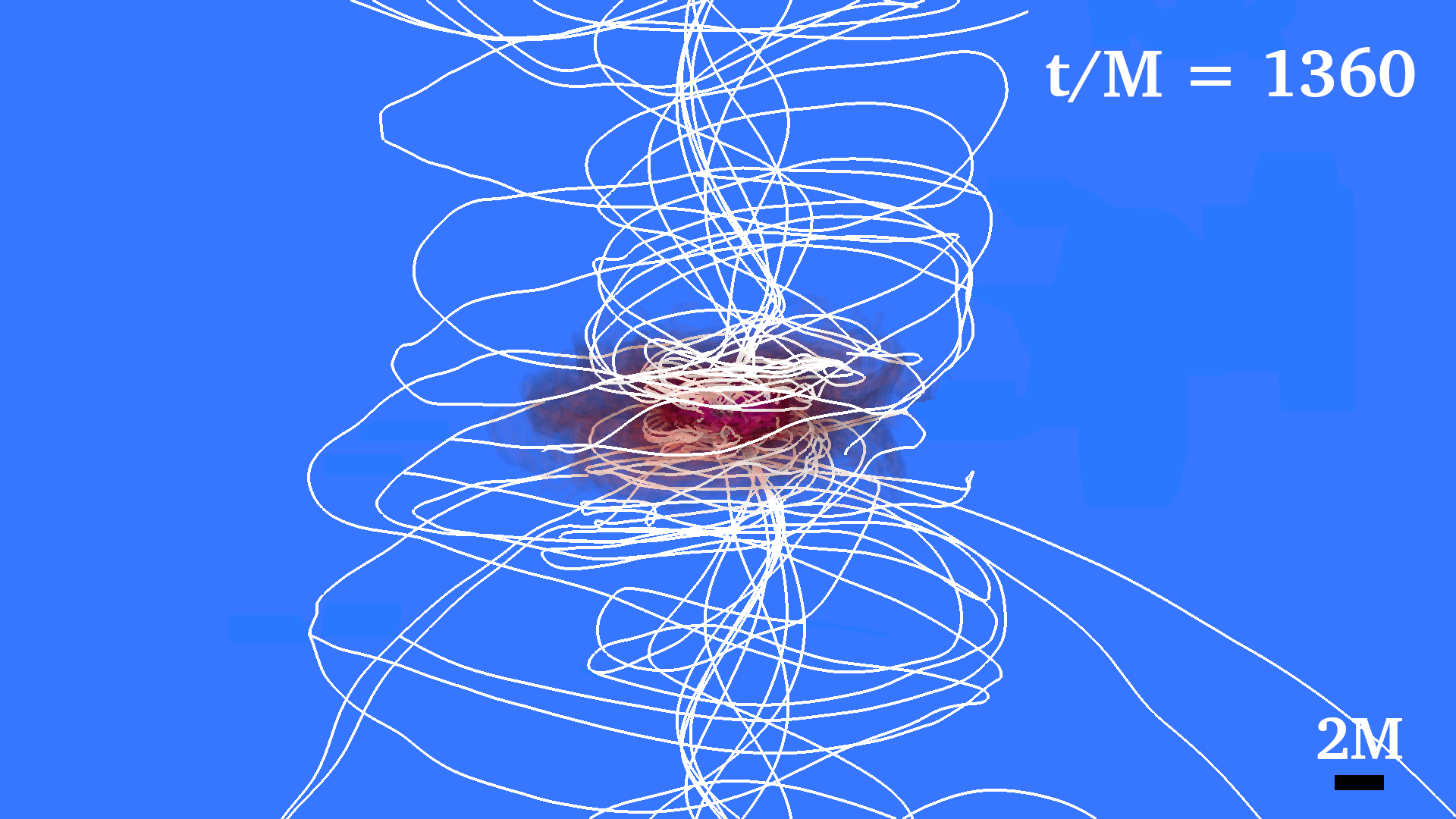}
    \includegraphics[scale=0.102]{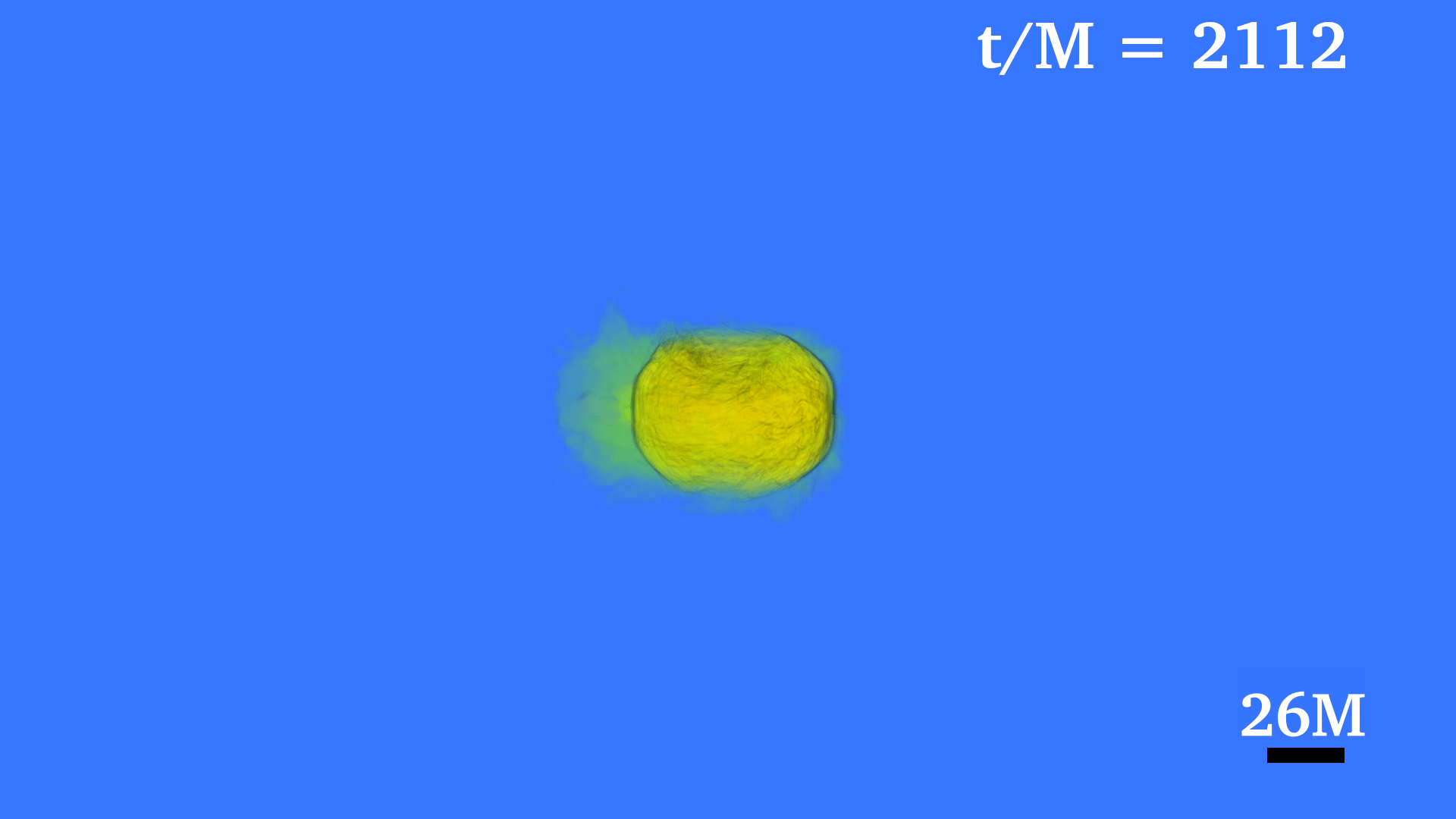}
    \includegraphics[scale=0.102]{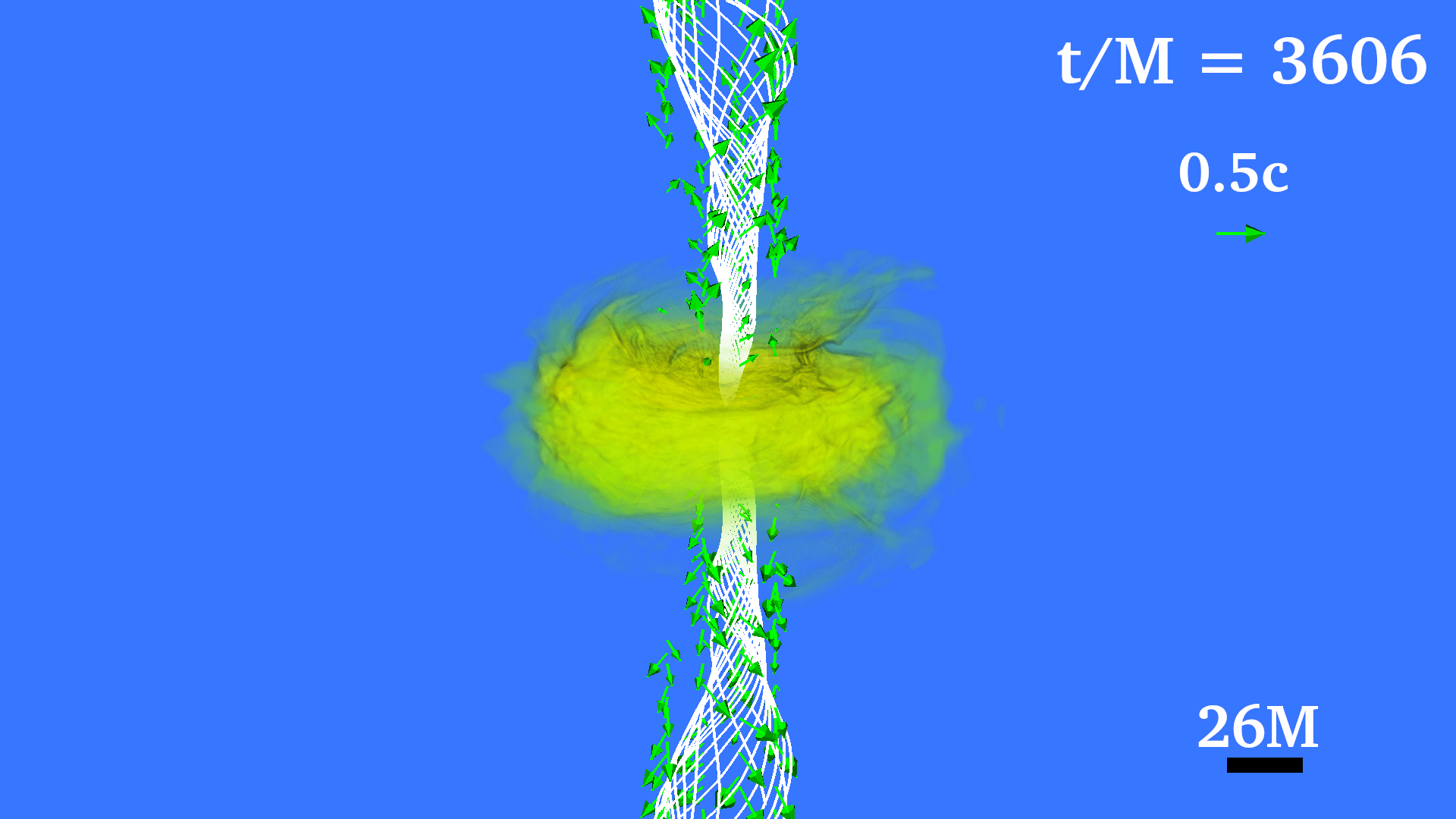}
    \includegraphics[scale=0.102]{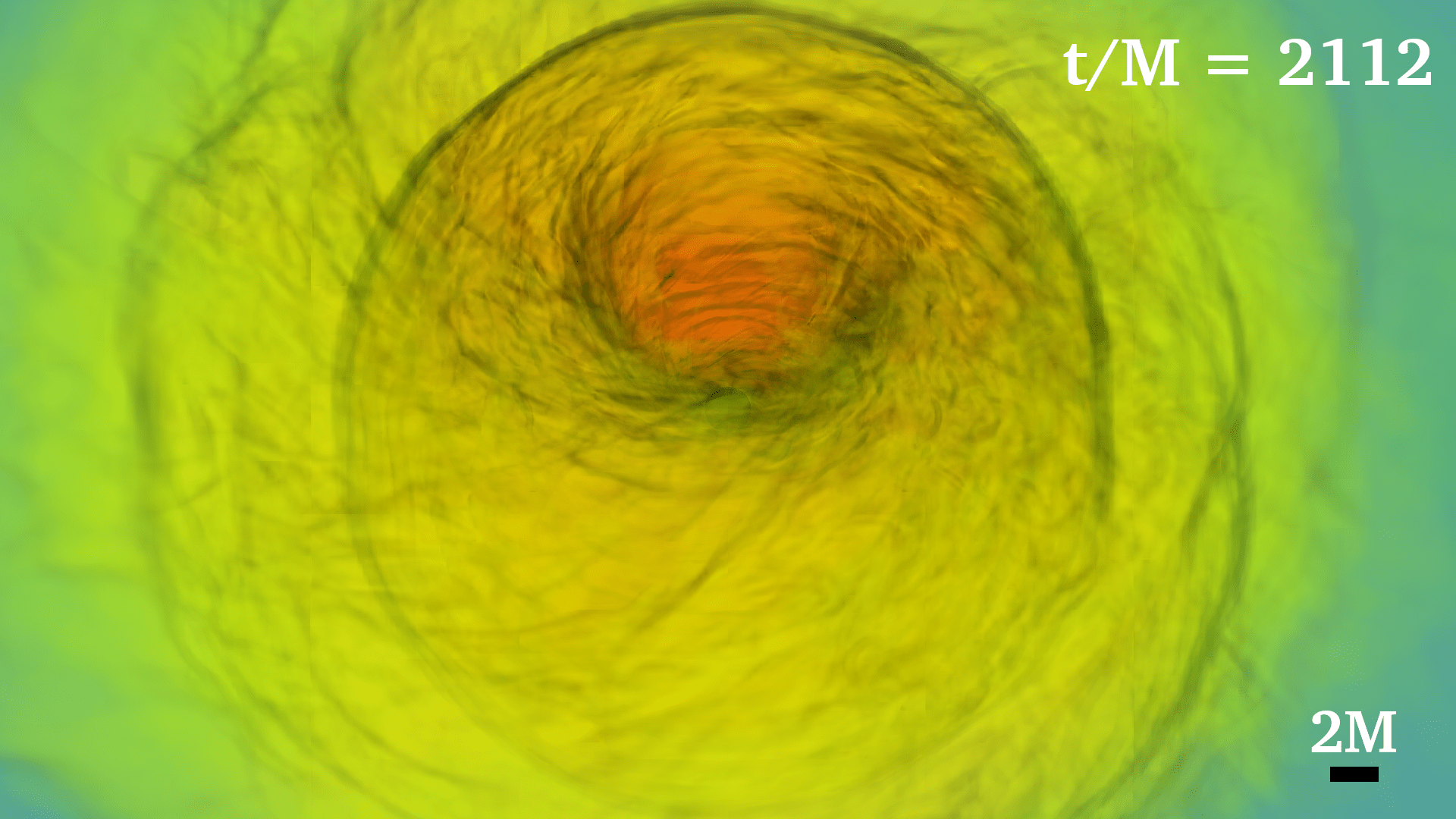}
    \includegraphics[scale=0.102]{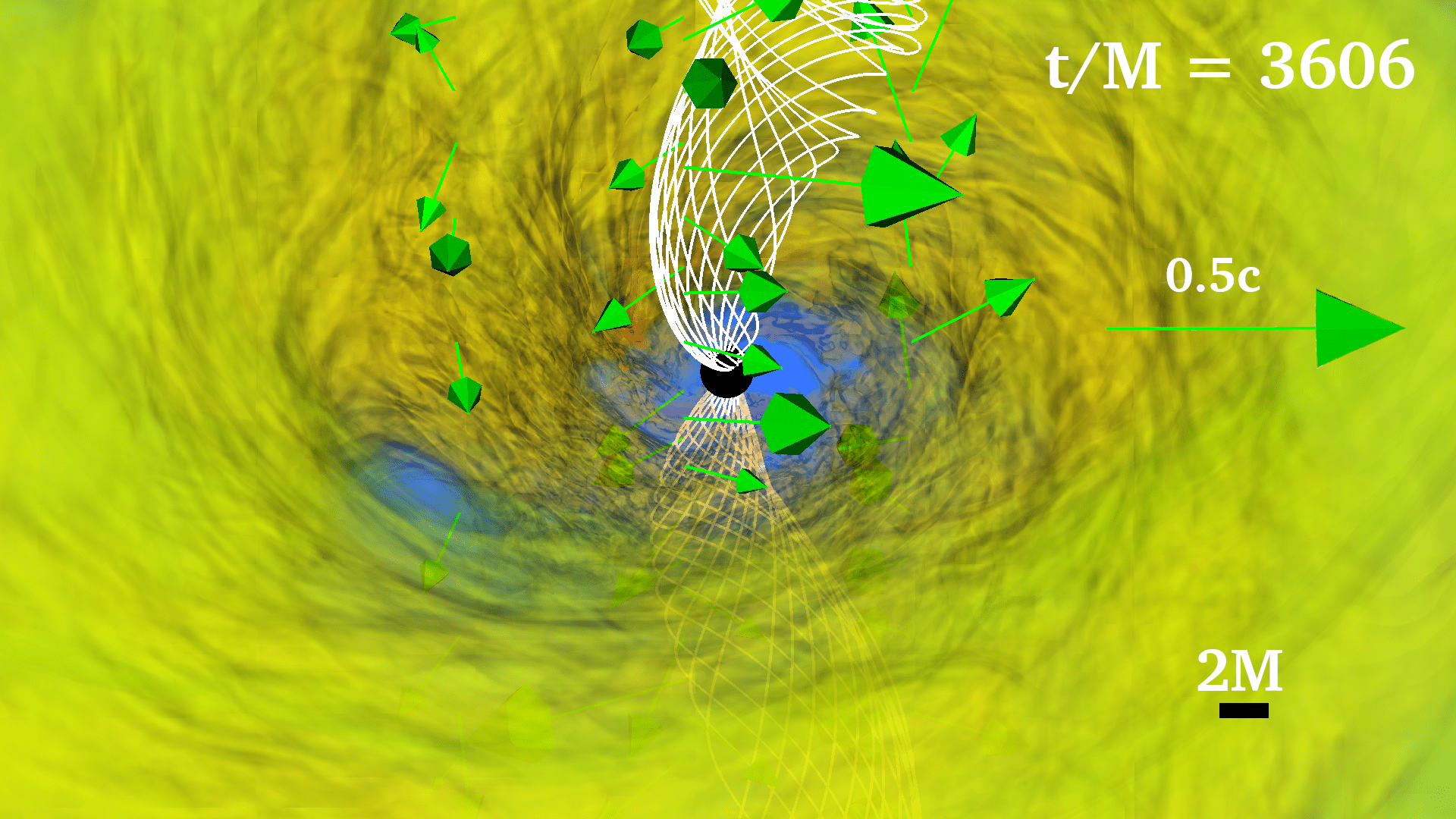}
    \caption{3D volume rendering of the rest-mass density $\rho_0$, normalized to its initial maximum value
      $\rho_{0,\rm{max}}=8.65\times 10^{14}\rm g\, cm^{-3}$ (log scale), at selected times
      for Unmag (left) and Mag (right) cases (see Table~\ref{table:Tab2}).
      The bottom right panel highlights the system after an incipient jet is launched.
      White lines shows the magnetic field lines emanating from the BH polar region and the arrows indicate plasma velocities.
      The BH apparent horizon is shown as a black sphere.
     Here $M=2.7 M_{\odot}= 0.0133 \,\rm ms = 3.99 \,\rm km$ is the ADM mass.}
    \label{fig:3Drho_1}
\end{figure*}

%%%%%%%%%%%%%%%%%%%%%%%%%%%%%%%%
%%%   fig 2: Radiation cases %%%
%%%%%%%%%%%%%%%%%%%%%%%%%%%%%%%%
\begin{figure*}
    \centering
    \includegraphics[scale=0.087]{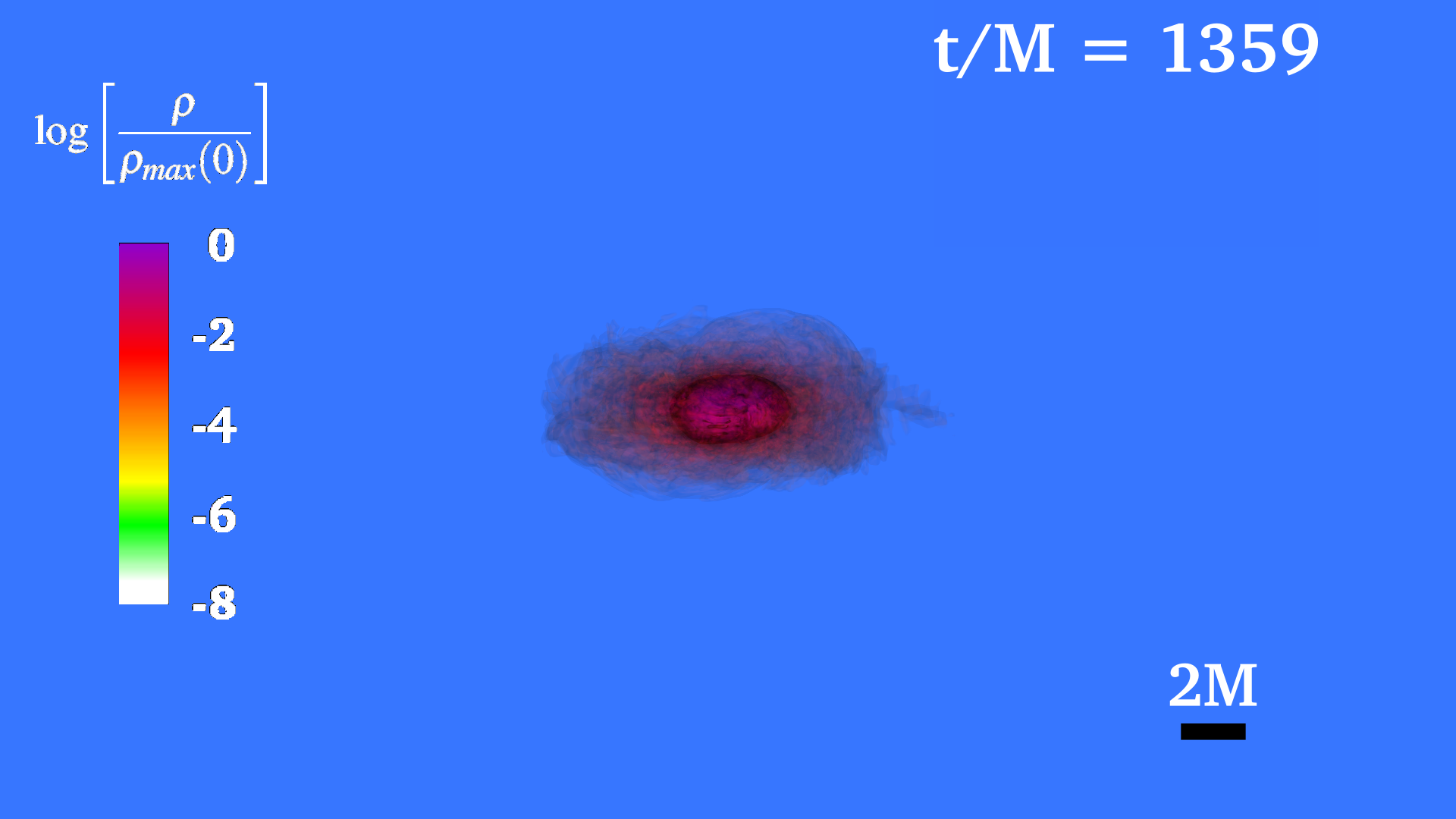}
    \includegraphics[scale=0.087]{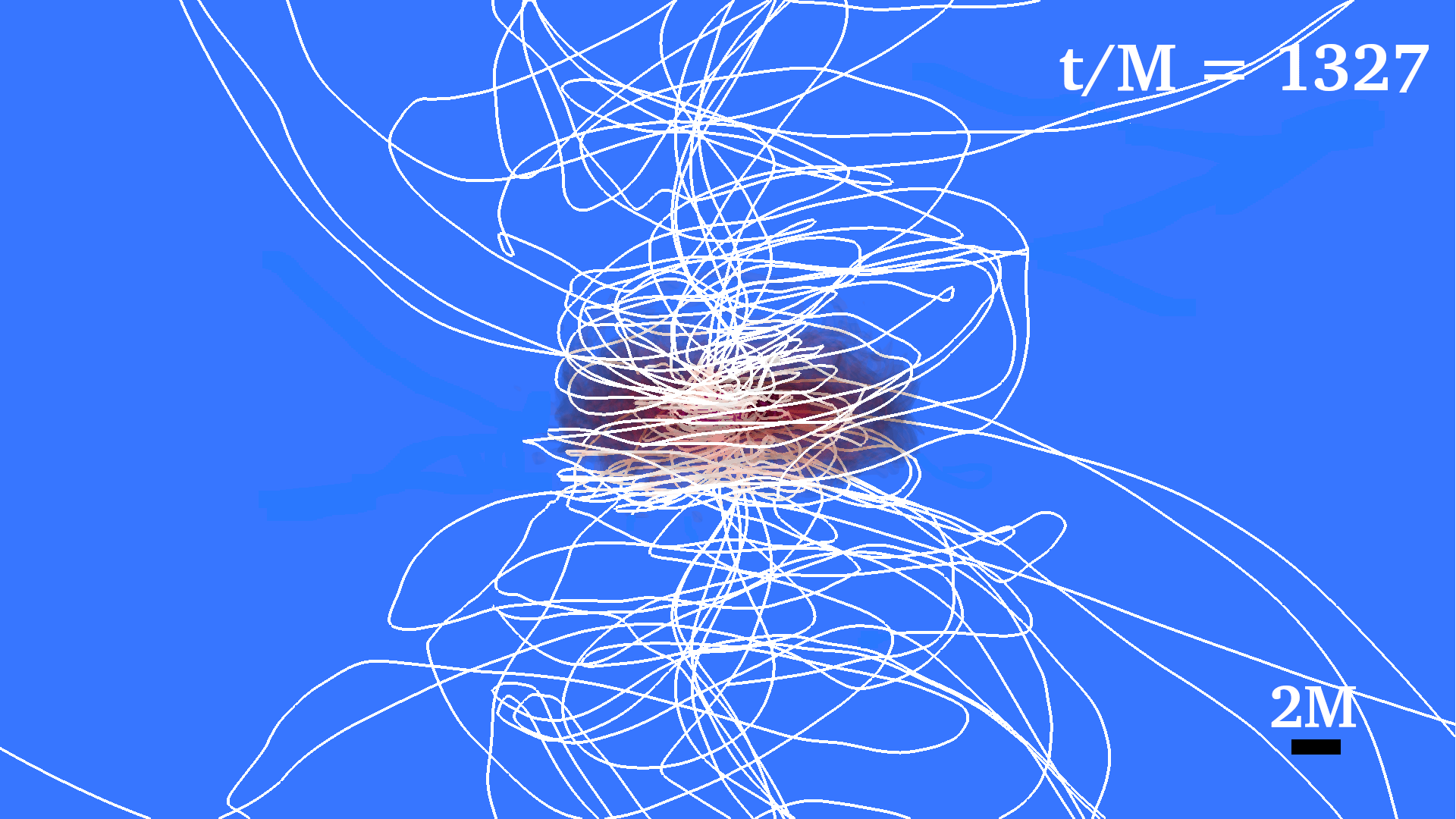}
    \includegraphics[scale=0.087]{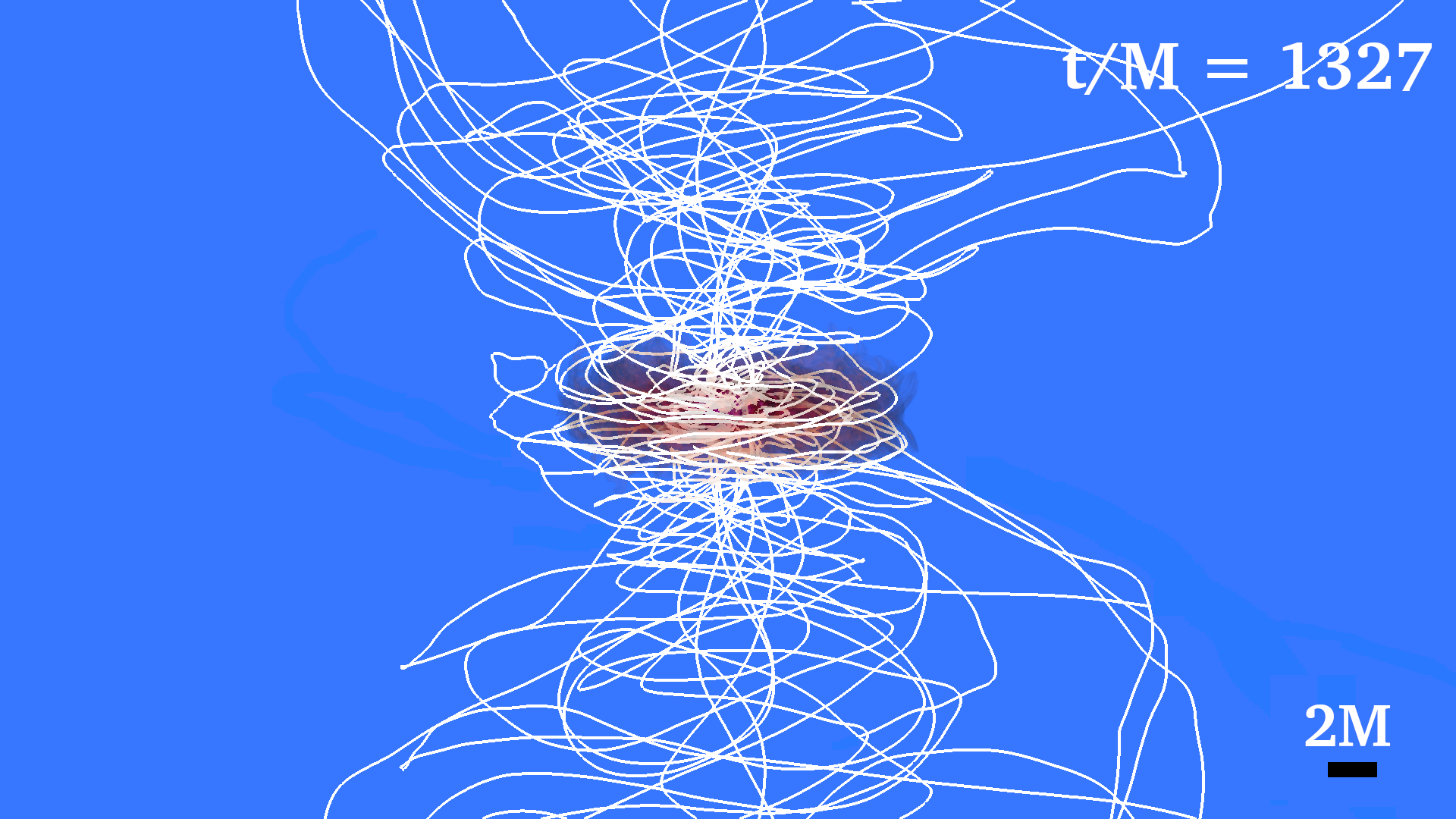}
    \includegraphics[scale=0.087]{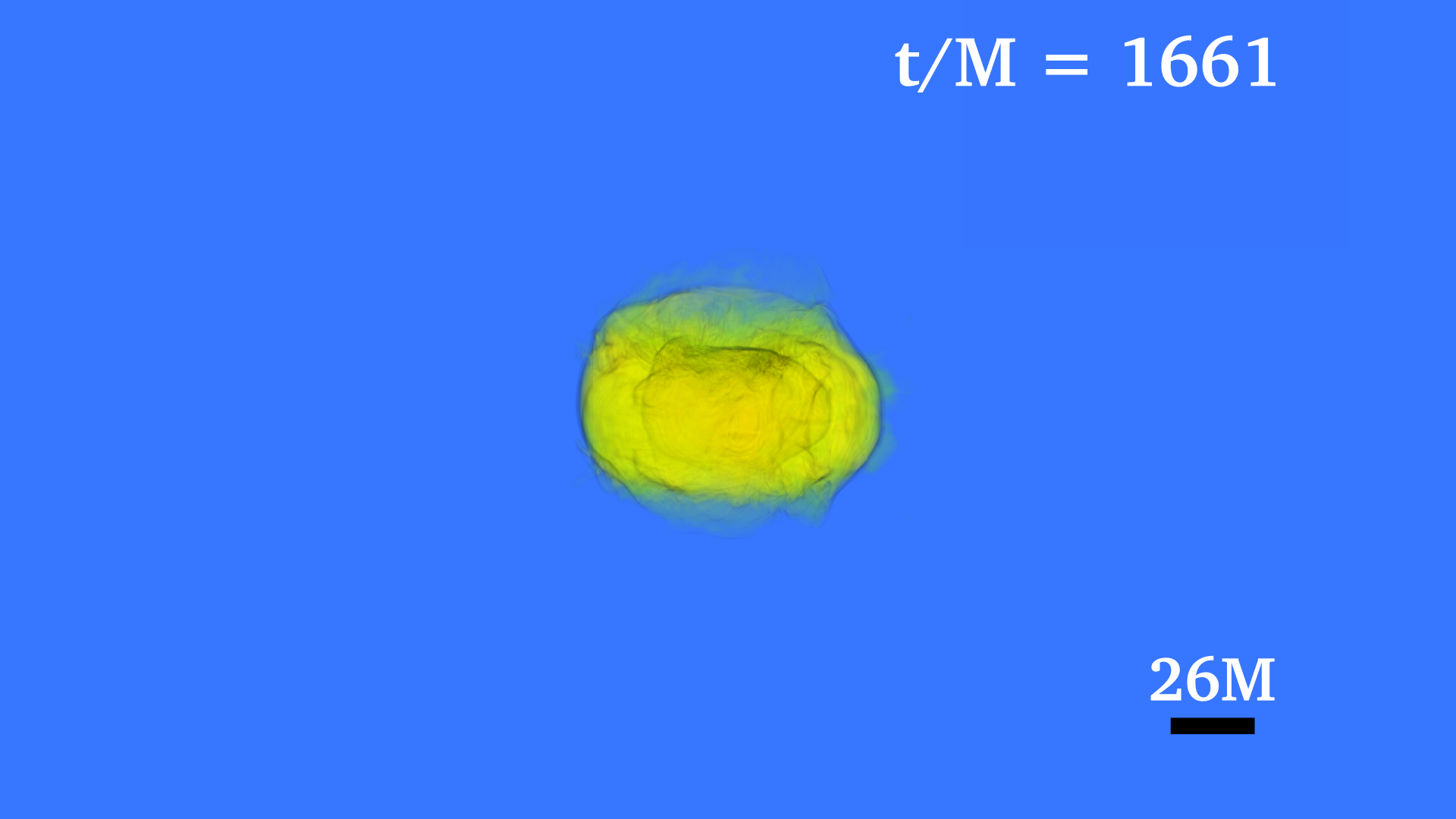}
    \includegraphics[scale=0.087]{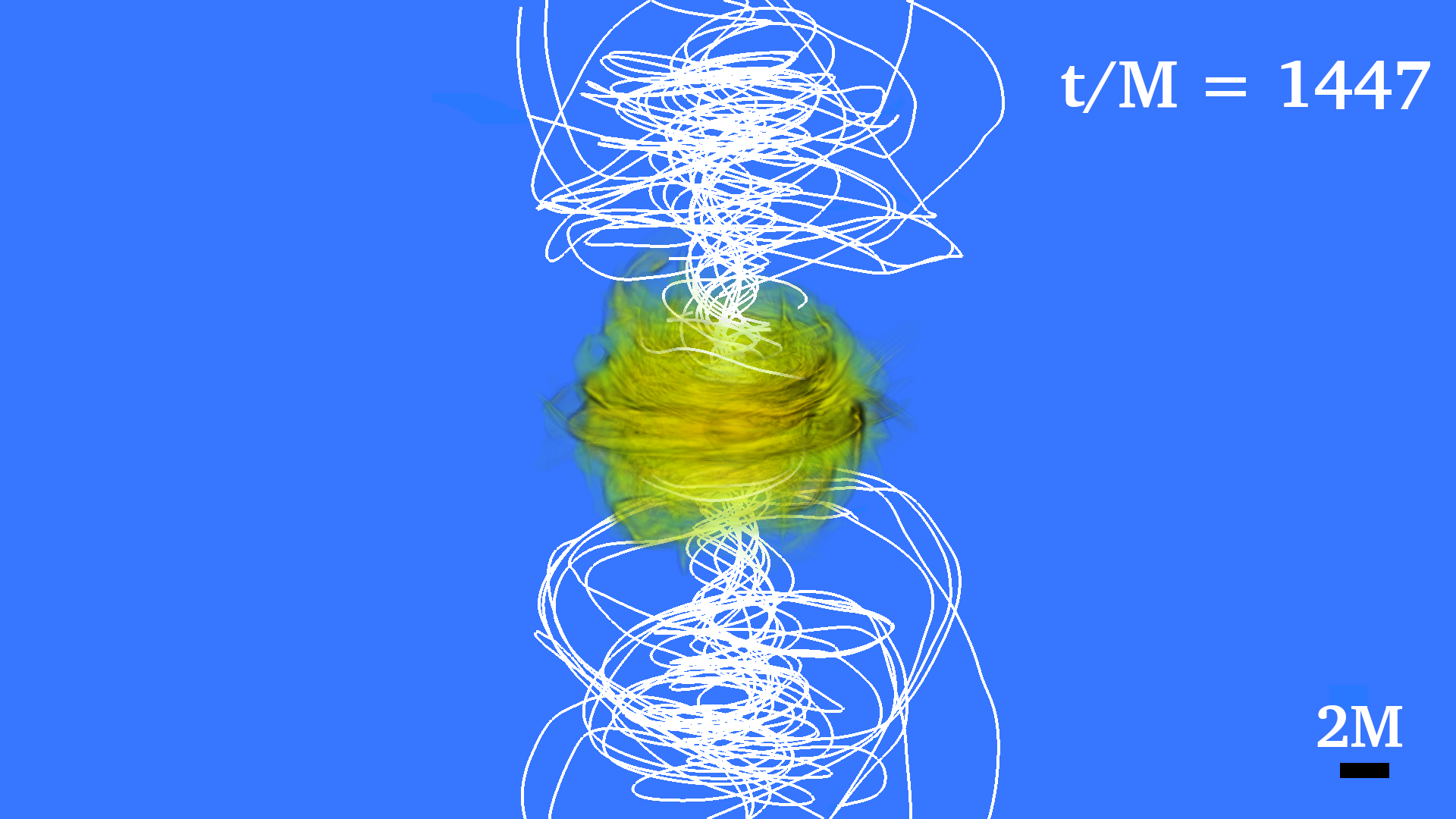}
    \includegraphics[scale=0.087]{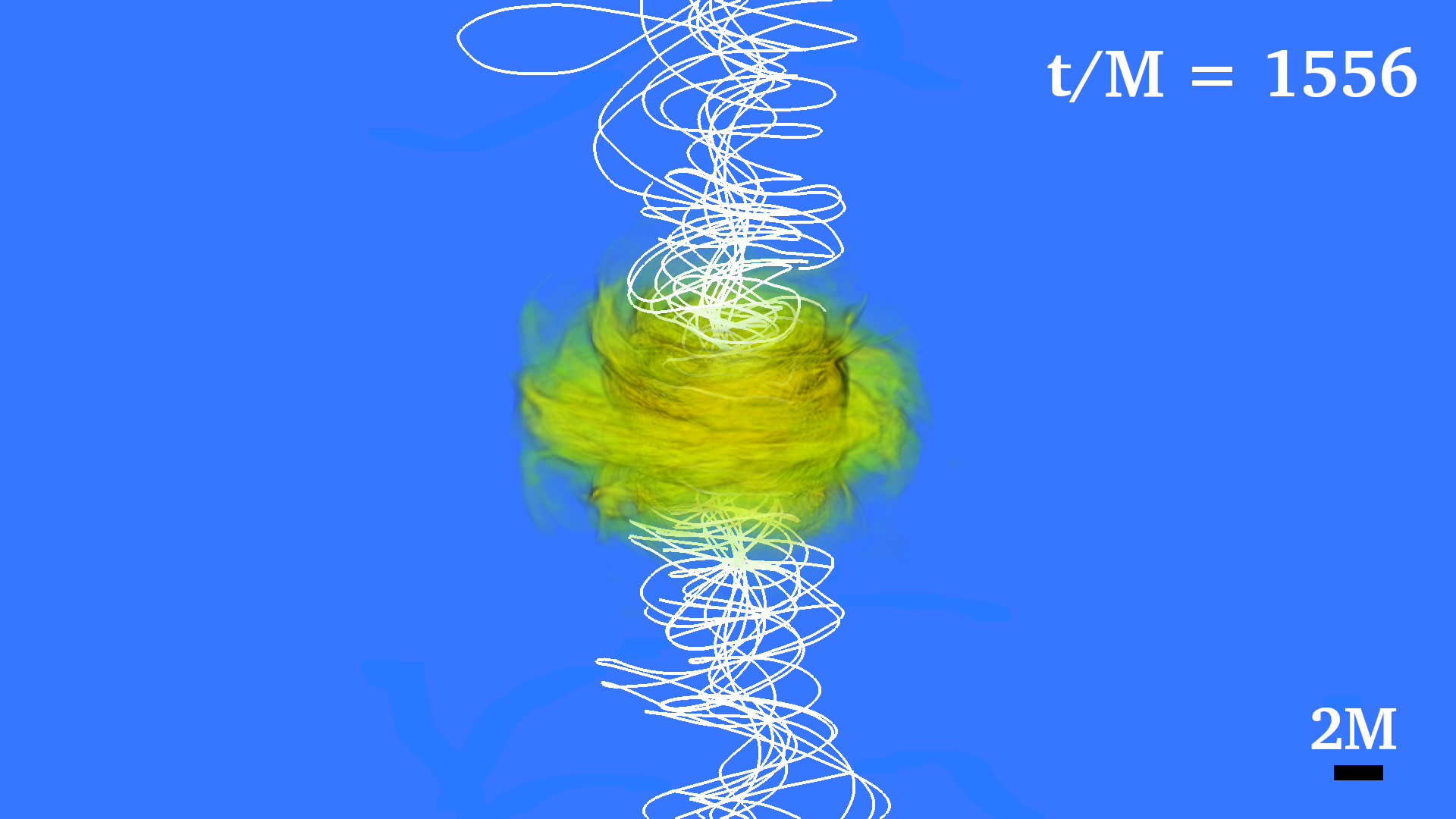}
    \includegraphics[scale=0.087]{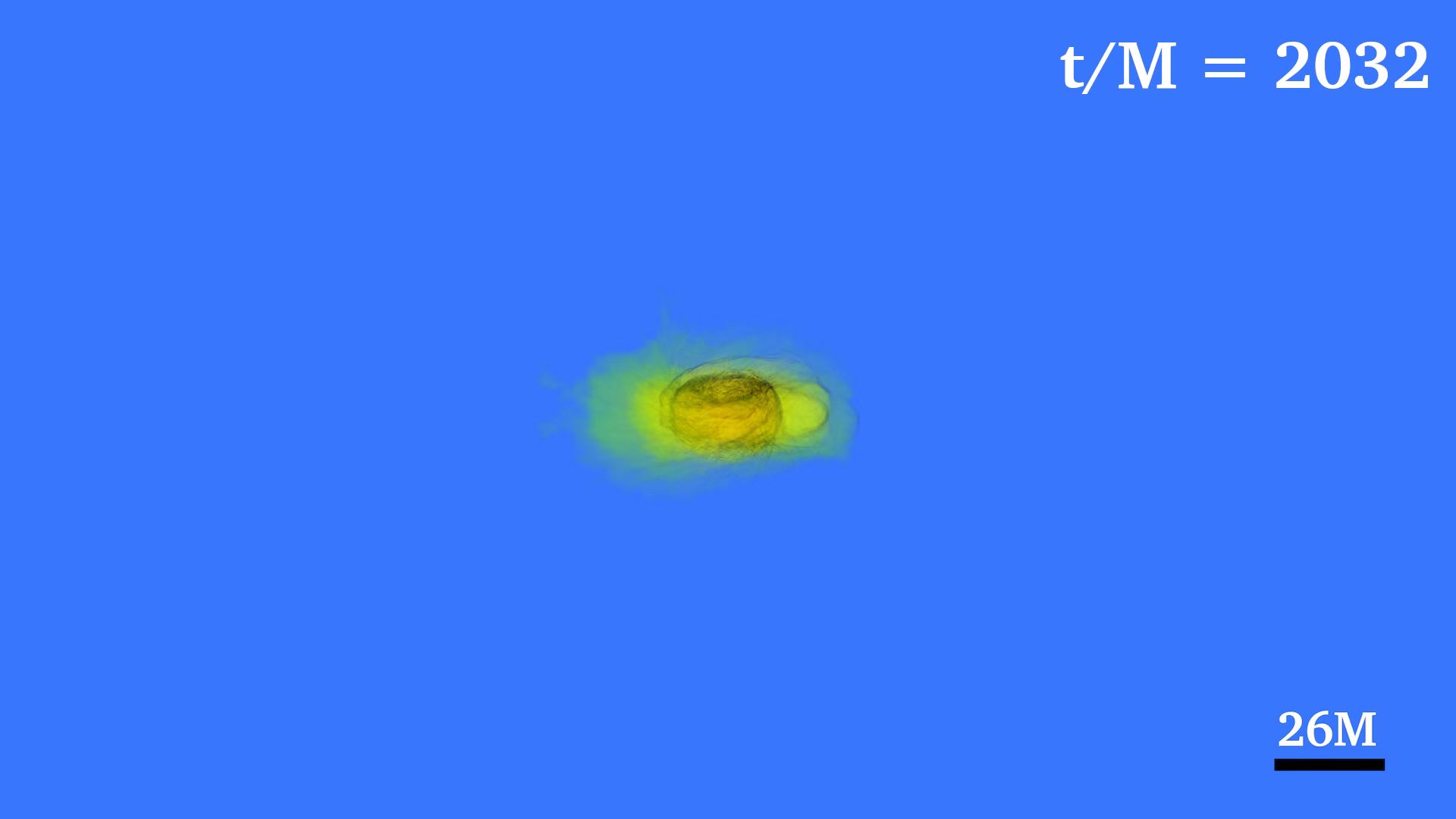}
    \includegraphics[scale=0.087]{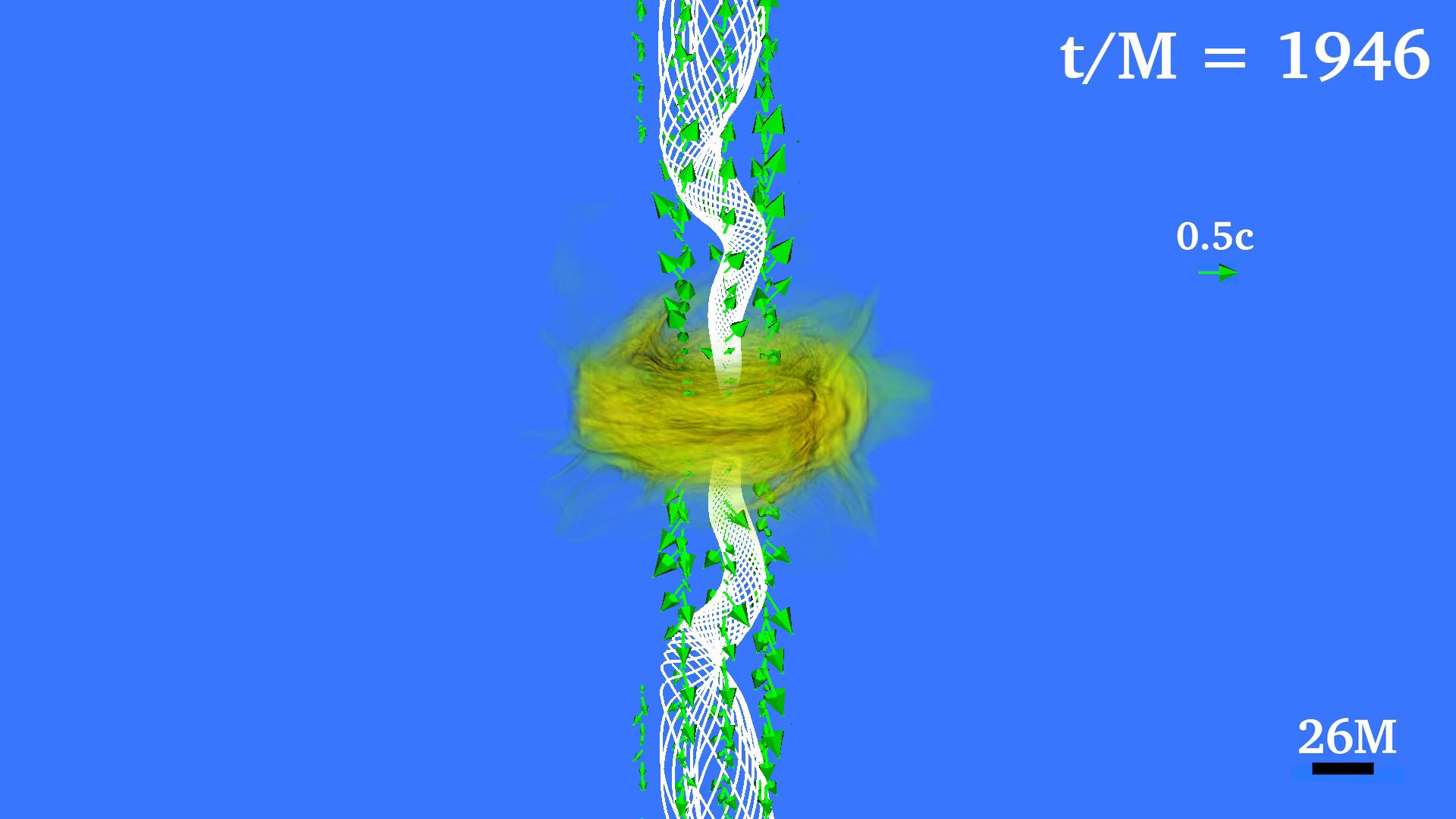}
    \includegraphics[scale=0.087]{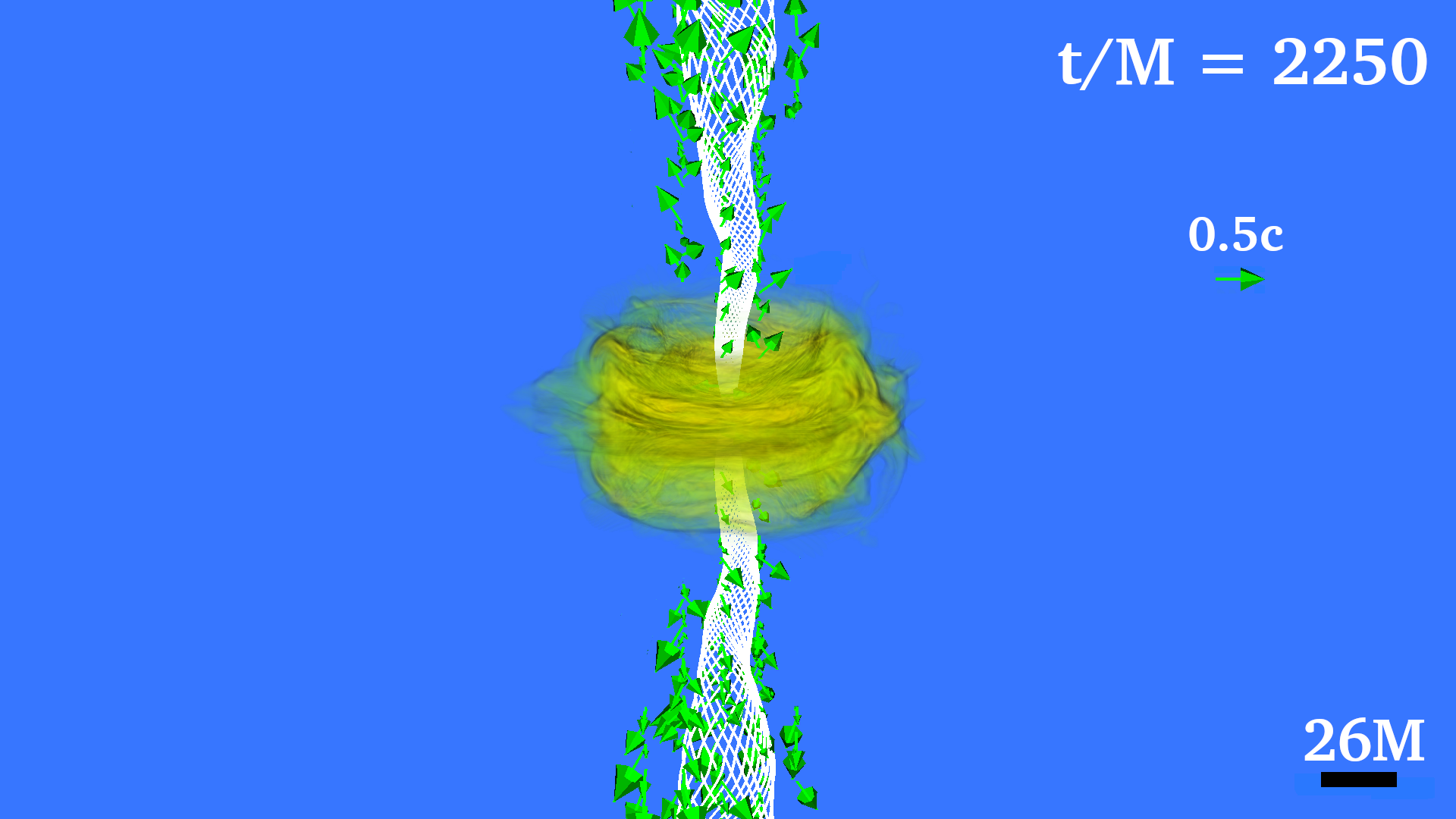}
    \includegraphics[scale=0.087]{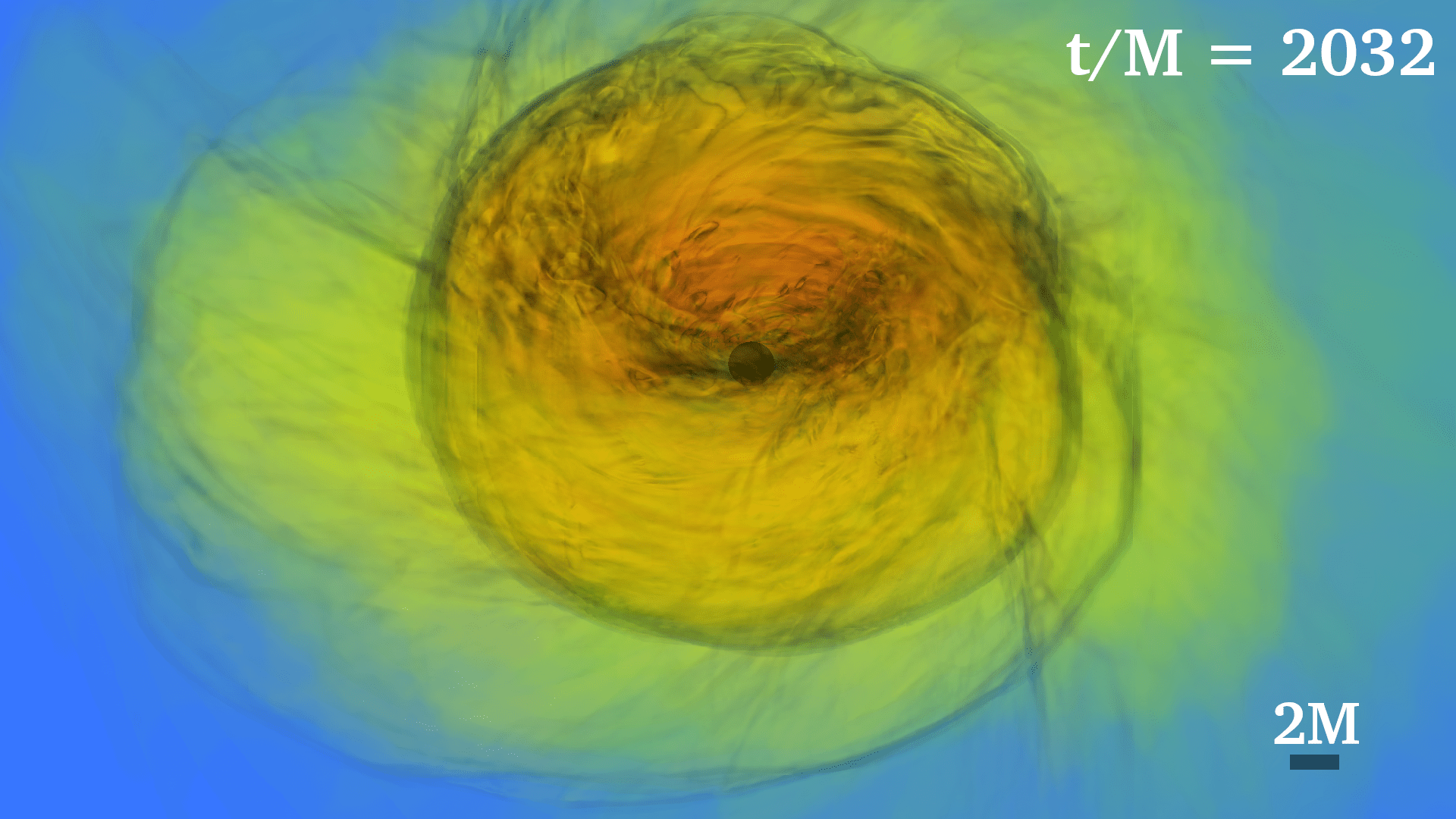}
    \includegraphics[scale=0.087]{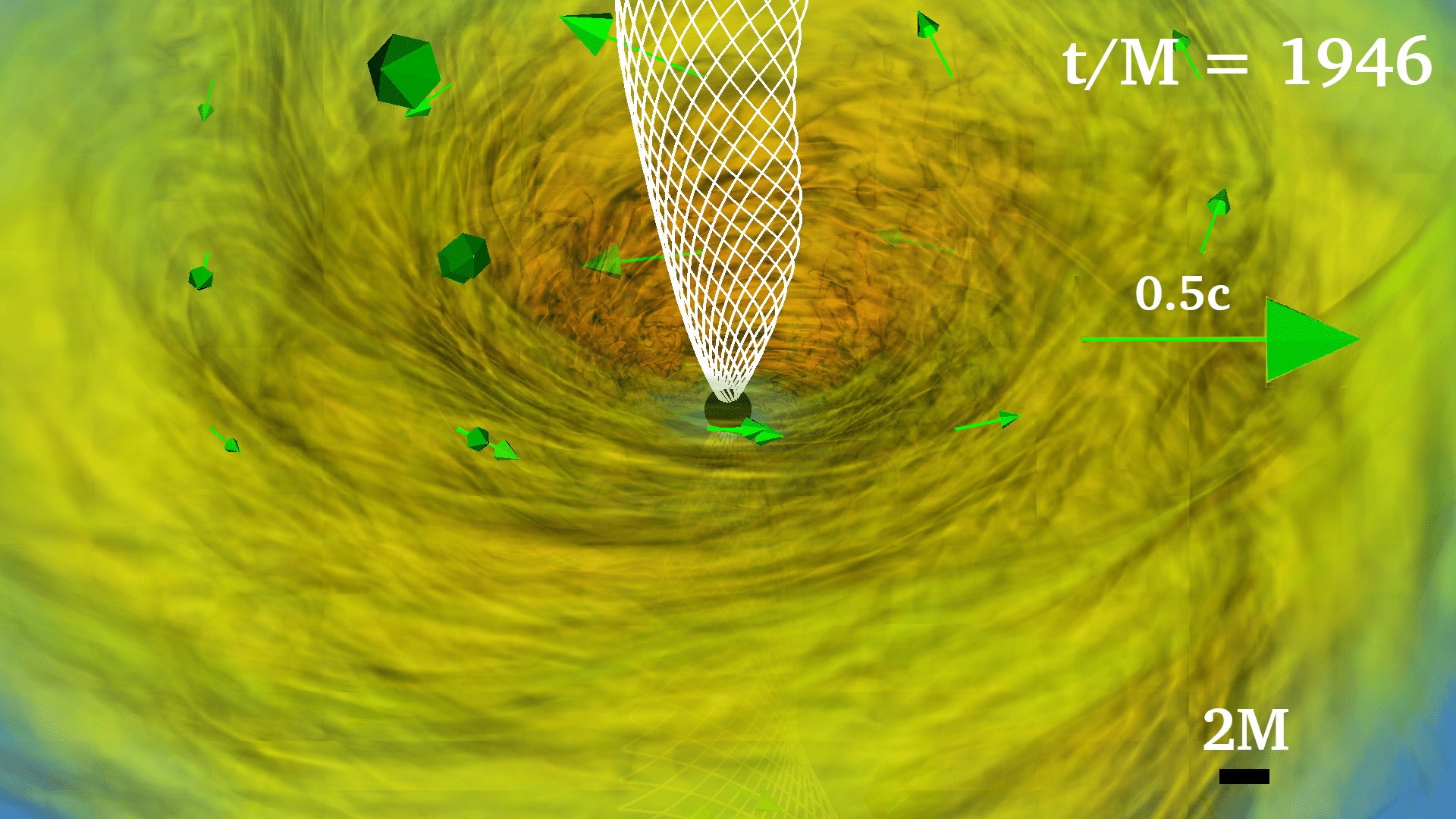}
    \includegraphics[scale=0.087]{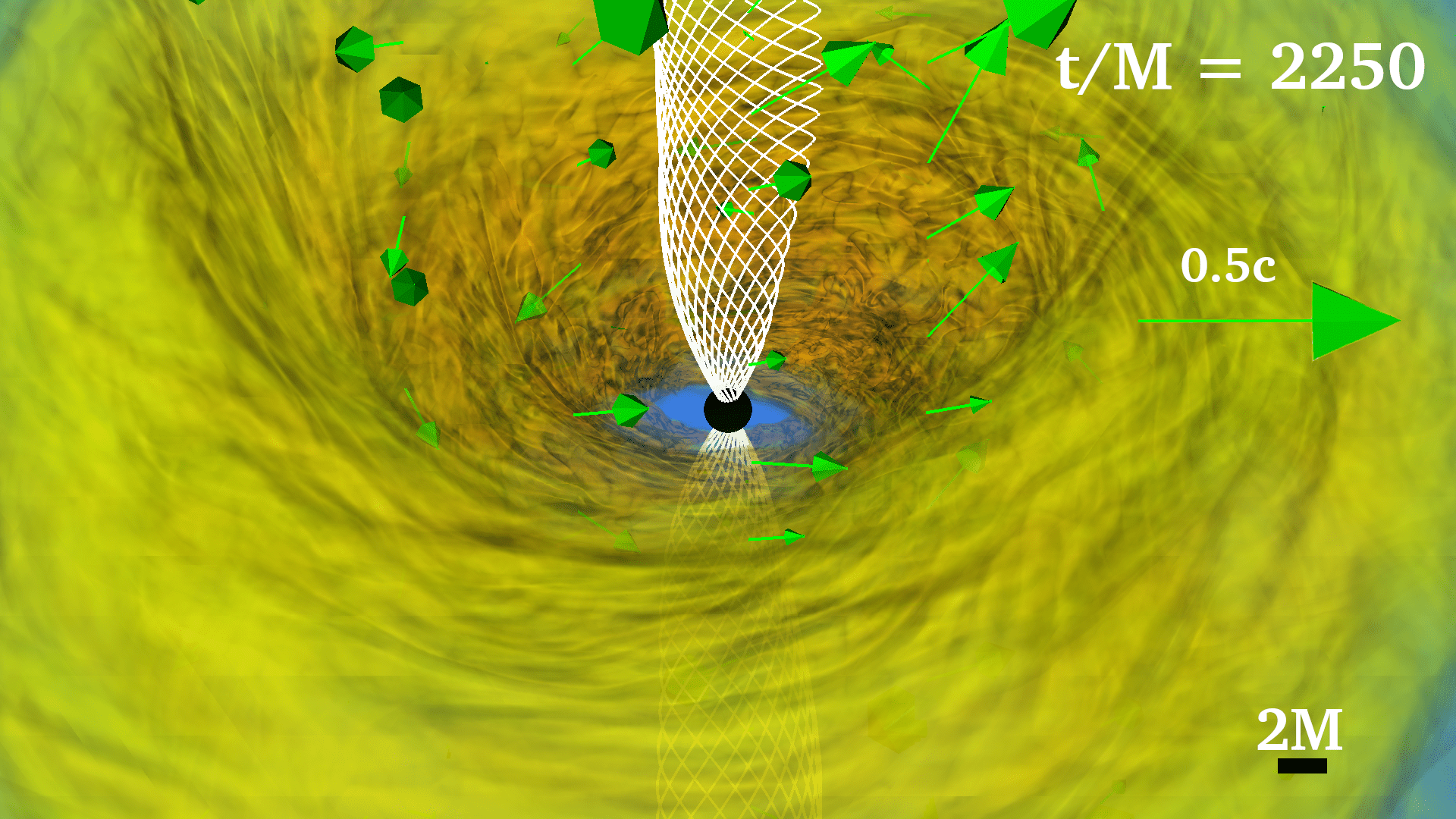}
    \caption{3D volume rendering of the rest-mass density $\rho_0$, normalized to its initial maximum value
      $\rho_{0,\rm{max}}=8.65\times 10^{14}\rm g\, cm^{-3}$ (log scale), at selected times
        for Unmag+Rad-Simp (left), Mag+Rad-Simp (middle), and and Mag+Rad-Full (right) cases
        (see Table~\ref{table:Tab2}). The bottom middle and right panels highlight the system
        after an incipient jet is launched. White lines shows the magnetic field lines emanating from the BH polar region and the arrows indicate plasma velocities.
      The BH apparent horizon is shown as a black sphere.
      Here $M=2.7 M_{\odot}= 0.0133 \,\rm ms = 3.99 \,\rm km$ is the ADM mass.}
    \label{fig:3Drho_2}
\end{figure*}

To track the conservation and transport of angular momentum, we compute the angular momentum carried off
by gravitational radiation $J_{\rm GW}$ using Eqs. (3.22)-(3.24) in~\cite{Ruiz08}. We also compute the angular
momentum carried off by fluid, EM field, and neutrinos via
\begin{eqnarray}
    J_{i, \rm fluid} &=& \int_{V} \epsilon_{ijk}\,x^j\,T^{kl}_{\rm fluid} \, d\mathcal{V}\,, \\
    J_{i, \rm EM} &=& \int_{V} \epsilon_{ijk}\, x^j\,  T^{kl}_{\rm EM} \, d\mathcal{V}\,,  \\
    J_{i, \nu} &=& \int_{V} \epsilon_{ijk}\, x^j\,  R^{kl} \, d\mathcal{V}\,.
\end{eqnarray}
Here $\epsilon_{ijk}$ is the 3D Levi-Civita symbol, and $T^{\alpha\beta}_{\rm fluid}$ and $T^{\alpha\beta}_{\rm EM}$
are the stress-energy tensors associated with the perfect fluid and the electromagnetic field given by Eqs. (44) and (42) in ~\cite{grrmhd},
respectively.
We monitor the conservation of the total mass and angular momentum $M_{\rm int}$ and $J_{\rm int}$ computed via Eqs.~(19)-(22)
in~\cite{Etienne:2011ea}, which coincide with the ADM mass and ADM angular momentum initially. In all our evolved configurations, we find
that the total mass and total angular momentum are conserved to
within~$\sim1\%$ and $\sim5\%$, respectively. In addition, we monitor the conservation of the rest-mass $M_0=\int \rho_* d^3x$, where
$\rho_* \equiv -\sqrt{\gamma} \rho_0 n_\mu u^\mu$, which is conserved to within~$\sim 0.6\%$. Moreover, we compute the outgoing EM Poynting and neutrino luminosities $ L_{\rm EM}=-\int T^{r({\rm EM})}_{~0}\,\sqrt{-g}\,d\mathcal{S}$ and
$ L_{\rm \nu}=-\int R^{r}_{~0}\,\sqrt{-g}\,d\mathcal{S}$, across spherical surfaces of coordinate radii between $r_{\rm ext}=50M\sim 200\,\rm km$ and
$350M\sim 1400\,\rm km$. For Mag+Rad-Full cases, we compute the neutrino luminosity for each species. To study the effect of neutrinos on MRI in our
evolution, we compute the number of grid points resolved by the fastest growing MRI mode $\lambda_{\rm MRI}$~\cite{Etienne:2012te},
namely the $\lambda_{\rm MRI}$-quality factor $Q_{\rm MRI}\equiv\lambda_{\rm MRI}/dx$, where $dx$ is the local grid spacing. A successful
capture of MRI must satisfy the condition $Q_{\rm MRI}\gtrsim 10$ and $\lambda_{\rm MRI}$ must fit inside the remnant
\cite{Sano:2003bf,Shiokawa:2011ih}. Lastly, to study the effective turbulent viscosity due to the magnetic field and compare it to the effective
viscosity induced by neutrinos (see Eq.~(\ref{eq:nu_neutrino})), we compute the effective Shakura--Sunyaev $\alpha_{\rm SS}$ parameter~\cite{Shakura73} by
$\alpha_{\rm SS} \equiv \left\langle T^{\rm EM}_{\hat{r}\hat{\phi}}/\langle P\rangle\right\rangle_t$
(see~Eq.~(26)~in~\cite{FASTEST_GROWING_MRI_WAVELENGTH}), where the brackets denote an time-averaged quantity.

%%%%%%%%%%%%%%%%%%%
%%%   Results   %%%
%%%%%%%%%%%%%%%%%%%
\section{Evolution Results}
\label{sec:Evolution}

\subsection{General Properties of Merger}
\label{sec:4-1}
The inspiral phase of the NSNS models listed in Table~\ref{table:Tab2}
is summarized in the first rows of Fig~\ref{fig:3Drho_1}. The frozen-in
magnetic field is simply advected with the matter while GWs carry away
angular momentum, causing the orbital separation to shrink. We observe that
all our cases merge roughly at $t_{\rm merge}\sim 15\,\rm ms$ (see second row and Table III in~\cite{Ruiz2021}).
Here, we define the merger time $t_{\rm merge}$ as the
time of peak GW amplitude. Following merger, a transient remnant forms with two, massive central
cores rotating about each other that gradually coalesce. They form a highly differentially rotating HMNS,
surrounded by a low-density cloud of matter from fall-back tidal debris. 

As we are primarily concerned with the influence of neutrino emission on the launching of jets, we
inserted neutrinos in Unmag and Mag at $t-t_{\rm merge}\approx 226\sim 3\,\rm ms$  following merger
(see first row in Fig.~\ref{fig:3Drho_2}). In the following sections, we describe the final outcome of
our simulations with different magnetic field content, neutrino transport, and the two microphysics versions.
For comparison, we also include the SLy cases without neutrinos previously treated in~\cite{Ruiz2021}.

As mentioned before, we denote the unmagnetized case as ``Unmag'' and its simplified neutrino radiation
counterpart as ``Unmag+Rad-Simp". The magnetized case is denoted as ``Mag''  and
its neutrino radiation counterparts as ``Mag+Rad-Simp" and ``Mag+Rad-Full", based on their microphysical
schemes~(see Table \ref{Tab:Scheme}). Important quantities from these simulations are summarized in
Table~\ref{table:Tab2}. 
%
%%%%%%%%%%%%%%%%%%%%%%%%
%%%  Accretion plot  %%%
%%%%%%%%%%%%%%%%%%%%%%%%
%
\begin{figure}
\centering
    \includegraphics[scale=0.59]{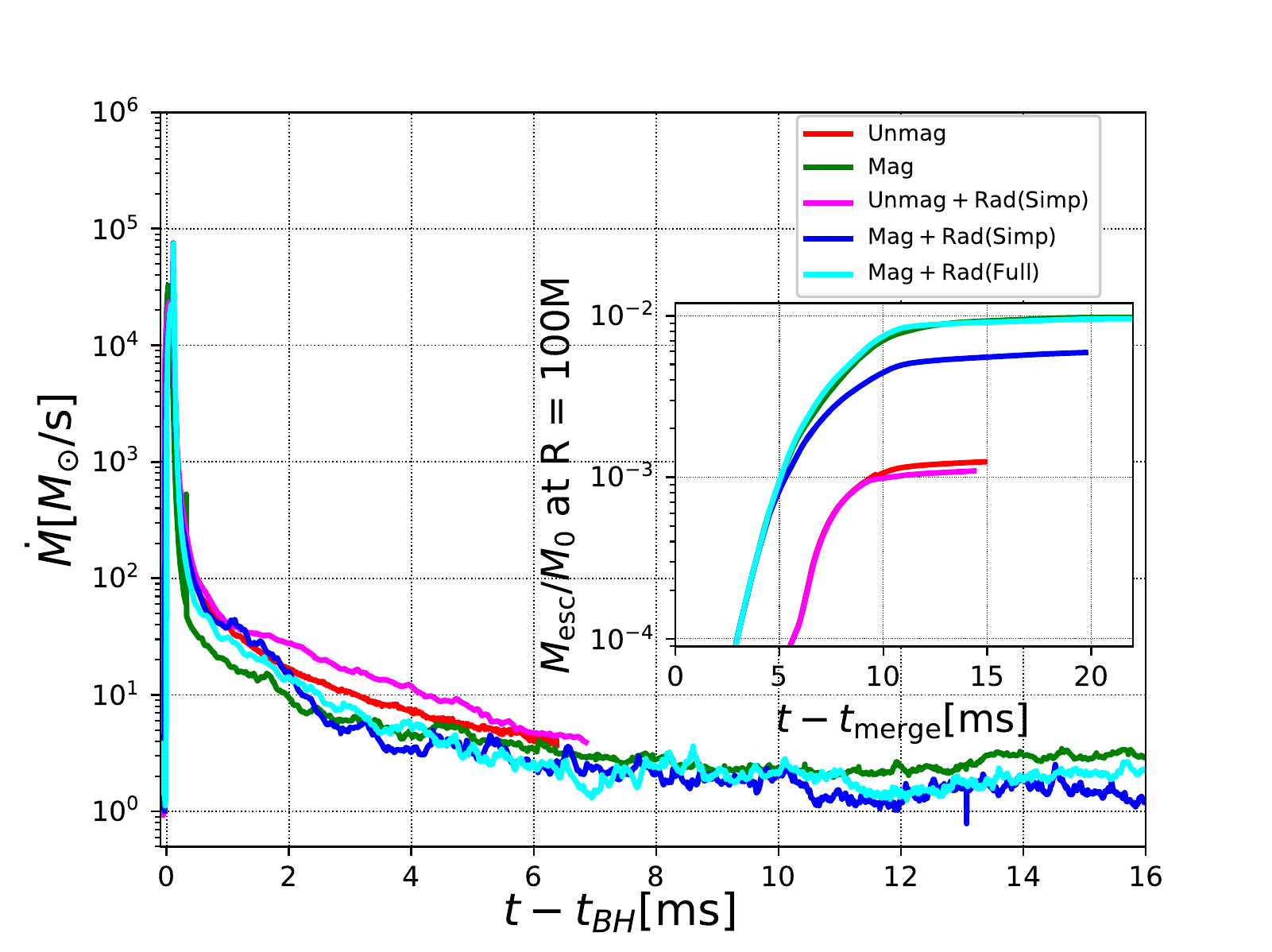}
    \caption{Rest-mass accretion rate as a function of time for all cases in Table~\ref{table:Tab2}.
      The insertion displays the ejected rest-mass fraction $M_{\rm esc}/M_0$ measured on a coordinate sphere
      of radius $r = 100 \rm M \approx 400~\rm km$ following merger.}
    \label{fig:Mdot}
\end{figure}
%
%%%%%%%%%%%%%%%%%%
%%%  MRI plot  %%%
%%%%%%%%%%%%%%%%%%
%
\begin{figure}
  \centering
  \includegraphics[width=0.52\textwidth]{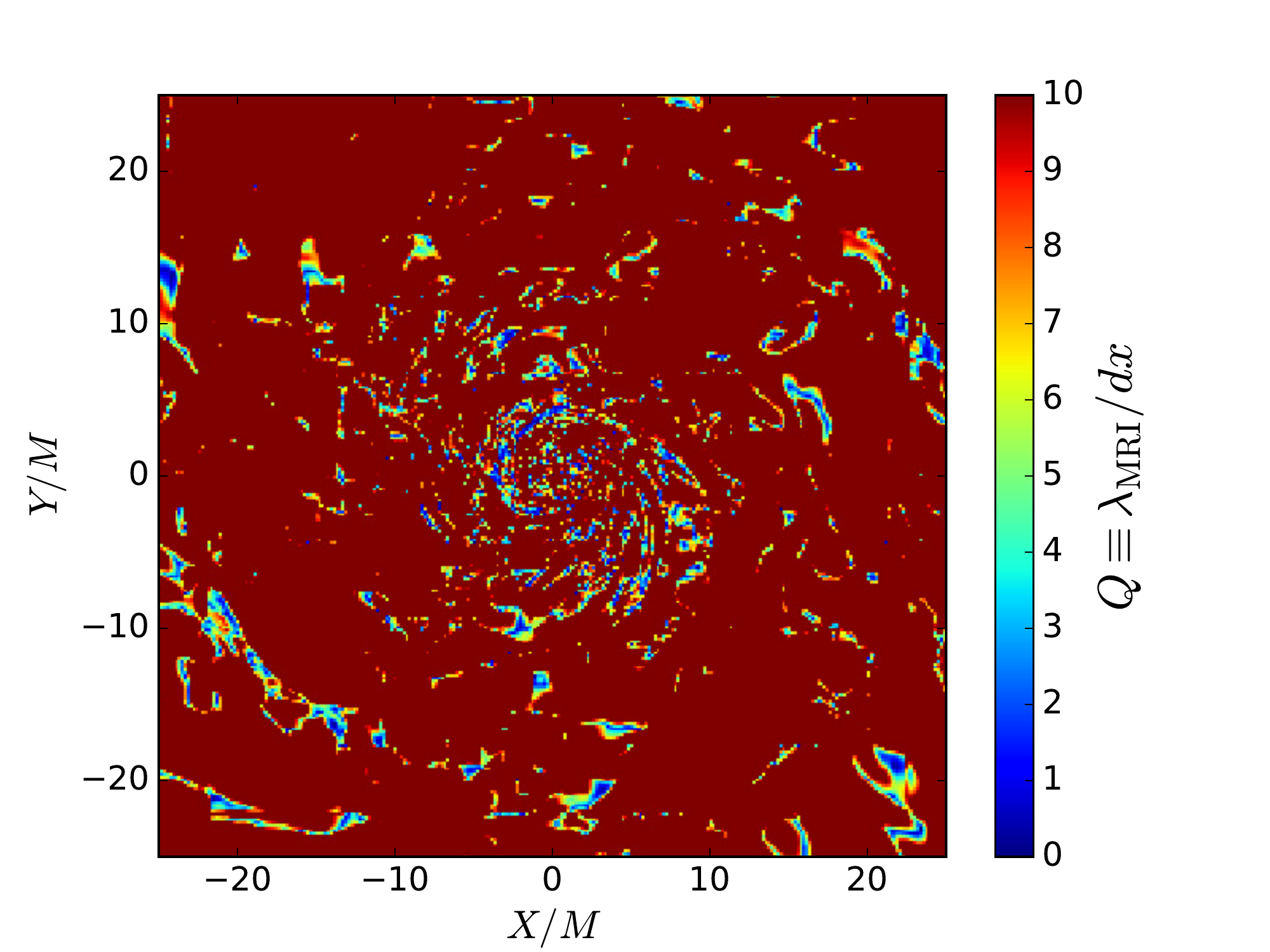}
  \includegraphics[width=0.52\textwidth]{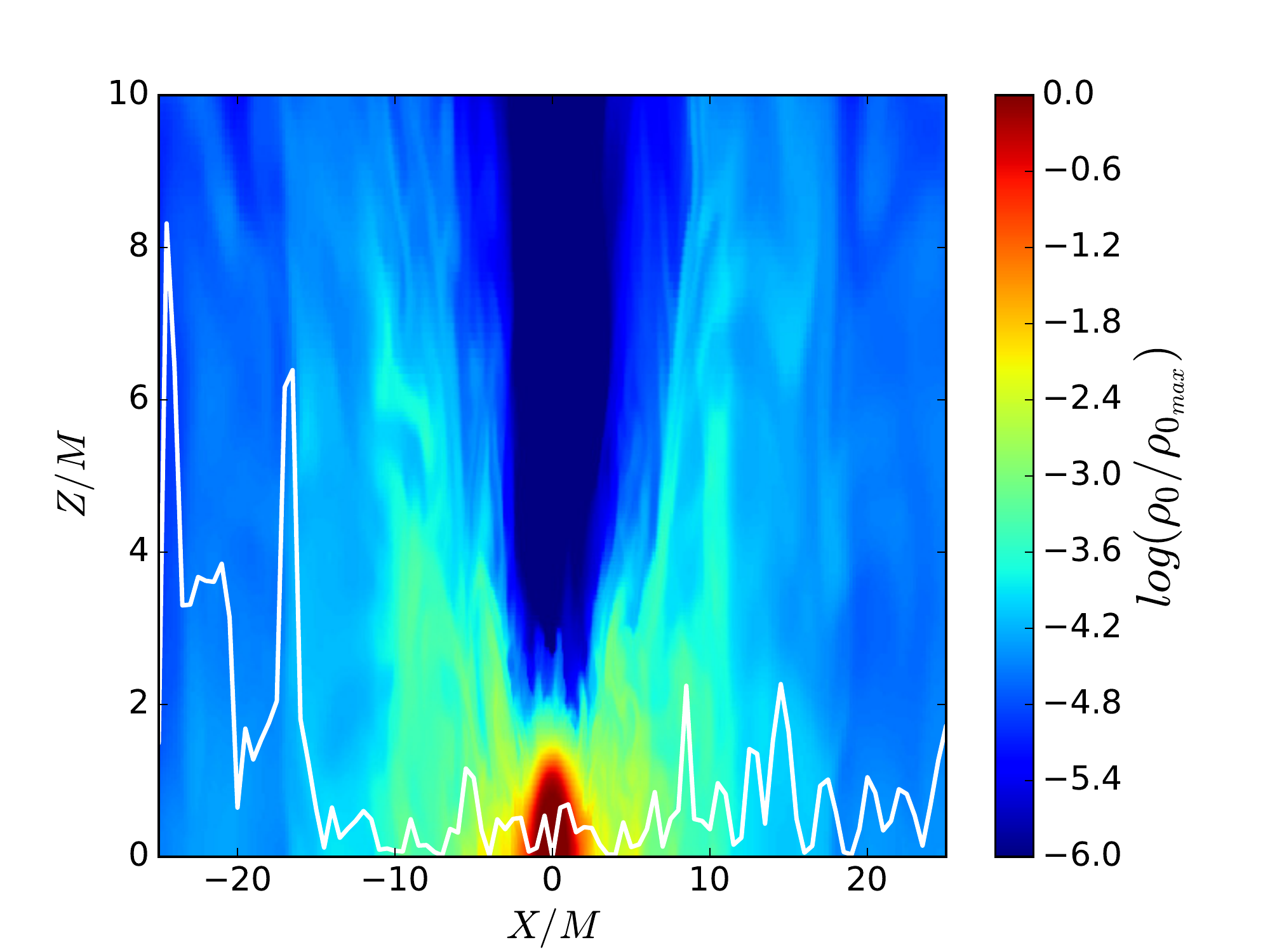}
  \caption{
    Contours of the quality factor $Q=\lambda_{\rm MRI}/dx$ on the equatorial plane (top),
    and the rest-mass density of the transient HMNS normalized to its initial maximum value
    (log scale) along with $\lambda_{\rm MRI}$ (white line) on the meridional plane (bottom) for
    Mag+Rad-Full at $t-t_{\rm GW}\approx 390M\sim 5.2\,\rm ms$. Similar behavior is observed in all
      magnetized cases in Table~\ref{table:Tab2} (see Fig.~7 in~\cite{Ruiz2021}).
    \label{fig:mri_alig}}
\end{figure}

%%%%%%%%%%%%%%%%%%%%%%%%%%%%%%
%%%   Final configuration  %%%
%%%%%%%%%%%%%%%%%%%%%%%%%%%%%%
\subsection{Final Configuration}
\label{sec:4-2}

\subsubsection{Unmagnetized NSNS binaries}
The left column in Fig.~\ref{fig:3Drho_1} displays several key moments during the evolution
of Unmag, while its neutrino counterpart is shown in the left column in Fig.~\ref{fig:3Drho_2}.
The HMNS remnant in Unmag lasts for approximately~$t-t_{\rm merge}=700 M \approx 9.3~\rm ms$
(see left column of Fig.~\ref{fig:3Drho_1}) followed by the formation of a highly spinning BH
with a mass $M_{\rm BH} \simeq 2.55M_\odot$ and spin $a_{\rm BH}/M_{\rm BH} \simeq 0.70$  immersed
in an accretion disk of radius $\sim 25 M\approx 100~\rm km$ which contains $\sim 2.2 \%$ of the
total mass of the system (see Table~\ref{table:Tab2}).
 By contrast, the HMNS remnant in Unmag+Rad-Simp collapses at $550 M \approx 7.4~\rm ms$ forming
 a spinning BH  with mass $M_{\rm BH} = 2.59M_\odot$ and spin parameter $a_{\rm BH}/M_{\rm BH} = 0.71$
 immersed in an accretion disk that extends to $\sim 40 M \approx 160~\rm km$ and contains  
 $\sim 1.8 \%$ of the total rest-mass of the binary (see left column of Fig.~\ref{fig:3Drho_2}).
 We observe that neutrino radiation induces an effective viscosity  (see below) causing the
 angular velocity of the Unmag+Rad-Simp transient to become almost uniform in the inner core.
 This causes an earlier collapse compared with Unmag.
 In both cases the accretion rate $\dot{M}$  begins to settle after $t-t_{\rm BH}\sim 380M\simeq 5\,\rm ms$
 reaching a value of $\sim 4 M_{\odot} \rm \, s^{-1}$~(see~Fig.~\ref{fig:Mdot}). The insert displays the
 fraction of the dynamical ejection of rest-mass following merger. We find that neutrinos do not significant enhance
 the ejection of matter. In both case $M_{\rm esc}/M_0 \simeq 1\times 10^{-3}$. These values are consistent
 with those reported previously in~\cite{Radice_2018, Radice_2021}.
 We note that transient kilonova signatures powered by the radioactive decay of heavy elements formed via rapid
 neutron capture (r-process) nucleosynthesis can be detected if the ejected mass is greater than $10^{-3}M_\odot$
 \cite{Li:1998bw,Metzger:2016pju}. We compute the estimated peak bolometric luminosity and the rise time of
 kilonova for these cases using an analytical model recently derived in~\cite{Perego14}, which assumes that 
$M_{\rm esc}$ is spherically distributed and expanding homologously (see Eqs.~(8)-(10) in~\cite{Ruiz2021}).
We find that the peak kilonova luminosities of the two unmagnetized cases are $L_{\rm knova} \approx 10^{41.3 \pm 0.2} \,
\rm erg\,s^{-1}$, the rise time is $\sim 4-11\,\rm h$, and an effective temperature $T_{\rm peak} \sim 10^{3.4}
\rm K$ (see Table~\ref{table:Tab2}). Converting the peak temperature to peak wavelength using the relation $\lambda_{\rm peak} =  1.35 \times
10^3 \, \rm nm \, (T_{\rm peak}/10^{3.33} \, K)^{-1}$~\cite{Perego14}, we find $\lambda_{\rm peak} \sim 1150 \,
\rm nm$. This emission may be observed by current or future instruments such as ALMA or
the Vera C. Rubin observatory~\cite{Matthews18, Chen21}.
%
%%%%%%%%%%%%%%%%%%%%%
%%%  Omega plot   %%%
%%%%%%%%%%%%%%%%%%%%%
\begin{figure}
  \centering
  \hspace{-6mm}\includegraphics[width=0.51\textwidth]{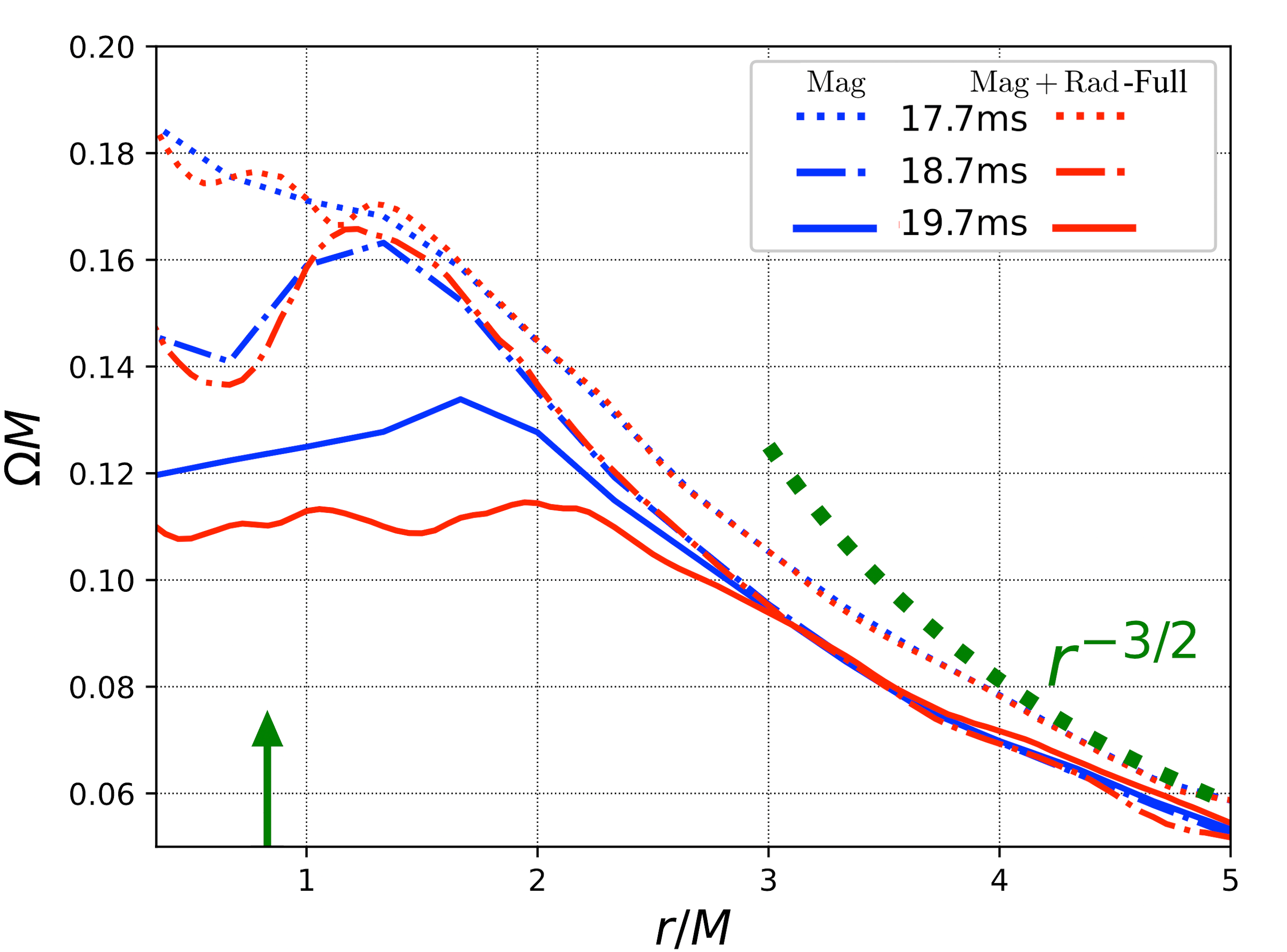}
  \caption{
    Average rotation profile of the HMNS (see Eq. 2 in~\cite{Ruiz:2019ezy}) for Mag (blue) and Mag+Rad-Full (red)
    on the equatorial plane following merger along with
    the Keplerian angular velocity profile. Times after neutrino inserion are included. The arrow marks the coordinate radius containing $50\%$ of the rest-mass
    of the HMNS.
    \label{fig:Omega}}
\end{figure}

%%%%%%%%%%%%%%%%%%%%%%%%%%
%%%  Magnetized NSNS   %%%
%%%%%%%%%%%%%%%%%%%%%%%%%%
%
\subsubsection{Magnetized NSNS binaries}
The basic dynamics and final outcome of the magnetized cases in Table~\ref{table:Tab2} are displayed in
the right column of Fig.~\ref{fig:3Drho_1} and in the middle and right columns of Fig.~\ref{fig:3Drho_2}
for case Mag, Mag+Rad-Simp, and Mag+Rad-Full, respectively. Consistent with our previous results
(see e.g.~\cite{Ruiz:2016rai,Ruiz2021}), in all cases we find that a magnetically-supported jet is
launched following the collapse of the transient HMNS. These results may indicate that incipient jets
are the typical outcome of NSNS undergoing delayed collapse.

Fig.~\ref{fig:mri_alig} shows that during the HMNS phase, the wavelength $\lambda_{\rm MRI}$ of the fastest-growing
mode is resolved by more than 10 grid points (top panel). In addition, $\lambda_{\rm MRI}$ fits in the bulk
of the HMNS (bottom panel), i.e. the region where the rest-mass density is $\rho_0\gtrsim 10^{11}\rm g/cm^3$,
well above of the floor density. Therefore, we conclude that MRI-induced turbulence is resolved and operating in our
systems. We observe that angular momentum is transferred from the inner to the outer
layers of the HMNS due both to magnetic winding and magnetic turbulence. Magnetic winding operates on a time scale of
\cite{s00, Sun:2018gcl}
\begin{eqnarray}
  \tau_{\text{\tiny A}}& \sim &\frac{R_{\text{\tiny HMNS}}}{v_{\text{\tiny A}}}\sim
  \\&&1\,{\rm ms}\,
  \left(\frac{\rho}{10^{15}\rm g/cm^3}\right)^{1/2}\,
  \left(\frac{|B|}{10^{15}G}\right)^{-1}\,
  \left(\frac{R_{\text{\tiny HMNS}}}{10^5\rm cm}\right)\,.
  \nonumber
\label{eq:alfven_timescale}
\end{eqnarray}
The transport of angular momentum induces the formation of a massive, nearly uniformly rotating core surrounded
by a Keplerian cloud of matter
(see Fig.~\ref{fig:Omega}). Here $v_{\text{\tiny A}}\sim|B|/\sqrt{4\pi\rho}$ is the  Alfv\'en speed, 
$B$ is the strength of the magnetic field, and $\rho$ and $R_{\text{\tiny HMNS}}$ the characteristic
rest-mass density and radius of the remnant, respectively. Similar results are reported in~\cite{Ruiz2021},
which, together with the discussion of neutrino viscosity below, suggest that
neutrino effects do not influence the growth of MRI. We note that the relativistic smoothed particle
hydrodynamics simulations
in~\cite{Guilet17} suggest that the growth rate of the MRI is significantly reduced by neutrino effects
if the magnetic field is lower than $\lesssim 10^{14} \rm G$ inside the HMNS. However, it is expected
that typical vales of the magnetic field in the HMNS are $\gtrsim 10^{15.5}\rm G$ due to the
KHI~\cite{Palenzuela:2021gdo,kkssw14}.

Calculating the effective Shakura–Sunyaev $\alpha_{\rm SS}$ parameter in the HMNS for the magnetized cases
in Table~\ref{table:Tab2}, we found that it ranges between $0.01$ and $0.09$. Similar values have
been reported in high-resolution NSNS merger simulations~(see e.g.~\cite{Kiuchi:2017zzg}). These results
explain why turbulent magnetic fields can redistribute angular momentum and damp the differential rotation
effectively on an effective turbulent magnetic viscous timescale due to MRI (see Eq.~(7)~in~\cite{Sun:2018gcl}):
%
%%%%%%%%%%%%%%%%%%%%%
%%%   rho0 on xz  %%%
%%%%%%%%%%%%%%%%%%%%%
%
\begin{figure}
\includegraphics[scale=0.47]{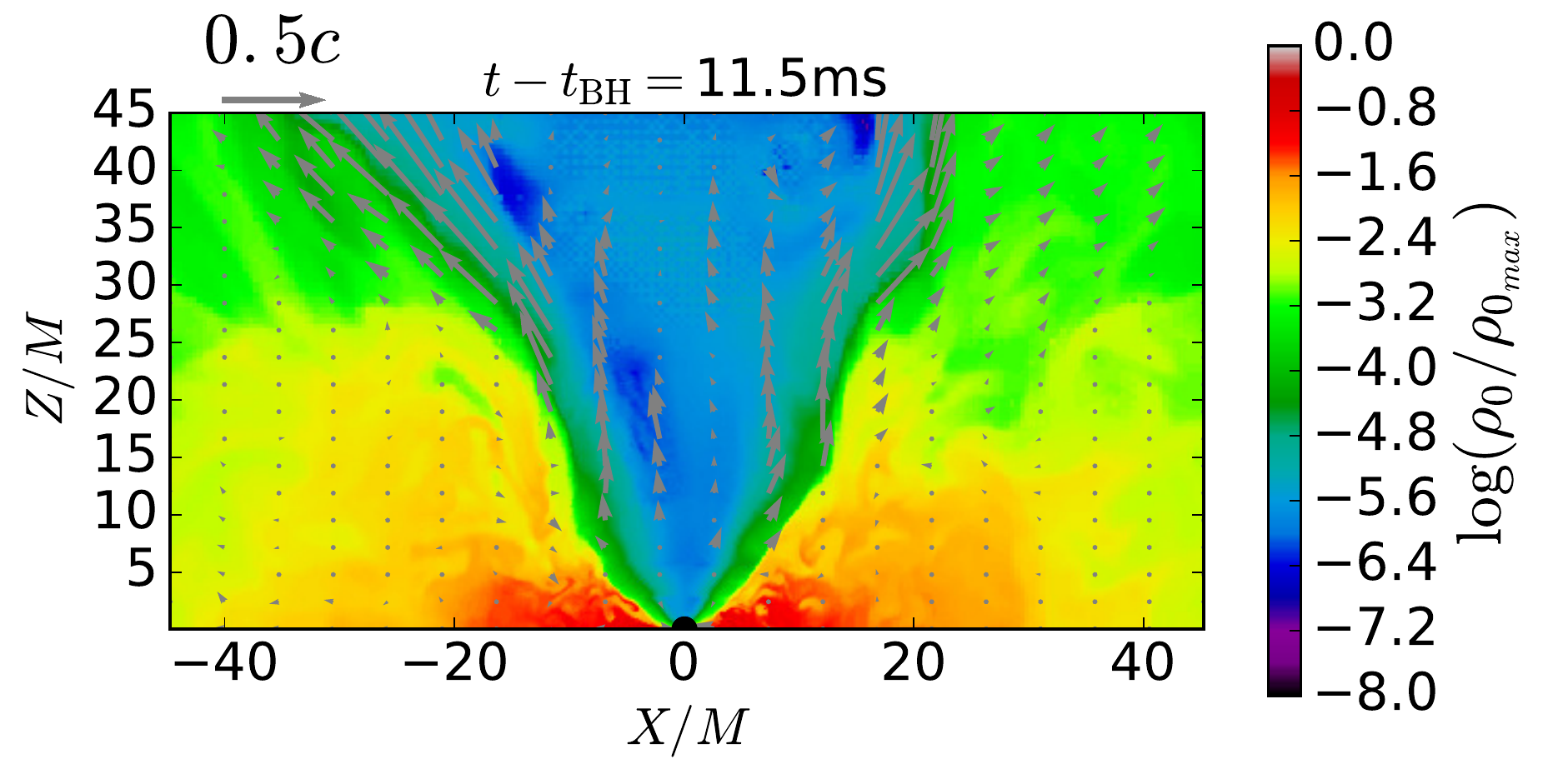}
\includegraphics[scale=0.47]{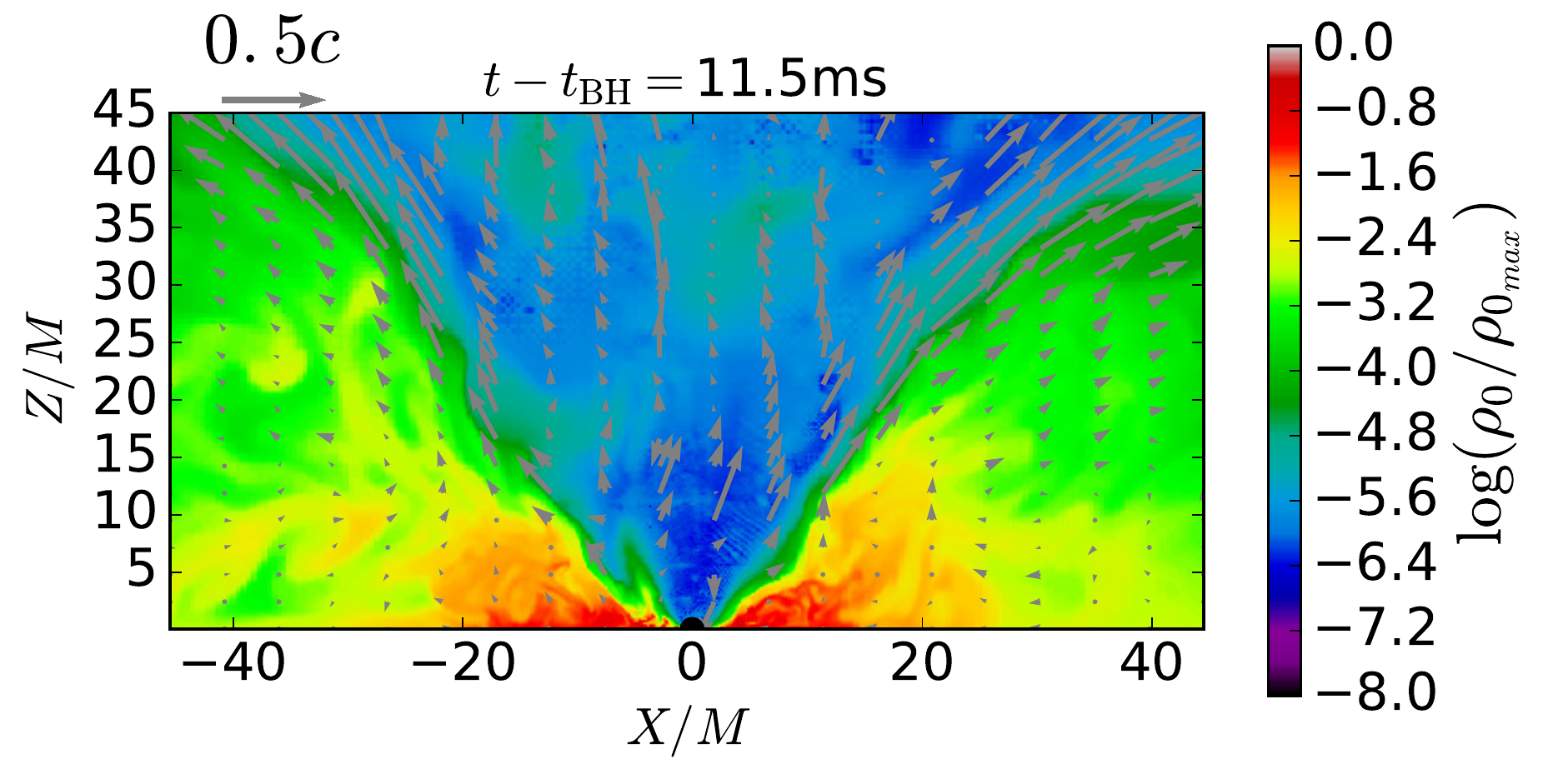}
\includegraphics[scale=0.47]{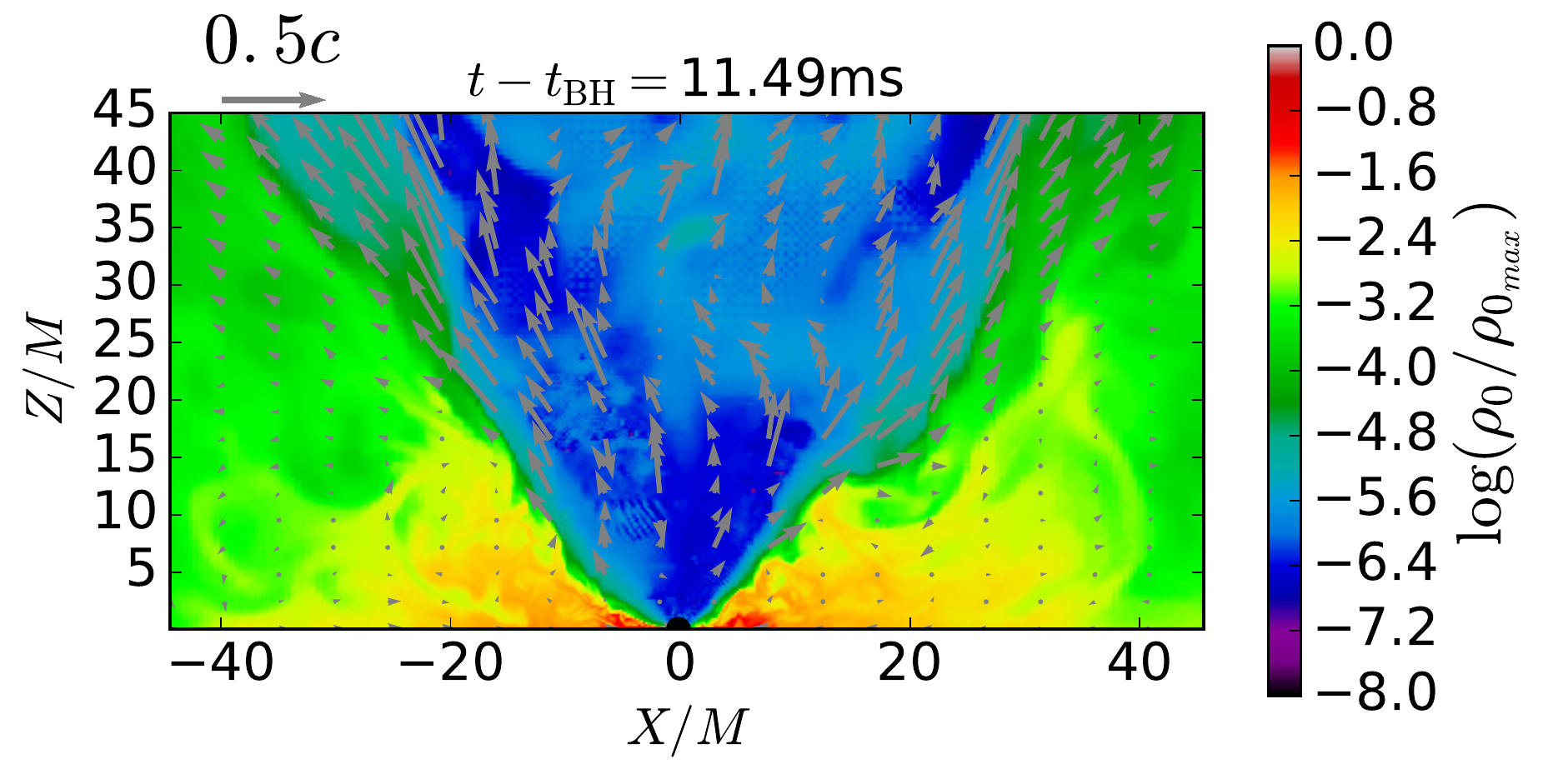}
\caption{\label{fig:rho2d} Rest-mass density normalized to its initial maximum value $\rho_{0,\rm max}$
  (log scale) on the meridional x-z plane for Mag (top), Mag+Rad-Simp (middle), and Mag+Rad-Full (bottom)
  cases  at $t-t_{\rm BH}\approx 11.5~\rm ms$. A BH apparent horizon is displayed as a black semi-circle, while
  arrows indicate fluid velocities.}
\end{figure}
\begin{eqnarray}
  \tau_{\rm vis} &\sim {(\alpha_{\rm SS}\,\Omega)}^{-1}& \,\sim {R_{\text{\tiny HMNS}}^{3/2}}\,{M^{-1/2}_{\text{\tiny HMNS}}
    \,\alpha_{\rm SS}^{-1}}\\
  &&\sim 1\,{\rm ms}\,
  \left(\frac{\mathcal{C}}{0.3}\right)^{-3/2} \,
 \left(\frac{M_{\text{\tiny HMNS}}}{3 M_\odot}\right)\,  
  \left(\frac{\alpha_{\rm ss}}{10^{-2}}\right)^{-1}
 \,,\nonumber
   \label{eq:magnetic_timescale}
\end{eqnarray}
where $M_{\text{\tiny HMNS}}$ is the characteristic mass of the HMNS and $\mathcal{C}=
M_{\text{\tiny HMNS}}/R_{{\text{\tiny HMNS}}}$ is its compaction.

As described in~\cite{Guilet17}, the diffusion of neutrinos trapped in the HMNS also induces an angular momentum transport. We observe (see~Fig.~\ref{fig:Omega}) that neutrino transport helps drive the central core of the HMNS to nearly uniformly rotation, causing the
HMNS in the Mag+Rad cases to collapse $\sim 4\,\rm ms$ faster than in Mag (see Table~\ref{table:Tab2}).
Due to this effect, more material from the external layer of the HMNS remains inside the
innermost stable circular orbit (ISCO) during the BH formation, which induces the formation of lighter accretion
disks ($\lesssim 25\%$) in the Mag+Rad cases than in the Mag case (see Table~\ref{table:Tab2}).
Note that neutrino radiation can also help trigger the collapse of the HMNS by
reducing the thermal support, though we do not observe a significant change in the gas temperature in our
simulations. We caution that the collapse time of a short-live HMNS depends on the strength of
the seed magnetic field (see e.g. Fig.~2~in~\cite{grb11}), as well as on the numerical resolution, even
in non-magnetized evolutions~\cite{Paschalidis:2015mla,East:2015vix}.

As described in~\cite{Guilet15, Guilet17}, for MRI wavelengths shorter than the neutrino mean free path, neutrino radiation induces a drag on the velocity with a damping rate independent of the wavelength. Here we study the viscosity and dragging effect due to neutrinos on the
suppression of MRI. Comparing the analytical estimates of the effective viscosity due to neutrino diffusion
$\nu_{\rm neutrino}$ (see Eq.~\ref{eq:nu_neutrino})
and the effective viscosity induced by MHD turbulence using an $\alpha$-disk model $\nu_{\rm MHD}$~\cite{Artymowicz94, MacFadyen_2008}, we find that $\nu_{\rm neutrino}$ does not have an significant effect on the suppression of MRI both inside
HMNS and in the disk. The magnetic field strength in the HMNS in all our simulations exceeds the critical strength of
the magnetic field, $B_{\rm crit} \sim 10^{14} \, \rm G$, below which the viscosity can significantly suppresses the growth of
the MRI~\cite{Guilet17}. The viscous effect is controlled by the Elsasser number $E_\nu \equiv v^2_A/(\nu_{\rm neutrino} \Omega)$
\cite{Pessah_2008, Longaretti10, Guilet15}, which is $E_\nu \gtrsim 10$ in our simulations. This value indicates that the effective viscosity induced by neutrinos has little effect on to the growth of MRI (see Fig.~\ref{fig:mri_alig} and Fig.~7
in~\cite{Ruiz2021}). On the other hand, when the wavelength of the fastest growing MRI mode is shorter than the mean free
path of heavy-lepton neutrinos (which gives the most restrictive constraint among all species), neutrinos can induce a drag force.
Using the scaling relation of the heavy lepton neutrino mean free path~\cite{Guilet15}
\begin{equation}
  l_\nu =10^4\,\left(\frac{\rho_0}{10^{13}\,\rm g\,cm^{-3}}\right)^{-1}\,
  \left(\frac{T}{10 \,\rm MeV}\right)^{-2}\,\rm cm\,,
\end{equation}
we observe that neutrino drag becomes significant only when the magnetic field is lower than the critical magnetic
field strength ($\lesssim 10^{14}\,\rm G$) in the HMNS, and hence neither the effective neutrino viscosity nor drag force
has a distinguishable influence on the development of MRI in our simulations during the HMNS phase

We also estimate analytically the shear viscosity due to nucleon-nucleon scattering using Eq. (14) in ~\cite{Cutler87}
as a function of $T$ and $\rho_0$. We note that this expression assumes non-superfluid matter and ignores the collisions
of electrons and muons. We find that it has a comparable magnitude as $\nu_{\rm neutrino}$ deep inside the HMNS but
decreases quickly as $r > 5~\rm km$ (see Fig.~\ref{fig:vis_vs_x} in Appendix~\ref{subapp:nu-vis}). Therefore, we conclude
that MHD-induced viscosity is the dominant viscosity source. The transport of angular momentum in the neutrino-involved
cases is more substantial due to both MRI
and neutrino-induced effective shear viscosity (see Appendix~\ref{subapp:nu-vis} for more detailed discussions regarding viscosities). Note
that we ignore the bulk viscosity due to Urca interactions and hyperons processes, since the shear viscosity is most
important in transporting and redistributing angular momentum in the HMNS, which leads to its collapse. Bulk viscosity
is mainly important during the late inspiral phase~\cite{Most22}.
%
%%%%%%%%%%%%%%%%%%%%%%
%%%   Luminosity   %%%
%%%%%%%%%%%%%%%%%%%%%%
%
\begin{figure}[]
\centering
\includegraphics[scale=0.20]{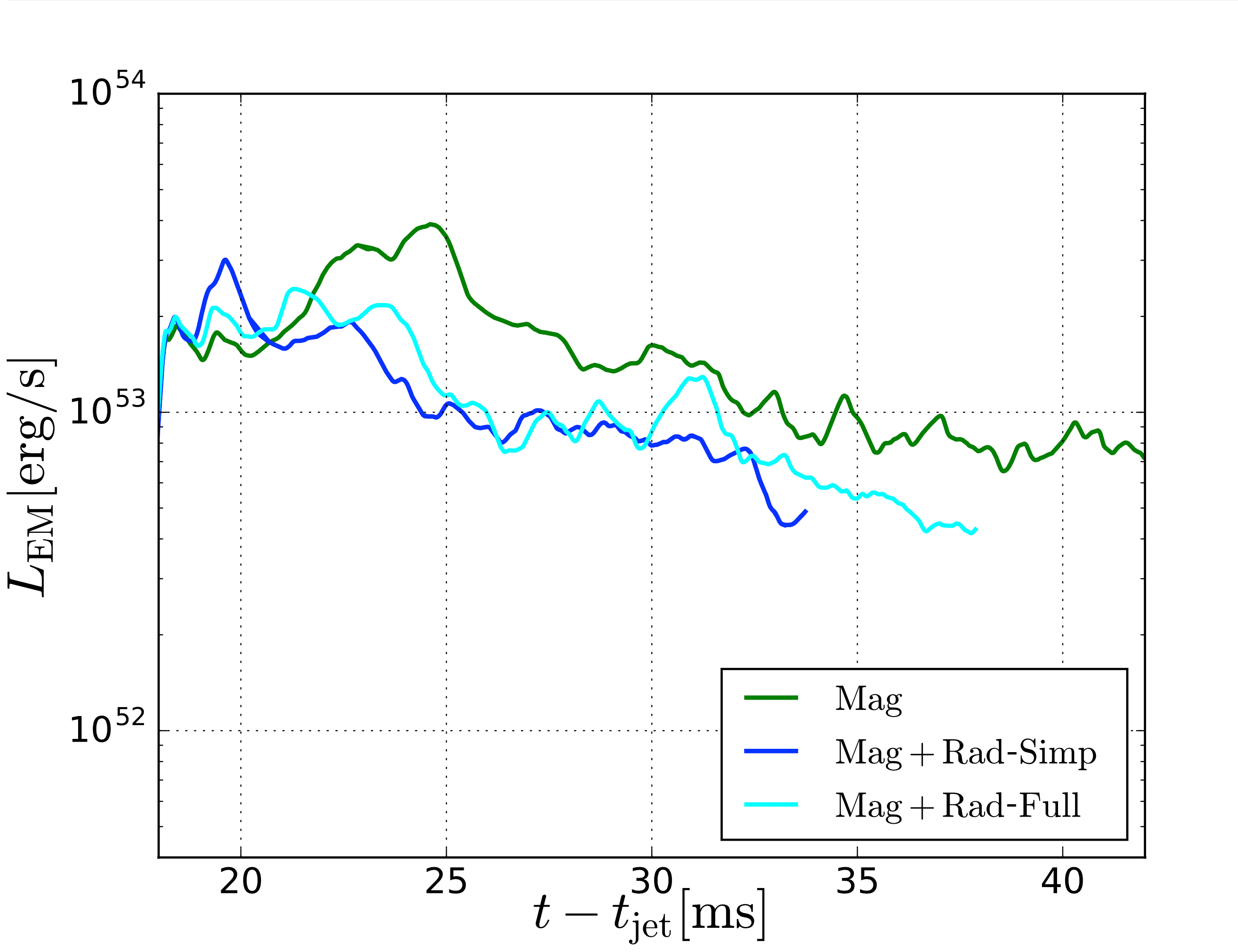}
\caption{\label{fig:L_em} Outgoing EM Poynting luminosities extracted at a coordinate sphere of
  radius~$r = 160M\sim 640\,\rm km$ as a function of $t - t_{\rm jet}$ for all magnetized cases in
  table~\ref{table:Tab2}.}
\end{figure}

The transient HMNS remnant collapses to a BH with mass $M_{\rm BH}\sim 2.4-2.6M_\odot$, and with spin
$a/M_{\rm BH} \sim 0.62$ for Mag and $a/M_{\rm BH} \sim 0.68$ for Mag+Rad cases (see Table~\ref{table:Tab2}). 
By $t-t_{\rm BH}\sim 450M\sim 6~\rm ms$ following BH formation, when the accretion rate $\dot{M}$ begins to settle
down (see Fig.~\ref{fig:Mdot}),  we observe that magnetic-dominated regions ($B^2/(8\,\pi\rho_0)\gtrsim 1$) above the BH poles
begin to expand and reverse the fall-back debris inflow. Magnetic winding above that BH poles, which began during
the HMNS phase (see third-row right panel in Fig.~\ref{fig:3Drho_1} and top panels in Fig~\ref{fig:3Drho_2}),
induces the formation of a tightly wound helical magnetic funnel. As the accretion near the equator proceeds
and neutrino processes take place, the baryon-loaded environment inside the funnel gradually becomes thinner,
allowing a magnetically-driven jet to emerge eventually~(see bottom panels in Figs.~\ref{fig:3Drho_1} and~\ref{fig:3Drho_2}).
We note that in the Mag+Rad
cases, a magnetically-supported jet is launched after $t-t_{\rm BH}\sim 750M\sim 10\,\rm ms$, while in Mag it is launched
at $t-t_{\rm BH}\sim t-t_{\rm BH}\sim 1800M \sim 24\,\rm ms$~\cite{Ruiz2021}. Fig.~\ref{fig:rho2d} displays the
rest-mass density on the meridional plane for the above cases. We observe that the rest-mass density inside
the funnel of the Mag+Rad cases is a factor of $\sim 10$ lighter than in Mag. This suggests that neutrino radiation
reduces the baryon-loaded environment in the polar region above the BH poles. As we do not observe an enhancement of
the magnetic field strength following BH formation, the funnel emptying due to neutrinos makes it easier for the magnetic
pressure gradients above the BH poles to overcome the fall-back ram pressure of the infalling material.
Similar results have also been reported in~\cite{M_sta_2020}. We note that by $t-t_{\rm BH}\sim 6\,\rm ms$, the
accretion rate is $\sim 2\,M_{\odot}/\rm s$ and the rest-mass of the accretion disk is $\sim 0.2M_\odot$ (see
Table~\ref{table:Tab2}). Therefore, the disk will be accreted in $\Delta t\sim M_{\rm disk}/\dot{M}\sim 0.1\,\rm s$,
roughly consistent with the lifetime of a sGRB central engine~\cite{Bhat:2016odd,Beniamini:2020adb}.

Fig.~\ref{fig:L_em} displays the outgoing Poynting luminosity $L_{\rm EM}$ as a function of $t - t_{\rm jet}$
for cases in Table~\ref{table:Tab2}. We observe that $L_{\rm EM}\sim10^{53}~\rm erg\, s^{-1}$. As pointed out
in~\cite{Ruiz2021}, the Poynting luminosity roughly agrees with the theoretical range we derived in~\cite{shapiro:2017cny}
for BH + disk + jet systems arising from compact binary mergers containing NSs, or from the magnetorotational collapse
of massive stars. The luminosities are also in accord with the narrow range characterizing the observed
luminosity distributions of over 400 GRBs~\cite{LiZhangLu16,Beniamini_2020}.

Following~\cite{Ruiz2021}, we assess if the BZ mechanism is likely operating in the BH + disk remnant.
We begin by comparing the outgoing Poynting luminosity $L_{\rm EM}$ in our simulations~(see~Fig.
\ref{fig:L_em}) with that from the BZ mechanism~\cite{Thorne86} 
\begin{eqnarray}
L_{\rm BZ}\sim 10^{52}\,\left(\frac{\tilde{a}}{0.75}\right)^2\,
  \left(\frac{M_{\rm BH}}{2.8 M_\odot}\right)^2\,|B_{\rm p}|_{16}^2\,\rm erg\, s^{-1}\,,
  \label{eq:LBZ}
\end{eqnarray}
where $|B_{\rm p}|_{16}\equiv|B_{p}|/10^{16}\rm G$ is the strength of the magnetic field at the BH
poles. The EM Poynting luminosity in our simulation is $\sim 10^{53}\,\rm erg\, s^{-1}$~(see Table~\ref{table:Tab2}).
Next, we estimate that $|B_{\rm p}|\sim~10^{16}\,\rm G$, and hence $L_{\rm BZ}\sim 10^{52}\rm\, erg\, s^{-1}$,
in roughly agreement. We also estimate the magnetic field-to-BH angular frequency ratio $\Omega_F/\Omega_H$
on a meridional plane passing through the BH centroid and along coordinate semicircles of radii $r_{\rm BH}$
and $2\,r_{\rm BH}$. We find that in all cases $\Omega_F/\Omega_H\sim 0.2-0.6$. As pointed out in~\cite{Ruiz2021},
deviation from the split-monopole value $\sim 0.5$~\cite{Komissarov2001} may be due to  artifacts such
as the deviation from strictly force-free conditions, deviations from monopole geometry and/or
lack of resolution.  Our above results suggest that the BZ mechanism is likely operating in our
system, as we concluded in~\cite{Ruiz:2016rai,Ruiz:2019ezy}.

Roughly $3\times 10^{-2}$ of the total initial mass escapes the system after steady state in Mag and Mag+Rad-Full cases
and $1.5 \times 10^{-2}$ of the total mass escapes in the Mag+Rad-Simp case. We estimate that the peak kilonova luminosity and
rise time of the case Mag+Rad-Simp are $\approx 10^{41.5} \, \rm erg \, s^{-1}$ and $\approx \, \rm 9.6 \, hours$, respectively. For Mag+Rad-Full,
the peak kilonova luminosity reaches $\approx 10^{41.6} \, \rm erg \, s^{-1}$ at around $\approx 11.1 \, \rm hours$.
Moreover, we estimate that the peak kilonova luminosity, the rise time, and the peak temperature for the magnetized cases are
similar to those in the unmagnetized cases (see Table~\ref{table:Tab2}). 

%%%%%%%%%%%%%
%%%    GW %%%
%%%%%%%%%%%%%%
\subsection{GW Signals}
%

%%%%%%%%%%%%%%%%%%%%
%%%  waveforms  %%%
%%%%%%%%%%%%%%%%%%%%
\begin{figure}[h]
\includegraphics[scale=0.37]{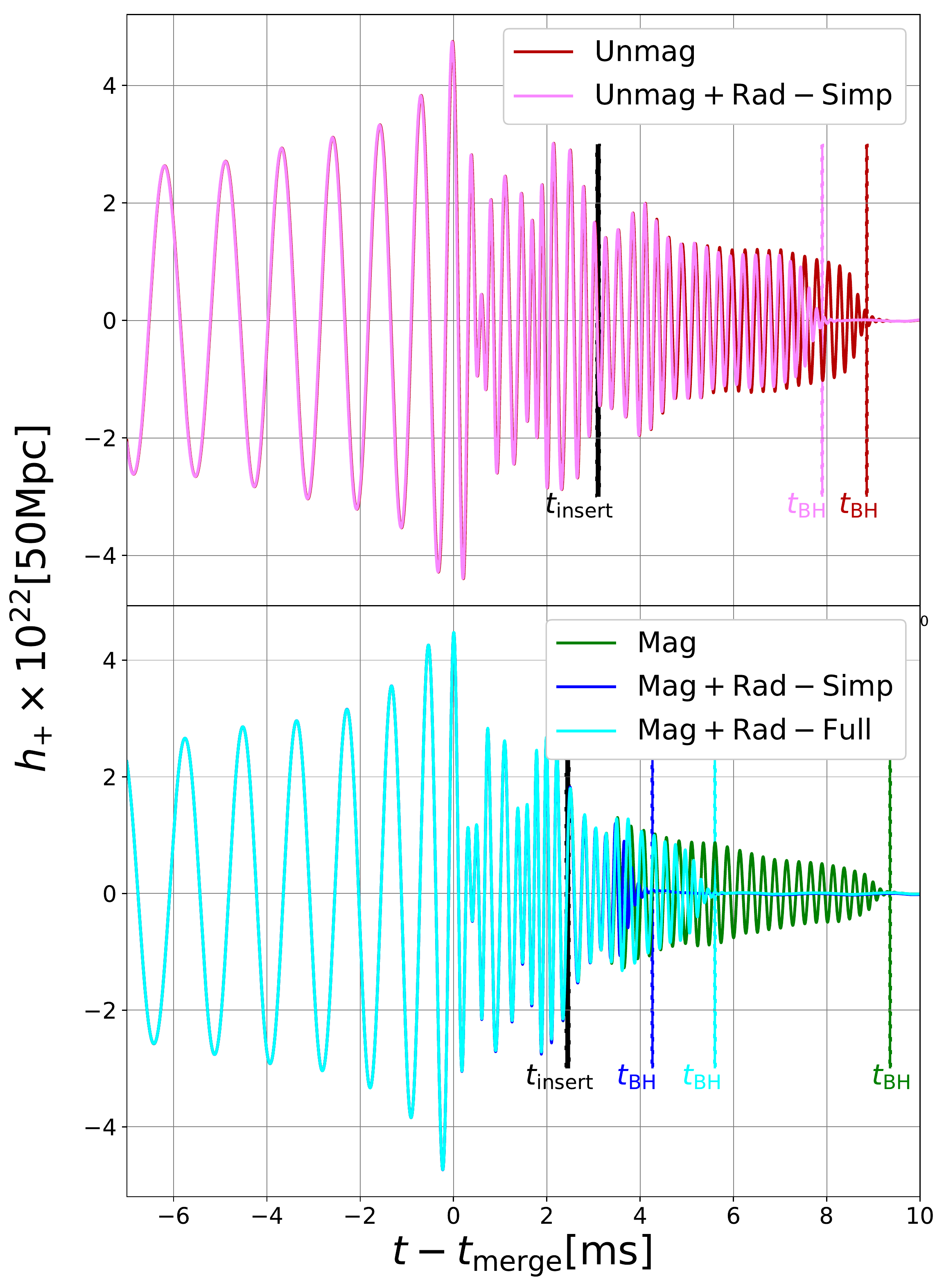}
\caption{\label{fig:hp_2} 
  GW strain of the dominant mode $h^{22}_+$  for unmagnetized (top panel) and magnetized (bottom panel) cases
  at a source distance of $50\,\rm Mpc$. Vertical lines mark the neutrino insertion (Mag + Rad cases), and BH
  formation times.}
\end{figure}
%
%%%%%%%%%%%%%%%%%%%%
%%%   spectrum   %%%
%%%%%%%%%%%%%%%%%%%%
%
\begin{figure}[h]
    \centering
    \includegraphics[scale=0.36]{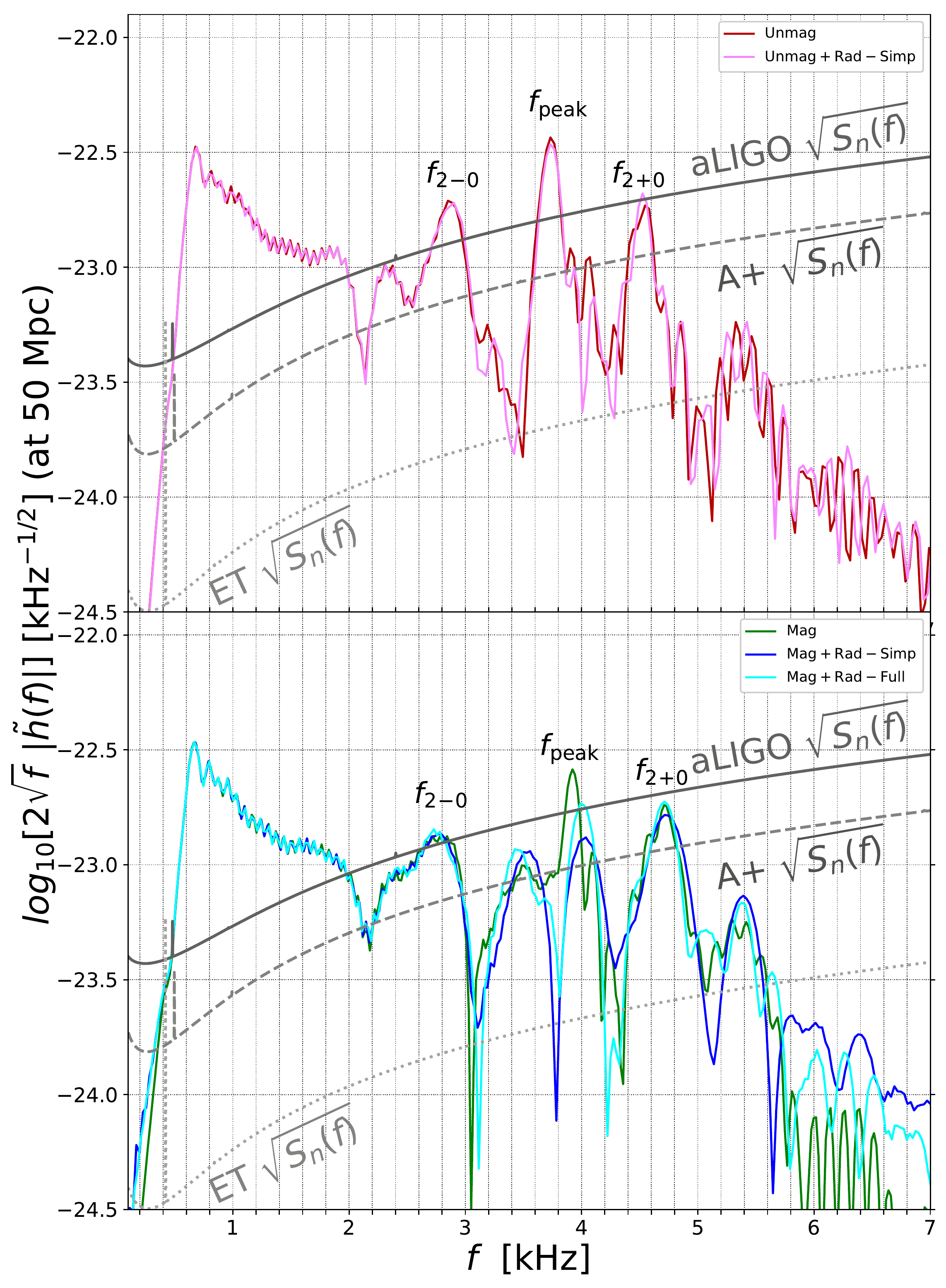}
    \caption{GW power spectrum of the dominant mode $(l, m) = (2, 2)$ at a distance of $50\,\rm Mpc$ for
      unmagnetized (top panel) and magnetized (bottom panel) cases. The solid-dark, dashed- and dotted-light gray
      curves displays the noise sensitivity curves of aLIGO (\url{ZERO_DET_HIGH_P} configuration), $A^+$,
      and the Einstein Telescope (ET-D)~\cite{Hild_2011}, respectively. Main spectral frequencies are denoted
      as $f_{\rm peak}$ and $f_{2\pm0}$.}
    \label{fig:strain_2}
\end{figure}

Fig.~\ref{fig:hp_2} displays the dominant $l = m = 2$ mode of the GW strain $h_+$ versus time, shifted by $t_{\rm merge}$
for the unmagnetized (top panel) and the magnetized (bottom panel) cases. Vertical lines mark the neutrino insertion and
BH formation time, respectively. We note that during the inspiral, the GW amplitude differences between these cases is
$\lesssim 3\%$, with the peak amplitudes shifted by $\sim 80M~\sim 1\,\rm ms$. This result is anticipated because the seed
magnetic field is initially dynamically unimportant. Following merger, nonaxisymmetric rotation and oscillation modes of the HMNS,
which persist until stellar collapse to a BH, trigger the emission of quasiperiodic GWs. We observe that
dissipation of energy and angular momentum due to GW radiation is more efficient in the unmagnetized cases,
where the GW amplitude is a factor of $\sim 1.5$ larger than that in the magnetized cases. As pointed out
in~\cite{Ruiz2021}, magnetic turbulence is an efficient mechanism in damping differential rotation and
driving the system into an almost axisymmetric configuration. We also note that neutrino transport of angular momentum (see Fig.
\ref{fig:Omega}) enhances the triggering the stellar collapse to a BH.

Fig.~\ref{fig:strain_2} displays the power spectrum at a distance of $50\rm\,Mpc$ for all cases in Table~\ref{table:Tab2} along
with the sensitivity curves  {\tt ZERO$\_$DET$\_$HIGH$\_$P} for aLIGO, $A^+$, and ET in~\cite{LIGOScientific:2016wof}.
The spectra  show three distinctive peaks, which are previously referred as $f_{2-0}, f_{\rm peak},$ and $f_{2+0}$ in~\cite{Bauswein15, Vretinaris20}
and also studied in~\cite{Takami2014, Takami2015, Rezzolla2016, Ruiz2021}. It has been suggested that the most prominent peak, $f_{\rm peak}$,
is due to the rotation of the bar-deformed HMNS~\cite{Sterg:11}. However, the origins of $f_{2-0}$ and $f_{2+0}$ are still debatable. One possible
explanation of the origin of $f_{2-0}$ is the nonlinear interaction between the quadrupole and quasiradial modes, and $f_{2+0}$ is the result
of the nonlinear interaction of $f_{\rm peak}$ with other nonquasiradial modes~\cite{Sterg:11}. Another possible origin of these two modes
are the nonlinear oscillations of the two repeatedly colliding and bouncing stellar cores~\cite{Takami2015}. 
As a result, it is expected that $f_{\rm peak} \approx (f_{2-0} + f_{2+0})/2$ for typical HMNS remnants~\cite{Takami2015}.
In our cases, we note that the three peaks $(f_{2-0}, f_{\rm peak}, f_{2+0})$ are located at $(2.8 \pm 0.05, 3.8 \pm 0.1,
4.7 \pm 0.1$ kHz), and hence roughly agree with this relation. Moreover, the location of $f_{\rm peak}$ agrees with the rotation
half-period of the bar-mode in HMNS at $\sim 0.26 \pm 0.02 \rm \, ms$, confirming the bar-mode origin of $f_{\rm peak}$. 
We also note that the  main frequency $f_{\rm peak}$ in  the unmagnetized cases is well above the  sensitivity curve of aLIGO,
and therefore it may be detected with the current sensitivity. By contrast, the main frequency in the magnetized cases is
only marginally above it. The other peak frequencies are either marginally above or below the sensitivity curve of aLIGO.
These results suggest that the next generation GW observatories, such as $A^+$ or the ET (see Fig.~\ref{fig:strain_2}),
are required to characterize the GW signals from GW170817-like remnant events.  Furthermore, the majority of the spectrum
between 0 and $5\,\rm kHz$ can be detected by both $A^+$ and the ET, which suggests the potential successful detection
of such a source.

As shown in the top panel in Fig.~\ref{fig:strain_2}, The spectra of the two unmagnetized cases show a high resemblance,
with several lower dips in the Unmag+Rad-Simp cases as a result of the slightly shorter HMNS lifetime.
The main peak $f_{\rm peak}$ for the Mag case lies above aLIGO noise curve, while the two cases with neutrinos have
significantly reduced main peaks. On the other hand, the $f_{2-0}$ and $f_{2+0}$ peaks are relatively well-preserved.
This suggests that the reduced ringdown phase in the neutrino cases may have a contribution to $f_{\rm peak}$ but has little
impact on the other two peaks. The relationship between the ringdown waveform and $f_{\rm peak}$ can also be inferred from
comparing our waveforms with the results of the general-relativistic hydrodynamics simulations in~\cite{Takami2015, Rezzolla2016}
with the SLy EOS and an initial NS mass of 1.35 of $M_\odot$. We find that the frequencies of the three peaks in their models are
roughly consistent with our unmagnetized cases.

We also assess if any signature of the magnetic field or neutrino radiation can be observed in the GWs.
For this, we  compute the match function $\mathcal{M}_{\text{\tiny{GW}}}$ defined in~\cite{Allen2012}
\begin{equation}
\mathcal{M}_{\text{\tiny{GW}}} =  \underset{(\phi_c,t_c)}{{\rm max}}
\frac{\left<{h}_1|{h}_2(\phi_c,t_c)\right>}{\sqrt{\left<{h}_1|{h}_1\right>
    \left<{h}_2|{h}_2\right>}}\,,
\label{eq:match}
\end{equation}
where $\left<{h}_1|{h}_2\right>$ is the noise-weighted inner product defined as~\cite{Allen2012}
\begin{equation}
\left<h_1|h_2\right>= 4\,{\rm Re}\int_0^\infty\frac{\tilde{h}_1(f)\,\tilde{h}^*_2(f)}
  {S_h(f)}\,df\,,
\end{equation}
where $h=h_+-i\,h_\times$, $\tilde{h}$ is the Fourier transform of the
strain amplitude $\sqrt{\tilde h_+(f)^2 +\tilde h_{\times}(f)^2}$ of the dominant mode $(l,m)=(2,2)$,
and $S_h(f)$ is a given detector sensitivity. We note that the value of $\mathcal{M}_{\text{\tiny{GW}}}$
at which two GW waveforms can be distinguishable potentially by instruments depends on the signal-to-noise
ratio (SNR). In particular, two signals are indistinguishable when~$\mathcal{M}_{\text{\tiny{GW}}}\gtrsim0.9978$
for a SNR of 15 or when $\mathcal{M}_{\text{\tiny{GW}}}\gtrsim 0.9992$ for a SNS of 25~\cite{Harry2018}.
Using the sensitivity curves in~\cite{aLIGO2009} and assuming a source distance of 50 Mpc, we find that:
\begin{itemize}
\item $\mathcal{M}_{\text{\tiny{GW}}} =0.9708$ between Unmag and Mag             with a SNR of $\sim 3$ for aLIGO, $\sim 1$ for KAGRA, $\sim 7$ for $A^+$, and $\sim 30$ for ET,
\item $\mathcal{M}_{\text{\tiny{GW}}} =0.9998$ between Unmag and Unmag+Rad-Simp with a SNR of $\sim 3$ for aLIGO, $\sim 1$ for KAGRA,  $\sim 7$ for $A^+$, and $\sim 30$ for ET,
\item $\mathcal{M}_{\text{\tiny{GW}}} =0.9993$ between Mag   and Mag+Rad-Simp   with a SNR of $\sim 3$ for aLIGO, $\sim 1$ for KAGRA,  $\sim 7$ for $A^+$, and $\sim 30$ for ET,
\item $\mathcal{M}_{\text{\tiny{GW}}}=0.9989$  between Mag   and Mag+Rad-Full   with a SNR of $\sim 3$ for aLIGO, $\sim 1$ for KAGRA,  $\sim 7$ for $A^+$, and $\sim 30$ for ET.
\end{itemize}
Notice that a SNR of $\sim 30$ requires a GW event at a distance of
$\sim  2\,\rm Mpc$ for KAGRA, $\sim 6.0\,\rm Mpc$ for aLIGO, or $\sim 12.0\,\rm Mpc$ for $A^+$.
GW170817, the closest GW signal detected to date, had a luminosity distance of $40^{+8}_{-14}\,\rm
Mpc$~\cite{TheLIGOScientific:2017qsa}. Therefore, it is unlikely that the current GW detectors can 
discern any signature from magnetic fields or neutrinos. However, next generation observatories (e.g.~ET) can easily
observe their imprints on the GWs.

%%%%%%%%%%%%%%%%%%%%%%%%%%%%%%
%%%   Neutrino Emission    %%%
%%%%%%%%%%%%%%%%%%%%%%%%%%%%%%
\subsection{Neutrino Emission}
\label{sec:4-5}
\begin{figure}[]
  \includegraphics[scale=0.19]{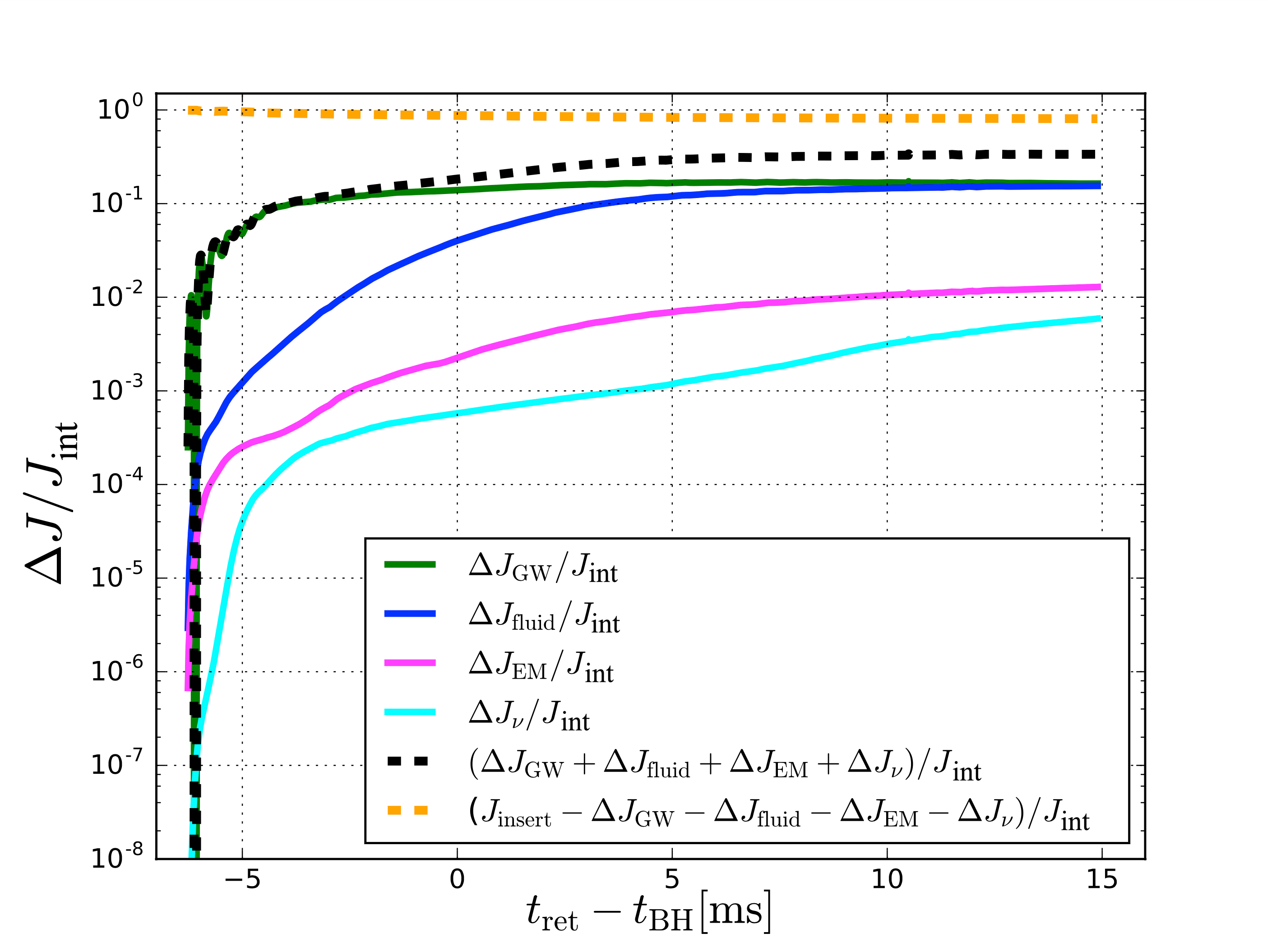}
  \caption{\label{fig:J_adm} Different components of angular momentum as functions of
    $\Delta t= t_{\rm ret}-t_{\rm BH}$ for Mag+Rad-Full, measured on a  spherical surface
    with coordinate radius $R = 300M\approx 1200\rm\,km$. Here $J_{\rm int}$ is the total interior angular momentum,
    $\Delta J_{\rm GW}$, $\Delta J_{\rm fluid}$, $\Delta J_{\rm EM}$, and $\Delta J_{\nu}$ denotes the angular
    momentum carried by GWs, fluid, electromagnetic fields, and neutrinos, respectively.}
\end{figure}

We calculate the angular momentum carried off by neutrinos after merger.
Our numerical results suggest that  the angular momentum loss due to neutrino
emission is negligible. Fig.~\ref{fig:J_adm} shows the evolution of each component of
angular momentum as a function of $\Delta t = t_{\rm ret}-t_{\rm BH}$ for  Mag+Rad-Full
(see Table~\ref{Tab:Scheme}). The quantities are measured on a spherical surface with coordinate
radius $R = 300M\approx 1200\rm\,km$. We observe that, the angular momentum carried off
by neutrinos $\Delta J_{\nu}$ is less than $1\%$ of the total angular momentum $J_{\rm int}$ and less than $10\%$ of
angular momentum loss due to GW radiation and escaping fluid matter. In the other two remaining 
cases, $\Delta J_{\nu}$ is even smaller as shown in Table~\ref{table:Tab2}. Therefore, we conclude
that angular momentum loss due to neutrinos is negligible. This result is consistent with
the calculation in~\cite{bs98a}, where it has been found that the emission of neutrinos is very
inefficient in carrying off angular momentum and may even increase the angular momentum if the
polytropic index is less than the critical value $0.45$ for a slowly rotating NS. For rapidly
rotating stars, the critical polytropic index may increase. In our simulations the SLy EOS is
parameterized with  polytropic indices ranging between $0.48-0.55$~\cite{Read:2008iy}, which is
close to the critical value. Therefore, it is expected that neutrino
emission has negligible effect on carrying off angular momentum. Moreover, we find that the angular
momentum loss rate $d\Delta J_{\nu}/dt$ is around $10^{48} \, \rm g \, cm^2 \, s^{-2}$, which agrees with the
estimate in~\cite{Fujibayashi_2020} using the general-relativistic momentum formalism~\cite{ShibataM1}.
Note that Fig.~\ref{fig:J_adm} also displays the quantity $J_{\rm insert} - \Delta J_{\rm GW} - \Delta
J_{\rm fluid} - \Delta J_{\rm EM} - \Delta J_{\nu}$, where $J_{\rm insert}$ is the total angular momentum
$J_{\rm int}$ at $t_{\rm insert}$, to track the conservation of the total interior angular momentum $J_{\rm int} (t) =
J_{\rm insert} - \Delta J_{\rm GW} - \Delta J_{\rm fluid} - \Delta J_{\rm EM} - \Delta J_{\nu}$.
We observe that $J_{\rm int}$ is conserved within $\sim 1\%$ after the system reaches a steady state.

%%%%%%%%%%%%%%%%%%%%%%%%%%%%%%%%%
%%%    Angular momentum plot  %%%
%%%%%%%%%%%%%%%%%%%%%%%%%%%%%%%%%

%%%%%%%%%%%%%%%%%%%%%%%%%%%%
%%%  Neutrino luminosity %%%
%%%%%%%%%%%%%%%%%%%%%%%%%%%%
\begin{figure}[]
  \includegraphics[scale=0.19]{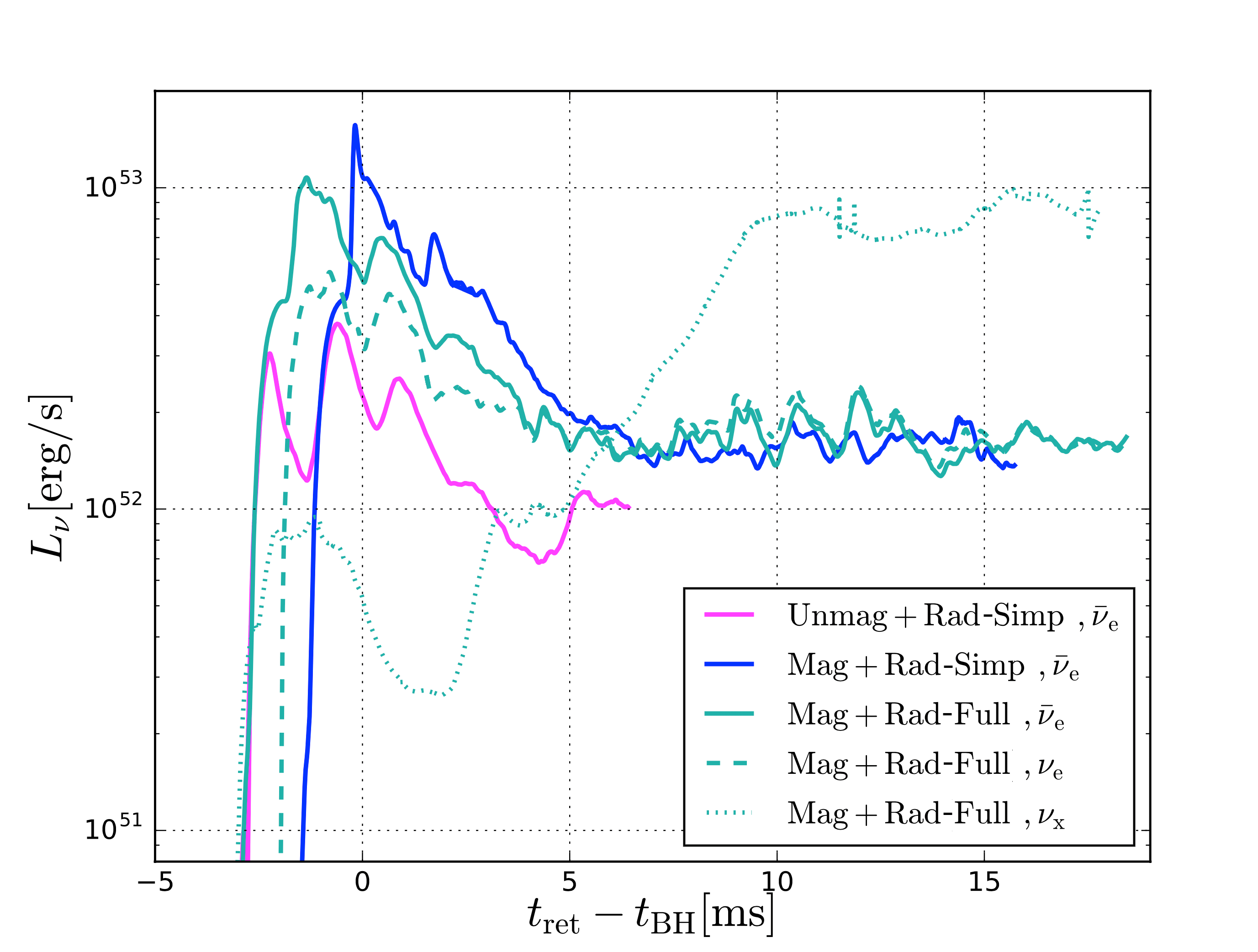}
  \caption{\label{fig:L_nue} Neutrino luminosities for the different neutrino species for Rad cases in~Table~\ref{table:Tab2}
    as a function of $\Delta t= t_{\rm ret}-t_{\rm BH}$. The luminosities are measured at radius $R = 300\,\rm M \approx 1200
    \,\rm km$.}
\end{figure}

Fig.~\ref{fig:L_nue} displays the luminosities of $\bar{\nu}_e$ in Rad-Simp cases, and of the three species in
Mag+Rad-Full as a function of $\Delta t=t_{\rm ret}-t_{\rm BH}$. The luminosities are measured on a sphere with
coordinate radius $R = 300M\approx 1200\,\rm km$. We observe that the $\bar{\nu}_e$ neutrino luminosity in Unmag+Rad-Simp
peaks at $\Delta t = 0$ with the value $\sim 3 \times 10^{52} \, \rm erg\, s^{-1}$. Then it gradually decreases,
reaching a steady-state value of $\sim 10^{52} \,\rm erg\, s^{-1}$. By contrast, the neutrino luminosity in Mag+Rad-Simp
peaks at $\sim 2 \times 10^{53} \, \rm erg\, s^{-1}$ at $\Delta t \approx 135M\sim 1.8 ~\rm ms$ and settles down at
$\sim 2 \times 10^{53}\, \rm erg\, s^{-1}$ at $\Delta t \gtrsim 450M\gtrsim  6 \,\rm ms$. This difference in the
luminosity is likely due to the turbulent magnetic viscosity, which raises the temperature in the HMNS and the disk,
increasing the number and energy of neutrinos.
Similarly, the luminosities of $\bar{\nu}_e$ and $\nu_e$ in Mag+Rad-Full show analogous behavior to that in Mag+Rad-Simp
due to a similar thermal environment. However, we observe that the luminosity of $\nu_{\rm x}$ behaves differently in the
two cases, starting to rise quickly at $\Delta t \approx 300M\approx 4 \,\rm ms$ and reaching a peak value of $\sim 10^{53}
\, \rm erg\, s^{-1}$ at $\Delta t\approx 750M \approx 10\,\rm ms$. Differences between the two electron-type neutrinos
are negligible compared to differences between the electron-type and heavy-lepton neutrinos. Similar behavior has been also
reported in~\cite{Foucart_2020}. The dominance of the scattering opacity over the absorption opacity in a large region
of the system may cause a significant uncertainty in the heavy-lepton neutrinos luminosity. Due to the analytic closure scheme,
the neutrino diffusion rate depends heavily on the choice of energy spectrum, which depends on the fluid temperature.
Nevertheless, the electron-type neutrino luminosity obtained in our cases broadly agrees with previous M1 and MC
studies~\cite{Sekiguchi2016, Foucart_2020}.

%%%%%%%%%%%%%%%%%%%%%%
%%%   Conclusions  %%%
%%%%%%%%%%%%%%%%%%%%%%
\section{Conclusions and Limitations}
\label{sec:Conclusion}
To understand multimessenger observations, numerical simulations of compact binary mergers
involving magnetic fields, neutrinos and detailed microphysics are required. As another step toward this goal, we reported here our implementation of a radiative transport scheme for neutrino processes in our {\tt Illinois GRMHD} code.
This implementation uses the general-relativistic, truncated moment (``M1") formalism in which the first two
moments of radiation, i.e. energy density $E$ and the fluxes $F^\alpha$, are evolved. To close the evolution equations,
the radiation stress $\mathcal{P}^{\alpha \beta}$ is interpolated between the optically thick and thin limits
using the analytic expression in~\cite{MINERBO1978541}(see Appendix~\ref{App:A} below). The code has been tested in multiple regimes, including
those handled by our earlier radiation-GRMHD code in~\cite{grrmhd}. Here we presented the nontrivial, strong gravity test of ``thermal
Oppenheimer-Snyder collapse'' in Appendix~\ref{App:OS} to assess its capability of evolving radiation fields
and computing interactions between radiation and matter in optically thick, thin, and transition regions in a
strong gravitational field containing matter and a BH.
  
We used this radiation evolution module to incorporate neutrino interactions in GRMHD simulations of NSNS mergers
that undergo delayed collapse to a BH. In particular, we considered binaries modeled using a piecewise representation
of the cold SLy nuclear EOS previously reported in~\cite{Ruiz2021}, augmented by a nonzero temperature contribution appropriate for semi-degenerate neutrons. The neutrinos are inserted shortly after the binary merger ($t-t_{\rm merge}\sim 225M\sim 3\rm\,ms$). To probe the effects
of neutrino radiation, we adopted two version for the microphysics. In the first ``warm-up'' version, denoted as Rad(Sim), we only considered
the interactions of $\bar{\nu}_e$, setting the electron fraction $Y_e=0$. In the second version, denoted as
Rad-Full, we evolved three neutrino species $\nu_e$, $\bar{\nu}_e$, and $\nu_x$ and their anti-neutrinos,
and took all related interactions into account. We found that during the HMNS phase, the remnant is subjected to  an effective viscosity due to turbulent magnetic fields which induces the formation a massive central core
surrounded by a Keplearian cloud of matter. We noted that  the MRI would not be affected by neutrino-induced effective
viscosity and drag forces as discussed earlier~\cite{Guilet15}, since in our simulations the magnetic field in the
HMNS is $B\gtrsim 10^{15}\,\rm G$ and exceeds a critical field strength. However, our simulations indicate that neutrino advective transport may assist in reducing the differential rotation of the HMNS. These processes
reduce the lifetime of the remnant by $\gtrsim 4\rm\, ms$ compared to that in the neutrino-free cases.

Following BH formation, we observed that the BH remnants are more massive and have higher spins when neutrinos are
included, while the disk accretion rate  ($\dot{M}\sim \rm 2\,M_{\odot}/s$) and the fraction of escaping mass
($M_{\rm esc}\sim 8 \times 10^{-3} M_{\odot}$) show no noteworthy differences after a quasi-steady state is achieved.
We note that the latter is roughly consistent with the ejecta mass reported by a previous hydrodynamic SPH simulation
of NSNS mergers with a neutrino leakage scheme~\cite{Perego14}, as well as another GRMHD simulation of NSNS
mergers~\cite{Radice_2018, Radice_2021}. We also estimated the peak kilonova luminosity, the rise time, and the effective
temperature/wavelength inferred by our simulations and concluded that the signals may be observed by current
or future instruments such as ALMA or the Vera C. Rubin observatory~\cite{Matthews18, Chen21}.

In all cases, a magnetically-supported jet is launched after $t-t_{\rm BH}\gtrsim 750M\gtrsim 10\rm\, ms$. However,
neutrino processes reduce the baryon-loaded environment in the polar region above the BH poles,
allowing the emergence of the jet $\sim 1000M\sim 15\,\rm ms$ earlier than in cases without neutrinos.
The disk accretion time (jet's fuel) is $\sim 0.1 \, \rm s$, which is consistent with the lifetime of the
sGRBs central engine.  We also found strong evidence that the Blandford–Znajek  mechanism  for launching jets
is operating in our systems.  

In our simulations, we found that neutrinos are inefficient in carrying off angular momentum,
which is in agreement with previous analytic studies~\cite{bs98a}. The angular momentum loss rate due
to neutrino emission is $\sim 10^{48} \, \rm g \, cm^2\, s^{-2}$, which is consistent with the previous
GR simulations~\cite{Fujibayashi_2020}. We found that neutrino luminosity is approximately $10^{52-53}
\, \rm erg\, s^{-1}$ in our cases, which agrees with the previous compact binary merger simulations
using M1 or MC transport methods~\cite{Sekiguchi2016, Foucart_2020}.

To probe if magnetic field and neutrino signatures can be observed by current and/or future GW observatories,
we computed the GW power spectrum of the dominant mode assuming a source distance of $50\,\rm Mpc$.
We found that only its main frequency $f_{\rm peak}$ is well above the sensitivity curves of aLIGO and A+.
By contrast, in the magnetized cases the other key frequencies are at most marginally above these curves.
These results suggest that only next generation GW observatories, such as the Einstein Telescope, are required
to characterize fully the GW signals from the remnant of GW170817-like events. In addition, we computed the match function
between waveforms from systems with different magnetic field and neutrino content. We
found that at a distance of 50 Mpc, only the next generation of based-ground GW detectors, such as the Einstein
Telescope, can observe imprints of the magnetic field and neutrinos.

We also realize several limitations to our implementation and suggest some future improvements. Firstly, the calculation
of temperature is based on an approximate analytic expression, which in turn determines the net-electron fraction,
opacities and emissivities. Currently several more realistic, finite-temperature EOSs have been proposed. They are
based on a finite-temperature liquid drop model with a Skyrme nuclear force~\cite{Lattimer91}, on the relativistic
mean-field (RMF) model extended with the Thomas-Fermi approximation~\cite{Shen98}, or on a statistical model that
consists of an ensemble of nuclei and interacting nucleons in nuclear statistical equilibrium~\cite{Hempel10}.
These EOSs are in tabular form, which is rendered as functions of baryon density, temperature, and the net electron
fraction. We hope to implement these in future investigations. Also, even though the M1 method provides a reasonable
first approximation to radiative transport~\cite{Foucart_M1_2015}, neutrino
luminosities and the outflow composition have a non-negligible dependence on the analytic closure schemes~\cite{Foucart_2016,
  Foucart_2018}. Moreover, the energy-integrated M1 scheme is unable to provide information about the neutrino energy
spectrum. However, an energy-dependent transport scheme is computationally expensive for a full merger simulation,
especially when magnetic fields are present. Monte Carlo (MC) techniques have been used in neutron star
simulations~\cite{Foucart_2020} and been found to yield similar results to M1 schemes. The major disagreement comes
from the luminosity of heavy-lepton neutrinos. In regions where heavy lepton neutrinos have high scattering 
and low absorption opacities, the M1 scheme has greater errors due to the difficulty of predicting the correct neutrino
energy spectrum. We hope to explore improvements to our neutrino scheme in the future.

%%%%%%%%%%%%%%%%%%%%%%%%%%
%%%   Acknowledgments  %%%
%%%%%%%%%%%%%%%%%%%%%%%%%%
\begin{acknowledgments}
  We thank T. Baumgarte, F. Foucart, and J. Noronha-Hostler for several useful discussions.
  We also thank members of the Illinois Relativity Undergraduate Research Team (M. Kotak,
  J. Huang, E. Yu, and J. Zhou) for assistance with some of the visualizations. This work was supported
  in part by National Science Foundation Grant PHY-2006066 and the National Aeronautics and Space
  Administration (NASA) Grant 80NSSC17K0070 to the University of Illinois at Urbana-Champaign.
  L.S. thanks also The Illinois Center for Advanced Studies of the Universe (ICASU) for 
  financial support.  This work made use of the Extreme Science and Engineering Discovery Environment
  (XSEDE), which is supported by National Science Foundation Grant TG-MCA99S008.
  This research is part of the Frontera computing project at the Texas Advanced Computing Center.
  Frontera is made possible by National Science Foundation award OAC-1818253.
  Resources were  also provided by the NASA High-End Computing Program through the NASA Advanced
  Supercomputing Division at Ames Research Center.
\end{acknowledgments}
%
%%%%%%%%%%%%%%%%%%%%%
%%%   Appendices  %%%
%%%%%%%%%%%%%%%%%%%%%
\appendix
%
%%%%%%%%%%%%%%%%%%%%%
%%%   Appendix A  %%%
%%%%%%%%%%%%%%%%%%%%%
%
\section{Radiation Closure Scheme}
\label{App:A}
We compute the radiation pressure using the M1 closure scheme, which is an
interpolation between the optically thin and optically thick limits. Following~\cite{ShibataM1}
and~\cite{Foucart_M1_2015}, we express the radiation pressure terms as
\begin{equation}\label{eq:Pab}
  \mathcal{P}^{\alpha\beta} = \frac{3\chi(\zeta) - 1}{2} \mathcal{P}^{\alpha\beta}_{\rm thin} +
  \frac{3\left[1-\chi(\zeta)\right]}{2} \mathcal{P}^{\alpha\beta}_{\rm thick}\,,    
\end{equation}
where the factor $\chi$ is given by the Minerbo
closure function~\cite{MINERBO1978541}
\begin{equation}\label{eq:Minerbo}
\chi(\zeta)  = \frac{1}{3} + \zeta^2 \frac{6-2\zeta + 6\zeta^2}{15}\,,
\end{equation}
where the quantity $\zeta = F^\alpha F_\alpha/ E^2$, with $F^\alpha$ the radiation flux and $E$ the
radiation energy density. We note that $\zeta$ goes to zero in the optically thick region, and
to one in the optically thin region. The optically thick limit of radiation pressure is given by~\cite{grrmhd}
\begin{equation}
\mathcal{P}^{\alpha \beta}_{\rm thick} = \frac{E}{3} h^{\alpha \beta}\,,
\end{equation}
where $h^{\alpha\beta} = g^{\alpha\beta} + u^\alpha u^\beta$ is the projection tensor onto
the orthogonal slices of the fluid four-velocity. In the optically thin limit, we have
\begin{equation}\label{eq:Pradthin}
P^{\alpha\beta}_{\rm thin} = \frac{F^\alpha F^\beta}{F^\mu F_\mu}E\,.
\end{equation}
Note that, Eq.~(\ref{eq:Pradthin}) reduces to the form adopted in~\cite{Foucart_M1_2015} in asymptotically
flat spacetimes, which is satisfied in our simulations at a large distance from our binaries.
%
%%%%%%%%%%%%%%%%%%%%%
%%%   Appendix B  %%%
%%%%%%%%%%%%%%%%%%%%%
%
\section{Recovering Primitive Variables}
\label{App:B}
The evolution of the radiation conservative variables $(\bar{\tau}, \bar{S}_i)$ (see Eqs.~(\ref{eq:taubar})
  and~(\ref{eq:Sbar})) involves the calculation of the radiation primitive variables $(E, F^i)$ at each iteration,
  which in turn requires a $4\times 4$ root finder given  the closure expressions (\ref{eq:Pab})
  and (\ref{eq:Minerbo}).
 In our code, we use a Newton-Raphson solver to find the roots of the set of functions $f_\mu(E, F^i)$
 \begin{equation}
  f_i =  \alpha \sqrt{\gamma} \left[E u^0 u_i + F^0u_i+ F_i u^0 + \mathcal{P}^{0}_{\,\,i}
    \right]/\bar{S}_i - 1\,,
\end{equation}
 and,
\begin{equation}
f_4  =  (\alpha^2 \sqrt{\gamma})[E (u^0)^2 + 2F^0 u^0 +  \mathcal{P}^{00} ]/\bar{\tau} - 1 \,,
\end{equation}
where $\bar{S}_i$ and $\bar{\tau}$ are computed at a given iteration, while $E$ and $F^i$ are known
values at the previous iteration. We apply Newton-Raphson method to solve for the set of equations $f_\mu(E, F^i)=0$.
The solver returns the primitive variables $(E, F^i)$ as values at the current iteration when all four functions
reduce below the tolerance $f_\mu < 10^{-12}$.

The first-order Newton-Raphson method requires a matrix of partial derivatives, namely the Jacobian
\begin{equation}
J_f  = \begin{bmatrix}
\frac{\partial \bar{S}_i}{\partial F_j}  &  \frac{\partial \bar{S}_i}{\partial E}\\
\frac{\partial \bar{\tau}}{\partial F_j} &  \frac{\partial \bar{\tau}}{\partial E}
\end{bmatrix}\,,
\end{equation}
in which the derivatives can be evaluated analytically. The specific components of $J_f$ are
\begin{equation}\label{dtdE}
\begin{aligned}
  \frac{\partial \bar{\tau}}{\partial E} & = \alpha^2\sqrt{\gamma} \left\lbrace
  \frac{4 (u^0)^2}{3}   + (A-B\zeta) \right. \\ & \left. \left[ \frac{(F^0)^2}{F^\alpha F_\alpha}
    - \frac{g^{00} + (u^0)^2}{3} \right] + \frac{g^{00}}{3} \right\rbrace\,,
\end{aligned}
\end{equation}

\begin{equation}\label{dSdE}
\begin{aligned}
  \frac{\partial \bar{S_i}}{\partial E}  &= \alpha^2\sqrt{\gamma} \left\lbrace
  \frac{4 u^0u_i}{3} + \right. \\ & \left. (A-B\zeta) \left[ \frac{F^0 F_i}{F^\alpha F_\alpha} -
    \frac{u^0u_i}{3} \right] \right\rbrace\,,
\end{aligned}
\end{equation}

\begin{equation}\label{dtdF}
\begin{aligned}
\frac{\partial \bar{\tau}}{\partial F_i}  &= 
\alpha^2\sqrt{\gamma} \left\lbrace  2(g^{i0} + v^ig^{00}) (u^0 + AE\frac{F^0}{F^\alpha F_\alpha}) \right.
\\ &  + (F^i -v^iF^0)  \left[ \frac{B}{\sqrt{F^\alpha F_\alpha}} \left(\frac{(F^0)^2}{F^\alpha F_\alpha} -
  \frac{g^{00} + (u^0)^2}{3} \right) \right. \\ & \left. \left. - AE \frac{2(F^0)^2}{(F^\alpha F_\alpha)^2}
  \right] \right\rbrace\,,
\end{aligned}
\end{equation}
and
\begin{equation}\label{dSdF}
\begin{aligned}
\frac{\partial \bar{S}_j}{\partial F_i}  &= 
\alpha\sqrt{\gamma} \left\lbrace   (u^0 + \frac{AE F^0}{F^\alpha F_\alpha}) \delta^i_j  \right.\\
& \left. + (g^{i0} + v^ig^{00}) (u_j + AE\frac{F_j}{F^\alpha F_\alpha})  \right. \\
&  + (F^i -v^iF^0) \left[ \frac{B}{\sqrt{F^\alpha F_\alpha}} \left(\frac{F^0F_j}
  {f^\alpha F_\alpha} -\frac{u^0u_j}{3}\right) \right.\\ & \left. \left. - AE \frac{2F_jF^0}{(F^\alpha F_\alpha)^2} \right]
\right\rbrace\,,
\end{aligned}
\end{equation}
where
\begin{equation}
A \equiv \frac{3\,\chi -1 }{2} = \frac{1}{5} (3\,\zeta^2 - \zeta^3 + 3\,\zeta^4)\,,
\end{equation}
and
\begin{equation}
B \equiv \frac{1}{5} (6\,\zeta - 3\,\zeta^2 + 12\,\zeta)\,.
\end{equation}
Once the Jacobian is computed, the solver performs LU-decomposition and back-substitution
in the Jacobian to find the direction of the Newton step. Then, the code uses line searches
and backtracking methods (see Sec. 9.7 in~\cite{1992nrcabookP}) to reduce the number of
iterations for convergence. After obtaining $(E, F^i)$, we impose a floor value of radiation
energy density $E_{\rm atm} = 10^{-9} \, E_{\,\rm max}$ and set the radiation fluxes to zero
if $E < E_{\rm atm}$ to reliably evolve in the regions with low radiation energy.
Knowing the values of the primitive variables, we then recompute the conserved variables and
impose the radiation contributions to the total stress-energy tensor in the BSSN equations. 
%
%%%%%%%%%%%%%%%%%%%%%
%%%   Appendix C  %%%
%%%%%%%%%%%%%%%%%%%%%
%
\section{\label{app:FTH}Finite Temperature EOS}
To treat neutrinos, it is necessary to evolve NSs with finite-temperature matter. However, in our
simulations we adopt an SLy nuclear EOS to model the cold  component, and employ an approximate
analytic expression for the hot component. Notice that  our simplified, analytic method of
computing the temperature and the opacity during the NS evolution can be straightforwardly
extended to realistic, tabulated, thermal EOSs. 

%%%%%%%%%%%%%%%%%%%%%%
%%%   Appendix C1  %%%
%%%%%%%%%%%%%%%%%%%%%%
%
\subsection{Thermal energy density and pressure}\label{subapp:TEP}
Our simplified, analytic hot nuclear EOS assumes that the total nucleon energy density and pressure
consists of two parts 
\begin{equation}
\mathcal{E}= \mathcal{E}_{\text{cold}} +  \mathcal{E}_{\text{th}}\,,
\end{equation}
\begin{equation}\label{eq:P_tot}
P = P_{\text{cold}} +  P_{\text{th}}\,,
\end{equation}
where $P_{\rm cold}$ and $\mathcal{E}_{\text{cold}}$ are the cold pressure.
On the other hand, the energy density and the thermal component are
given by the sum of the nucleon and thermal radiation components
\begin{equation}\label{eq:E_th}
\mathcal{E}_{\text{th}} = \mathcal{E}_{\text{nuc}} +  \mathcal{E}_{\text{rad}}\,,
\end{equation}
\begin{equation}\label{eq:P_th}
P_{\text{th}} = P_{\text{nuc}} +  P_{\text{rad}}\,.
\end{equation}
The nucleon part of the thermal energy density is based on a semi-degenerate, ideal neutron gas expression
that limits to a Maxwell-Boltzmann gas in the non-degenerate limit~\cite{BST96}
\begin{eqnarray}
  \mathcal{E}_{\text{nuc}} &=& \frac{3}{2}\,n\,k_B\,T \times\nonumber\\
  && {\rm min} \left\lbrace 1,
  \frac{2\,(3\,\pi^2)^{1/3}\,m_{\rm nuc}}{18(\hbar\,c)^2}\,n^{-2/3}\,k_B\,T\right\rbrace\,, 
\label{eq:E_nuc}
\end{eqnarray}
where $m_{\rm nuc}$ is the nucleon mass, $n = \rho_0/m_{\text{nuc}}$ is the nucleon number density, $k_B$ is Boltzmann's constant, and $T$ is the temperature of the fluid. This expression accommodates
both the low-temperature semi-degenerate, and the high-temperature Maxwell-Boltzmann limits. The
thermal nucleon part of the pressure is given by
\begin{equation}\label{eq:P_nuc}
P_{\text{nuc}} = (\gamma -1) \mathcal{E}_{\text{nuc}}\,,
\end{equation}
with $\gamma = 5/3$, as the nucleons are essentially nonrelativistic. We note that in most NSNS merger and post-merger
scenarios, the second term in the bracket in Eq.~(\ref{eq:E_nuc}) is greater than or comparable to 1 throughout
the matter. Therefore, for simplicity, we typically can adopt the simpler
expression $\mathcal{E}_{\text{nuc}}= (3/2)\, n\,k_B\,T$. The ``radiation" part of the energy density includes photons
and relativistic electron and positron pairs, and follows a Stefan–Boltzmann relation\footnote{
  Note that electron-positron pairs become relativistic and the ${7}a\,T^4/4$ is appropriate only when $k_B\,T >
  2m_e\,c^2 \sim 1 \rm MeV$. However, in NSNS mergers, the temperature is typical $\gtrsim$ 10 MeV and hence we always
  keep this term.}
\begin{equation}\label{eq:E_rad}
  \mathcal{E}_{\text{rad}} = \mathcal{E}_{\text{ph}} + \mathcal{E}_{\text{pair}} = \left(1 + \frac{7}{4}\right)\,
  a\,T^4=\frac{11}{4}\,a\,T^4\,,
\end{equation}
and
\begin{equation}\label{eq:P_rad}
P_{\text{rad}} = \frac{1}{3} \mathcal{E}_{\text{rad}}\,.
\end{equation}
Combining nucleon and radiation components, the thermal energy density and pressure, we typically can adopt the simplification
\begin{equation}\label{Eth}
\mathcal{E}_{\text{th}} = \frac{3}{2}\,n\, k_B\, T + \frac{11}{4}\, a\,T^4\,,
\end{equation}
\begin{equation}\label{Pth}
P_{\text{th}} = n\,k_B\,T + \frac{11}{12}\, a\,T^4\,.
\end{equation}
%
%%%%%%%%%%%%%%%%%%%%%%
%%%   Appendix C2  %%%
%%%%%%%%%%%%%%%%%%%%%%
%
\subsection{Numerical Implementation}
\label{subapp:NIPSM}
After the update of rest-mass density $\rho_0$ and the total specific internal energy $\epsilon$ at each iteration, we obtain the thermal part of the specific internal energy according to
\begin{equation}
\epsilon_{\text{th}} = \epsilon-  \epsilon_{\text{cold}}\,,
\end{equation}
and the thermal energy density as
\begin{equation}
  \mathcal{E}_{\text{th}} = \rho_0\, \epsilon_{\text{th}}\,.
  \label{eq:ethdef}
\end{equation}
The cold part of the specific internal energy and pressure for SLy can be modeled by a 4-piece piecewise polytropic EOS
\begin{equation}
P^i_{\text{cold}} = \kappa^i \rho_0^{\Gamma_i}\,,
\end{equation} 
\begin{equation}
  \epsilon^i_{\text{cold}} = \frac{P^i_{\text{cold}}}{\rho_0 (\Gamma_i -1) } = \frac{\kappa^i \rho_0^{\Gamma_i - 1}}
          { \Gamma_i -1 }\,.
\end{equation}
Here $\kappa_i$ and $\Gamma_i$ are the corresponding polytropic constant and the polytropic exponent of the rest-mass density
in the range $\rho_{0,i-1}\leq \rho_0\leq \rho_{0,i}$, respectively~\cite{Read:2008iy}. Once the thermal energy density is computed,
we use Eqs.~(\ref{Eth}) and~(\ref{Pth}) to compute the temperature
and the thermal pressure.

We use the {\tt HARM 2D} primitive solver to compute the MHD variables~\cite{Gammie_2003, Noble_2006}. It
finds roots of a 2D-system with variables
\begin{equation}
v^2 \equiv v_iv^i,
\end{equation}
and 
\begin{equation}\label{eq:W}
W = w\gamma^2 = \frac{\rho_0\,(1+ \epsilon + {P}/{\rho_0})}{1-v^2}\,.
\end{equation}
In this setup, the total pressure $P$ needs to be expressed as functions of $v^2$ and $W$, while the
derivatives $dP/dW$ and $dP/dv^2$ are needed for the Jacobian. Let us begin with the temperature-dependent
thermal pressure and energy density. Simple algebraic manipulations of Eqs.~(\ref{Eth}) and~(\ref{Pth}) yield
\begin{equation}
\rho_0 \epsilon_{\rm th}  = 3\,P_{\rm th} -1.5 nT = 3\left(P - P_{\rm cold} -0.5\,\,T \right)\,.
\end{equation}
Next, using Eq.~(\ref{eq:W}) we obtain
\begin{eqnarray}
  P = 0.25\,&&\left[ W (1-v^2) - \rho_0 (1 + \epsilon_{\rm cold}) +\right. \nonumber\\
    &&3\left. P_{\rm cold} + 1.5\,n\,T  \right]\,.
\label{eq:PinW}
\end{eqnarray}
From Eq.~(\ref{eq:PinW}), we calculate the derivatives
\begin{equation}\label{dpdw}
\frac{dP}{dW} = 0.25 (1-v^2)\,,
\end{equation}
and
\begin{eqnarray}
  \frac{dP}{dv^2} &=  & 0.25 \left[\left( - 1 - \epsilon_{\rm cold}  + (3\Gamma_i -1)
    \frac{P_{\rm cold}}{\rho_0}\right)\frac{d \rho_0}{d v^2}  \right. \nonumber\\
    && + 1.5 \left.\frac{d(nT)}{dv^2} -W \right]\,.
  \label{dpdv2}
\end{eqnarray}
in which the derivative $d(nT)/dv^2$ is given by
\begin{eqnarray}
  \frac{d(nT)}{dv^2} &=& \frac{T}{m_n}\frac{d\rho_0}{dv^2}  + \frac{\rho_0}{m_n} \frac{dT}{dv^2} \nonumber\\
  & =& \frac{T}{m_n}\frac{d\rho_0}{dv^2}  + \frac{\rho_0}{m_n} \left(\frac{d T}{ d \rho_0}\frac{d\rho_0}{dv^2}
  \right)\,.
  \label{eq:dntdv2}
\end{eqnarray}
The first term in Eq.~(\ref{eq:dntdv2}) can be computed directly. To find the second term,
we begin setting the radiation constant $a=1$ in the thermal pressure~(Eq.~\ref{Pth}) and by taking a derivative with
respect to the temperature we have
\begin{equation}\label{dpthdT0}
\frac{ d P_{\rm th}}{d T} = \frac{11}{3}  T^3 + \frac{\rho_0}{m_n}.
\end{equation}
Next, we express $d P_{\rm th}/d T$ using $P$ and $P_{\rm cold}$ as
\begin{equation}\label{dpthdT}
  \frac{ d P_{\rm th}}{d T} = \frac{ d P_{\rm th}}{d \rho_0}\frac{d\rho_0}{dT} =
  \left(\frac{ d P}{d \rho_0} - \frac{ d P_{\rm cold}}{d \rho_0}\right)\frac{d\rho_0}{dT}\,,
\end{equation}
and, 
\begin{equation}
\begin{aligned}
  \frac{ d P}{d \rho_0} & = \frac{ d P}{d W}\frac{ d W}{d \rho_0}  =
  \frac{ d P}{d W} \left(\frac{1}{1-v^2} \frac{dw}{d\rho_0}\right) \\
  &= 0.25 \left[\frac{d(\rho_0 + \rho_0\epsilon + P)}{d \rho_0}\right]\,,
\end{aligned}
\end{equation}
where we used Eq.~(\ref{dpdw}) for $dP/dW$. Expanding this equation,
we obtain
\begin{equation}
  \label{dpdrho1}
\begin{aligned}
  3 \left(\frac{dP}{d\rho_0}\right)  = 1 + \rho_0\left[\frac{d(\epsilon_{\rm cold} +\epsilon_{\rm th}) }{d\rho_0}\right]
  + \epsilon_{\rm cold} +  \epsilon_{\rm th}\,.
\end{aligned}
\end{equation}
On the other hand, taking the derivative of the thermal energy density (Eq.~\ref{Eth}) with respect to temperature and using Eq.~(\ref{eq:ethdef}) we find
\begin{equation}
  \frac{d \epsilon_{\rm th}}{d\rho_0}  = \frac{1}{\rho_0} \left( 11T^3 + 1.5
  \frac{\rho_0}{m_n} - \frac{d\rho_0}{dT} \epsilon_{\rm th}\right) \frac{dT}{d\rho_0}\,.
\end{equation}
Inserting the above expression in Eq.~(\ref{dpdrho1}), we find
\begin{eqnarray}
3 \left(\frac{dP}{d\rho_0}\right)  
&=& 1 + \rho_0\left(\frac{d\epsilon_{\rm cold} }{d\rho_0}\right)  + \epsilon_{\rm cold} +\nonumber\\
&&  \left( 11T^3 +1.5\frac{\rho_0}{m_n} \right)\,\frac{dT}{d\rho_0}\,.
\end{eqnarray}
The last term in this equation can be computed using $dP/d\rho_0$ in Eq.~(\ref{dpthdT}) and equating it
with Eq.~(\ref{dpthdT0})
\begin{equation}
  \frac{d\rho_0}{dT} = \frac{1.5\frac{\rho_0}{m_n} }{1 + \epsilon_{\rm cold} + (1-3\Gamma_i) \frac{P_{\rm cold}}
    {\rho_0}}\,,
\end{equation}
which allow us to compute the second term in Eq.~(\ref{eq:dntdv2}). Finally, we use Eq.~(\ref{dpdv2}) to get
\begin{equation}
\frac{dP}{dv^2} = 0.25\left( -W + 1.5\frac{T}{m_n} \frac{d\rho_0}{dv^2}\right)\,,
\end{equation}
which completes the derivative terms in the Jacobian of the finite-temperature {\tt HARM 2D} solver.

%%%%%%%%%%%%%%%%%%%%%%
%%%   Appendix C3  %%%
%%%%%%%%%%%%%%%%%%%%%%
%
\subsection{Sound Speed}
\label{subapp:soundspeed}
\begin{figure}[h]
    \centering
    \hspace{-7mm}\includegraphics[scale=0.15]{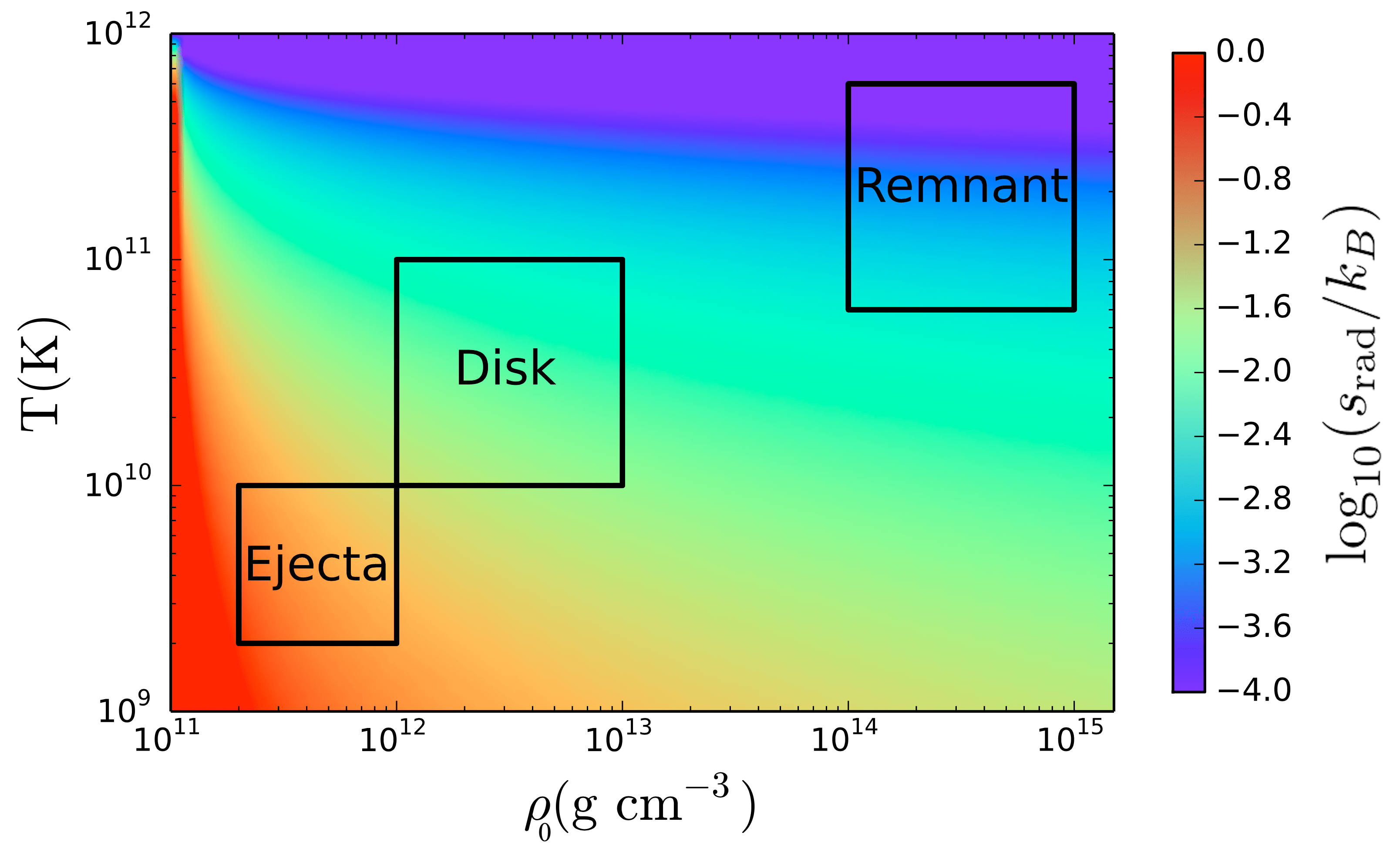}
    \caption{Ratio $s_{\rm rad}/k_B$ (log scale) as
      a function of $\rho_0$ and $T$. The boxes show the parameter space
      typical for the NSNS remnant, accretion disk, and ejecta.
    \label{fig:Sradoverkb}}
\end{figure}

The sound speed of the fluid $c_{\rm s}$ can be obtained using the derivative of $P$ with
respect to $\rho_0$ at constant entropy $c^2_{\rm s} = \left. \left(dP /d\rho_0 \right)/h \right|_{\rm s}$, where $h$ is the specific enthalpy.
Using the relations in Eqs.~(\ref{eq:P_th}), (\ref{eq:P_nuc}), and~(\ref{eq:P_rad}), we have
\begin{equation}
  c^2_{\rm s} = \frac{1}{h}\left(\frac{d P_{\rm cold}}{d \rho_0} + \frac{d P_{\rm nuc}}
  {d \rho_0} + \frac{d P_{\rm rad}}{d \rho_0}\right)_{\rm s}\,,   
\end{equation}
where
\begin{equation}
\frac{d P_{\rm cold}}{d \rho_0} = \Gamma_i \frac{P_{\rm cold}}{\rho_0}\,,
\end{equation}

\begin{equation}\label{eq:dP_nucdrho}
\begin{aligned}
  \frac{d P_{\rm nuc}}{d \rho_0} &= \frac{\partial P_{\rm nuc}}{\partial
    \rho_0} + \frac{\partial P_{\rm nuc}}{\partial T} \left. \frac{d T}{d \rho_0}\right|_{\rm s}\,,
\end{aligned}
\end{equation}
and
\begin{equation}\label{eq:dP_raddrho}
    \frac{d P_{\rm rad}}{d \rho_0} 
    = \frac{\partial P_{\rm rad}}{\partial T} \left. \frac{d T}{d \rho_0}\right|_{\rm s}\,.
\end{equation}
The term $\left. \left(d T /d \rho_0\right) \right|_{\rm s}$ is
calculated by using the isentropic condition, where the total entropy per baryon
$s_{\rm tot} = s_{\rm nuc} + s_{\rm rad}$ is constant, or $d s_{\rm tot} = d s_{\rm nuc} + d s_{\rm rad} = 0$. 
From Eqs.~(\ref{Eth}) and~(\ref{Pth}), we obtain
\begin{equation}
  s_{\rm nuc} = k_B \left[\frac{5}{2} + \ln\left[\frac{1}{n}
      \left(\frac{m_B k T}{2\pi \hbar^2}\right)^{3/2}\right] \right]\,,
\end{equation}
and,
\begin{equation}
  s_{\rm rad} = \frac{11}{3} \frac{m_B}{\rho_0} aT^3\,,
\end{equation}
which gives
\begin{equation}
    \begin{aligned}
      & d s_{\rm nuc} + d s_{\rm rad} = \\ &\left[k_B\left(-\frac{d n}{n}\right) + \frac{3}{2}k_B
        \frac{dT}{T}\right] + \left( 11 a \frac{T^2 dT}{n} + 11 aT^3 \frac{dn}{n^2}\right) \\
      & \left[k_B\left(-\frac{d \rho_0}{\rho_0}\right) + \frac{3}{2}k_B \frac{dT}{T}\right] +
      \left( 3 s_{\rm rad} \frac{dT}{T} - s_{\rm rad} \frac{d \rho_0}{\rho_0}\right) = 0\,.
    \end{aligned}
\end{equation}
Simplifying the above we obtain
\begin{equation}
  \label{eq:dTdrho_s}
\left.\frac{d T}{d \rho_0}\right|_{\rm s} = \left(\frac{2 \frac{s_{\rm rad}}{k_B} + 2}
{6\frac{s_{\rm rad}}{k_B} + 3}\right) \frac{T}{\rho_0}  = \sigma_{s2} \frac{T}{\rho_0}\,,
\end{equation}
where $\sigma_{s2}=({2\,{s_{\rm rad}}/{k_B} + 2})/({6{s_{\rm rad}}/{k_B} + 3})$.
Applying relation~(\ref{eq:dTdrho_s}) to~(\ref{eq:dP_nucdrho}) and~(\ref{eq:dP_raddrho}), the sound speed reads
\begin{equation}\label{eq:cs2_2}
  c^2_{\rm s} = \frac{1}{h}\left[ \Gamma_i \frac{P_{\rm cold}}{\rho_0} +
    \frac{2}{3} \varepsilon_{\rm nuc} \left(1 + \sigma_{s2}\right) + \frac{4}{3} \sigma_{\rm s2}
    \varepsilon_{\rm rad} \right]\,.
\end{equation}
In most regions of the NSNS remnant, the ratio $s_{\rm rad}/ k_B$ remains small. As shown in
Fig.~\ref{fig:Sradoverkb}, $s_{\rm rad}/ k_B$ stays  $\lesssim 10^{-3}$ for the NSNS remnant
and $\lesssim 10^{-2}$ in the accretion disk, even though it may reach $\sim 0.2$ in the ejecta.
Hence, for most regions, $\sigma_{s2} \rightarrow 2/3$, which reduces Eq.~(\ref{eq:cs2_2}) to
\begin{equation}\label{eq:cs2_3}
  c^2_{\rm s} \approx \frac{1}{h}\left[ \Gamma_i \frac{P_{\rm cold}}{\rho_0} +
    \frac{10}{9} \varepsilon_{\rm nuc} + \frac{8}{9} \sigma_{\rm s2} \varepsilon_{\rm rad} \right]\,.
\end{equation}
%
%%%%%%%%%%%%%%%%%%%%%%%%%% 
%%% ratio S_rad/S_nuc  %%%
%%%%%%%%%%%%%%%%%%%%%%%%%%
%
In our numerical implementation, we approximate the sound speed by
\begin{equation}\label{eq:code_cs2}
  c^2_{\rm s, code} = \frac{1}{h}\left[ \Gamma_i \frac{P_{\rm cold}}{\rho_0} + 
    \frac{8}{9}\left(\varepsilon_{\rm nuc} +\varepsilon_{\rm rad} \right) \right]\,,
\end{equation}
for efficiency. We note that this approximation of the isentropic sound
is valid when the $\varepsilon_{\rm nuc}$ is smaller than the other two terms
in~Eq.~(\ref{fig:Sradoverkb}). However, even if $\varepsilon_{\rm nuc}$ is larger, the above expression underestimates
the sound speed by at most $\sim 20\%$, which is smaller than the overestimation of computing
the characteristic speed $c^{\pm}$ for the HLL Riemann solver using the approximated general GRMHD dispersion
relation. According to ~\cite{Etienne:2015cea}, compared to the original, full dispersion relation, (Eq. (27)
of~\cite{Gammie_2003}), the simplified expression adopted in our code overestimates the maximum characteristic
speed by a factor $\leq 2$. Note that this
approximate dispersion relation has been widely used in various codes including {\tt Whisky-MHD}~\cite{whisky2007},
{\tt GRHydro}~\cite{GRHydro14}, and {\tt HARM 3D}~\cite{Noble_2009}.

%%%%%%%%%%%%%%%%%%%%
%%%   Apendix D  %%%
%%%%%%%%%%%%%%%%%%%%
%
\section{Neutrino Source Terms}
\label{App:source_term}

The calculation of the radiation four-force density $G^\alpha$ requires
three key quantities (see Eq~(\ref{eq:Gmu2})): 1) the neutrino emissivity $\eta$; 2) the absorption
opacity $\kappa_a$; and 3) the scattering opacity $\kappa_s$. Following~\cite{Foucart_M1_2015}, we first
compute the absorption opacity of electron neutrinos due to two charged-current interactions, which
are the absorption of $\nu_e$ onto neutrons according to Eq.~(A11) in~\cite{RF1996} (henceforth ``RJS'')
and the absorption of $\bar{\nu}_e$ onto protons according to Eq.~(A12) in RJS. The inverse processes
of the two interactions are responsible for the emission of electron-type neutrinos. In the optically
thick limit, we apply the energy-integrated Kirchhoff’s law~\cite{Foucart_M1_2015}
\begin{equation}
  \eta_{\rm Kir} \approx \int_0^\infty \kappa_a\,B_\nu\,
  (\varepsilon_\nu) d \varepsilon_\nu \approx\frac{7}{8}\,\mathcal{N}_\nu\,\kappa_a\,a\,T^4\,, 
\end{equation}
where the the specific intensity $B_\nu$ is the Fermi-Dirac intensity, and the absorption opacity
$\kappa_a$ is an energy-averaged (e.g. Rosseland mean) opacity. In the optically thin limit,
Eqs.~(B1) and (B2) in RJS give the free emission rates of the two charged-current interactions, from
which we obtain the emissivity in the free-streaming regime $\eta_{\rm fs}$. Hence, the total emissivity
from the charged-current interaction is given by the interpolation between $\eta_{\rm Kir}$ and
$\eta_{\rm fs}$ with the same method used to interpolate the radiation stress tensor
\begin{equation}
  \label{eq:eta_tot}
  \eta =  \frac{3\chi(\zeta) - 1}{2}\,\eta_{\rm fs} +
  \frac{3\left[1-\chi(\zeta)\right]}{2}\,\eta_{\rm Kir}\,,
\end{equation}
where $\chi$ and $\zeta$ are defined in Appendix~\ref{App:A}.
We then consider the free emission of three pair processes, including electron pair annihilation, plasmon
decay, and nucleon-nucleon bremsstrahlung. RJS gives the the emission rate due
to pair annihilation and plasmon decay for electron-type neutrinos in their Eqs.~(B8) and (B11), and those
for heavy-lepton neutrinos in their Eqs.~(B10) and (B12). The emission rate for nucleon-nucleon bremsstrahlung
is given in~\cite{Burrows2000}. Note that the expression of the emissivity for
the two heavy-lepton species and their anti-neutrinos ($n_\mu$, $\bar{\nu}_\mu$, $\nu_\tau$, and
$\bar{\nu}_\tau$) are the same, so we can merge the 4 types of particles as $\nu_x$ and evolve
them together as a single variable. Next, the inverse of the pair processes, contributes to the absorption
opacities, and is again computed using Kirchhoff's law. Note that, for heavy lepton neutrinos,
$\mathcal{N}_\nu = 4$ as a result of lumping 2 species (plus anti-neutrino) together, while
$\mathcal{N}_\nu = 1$ for $\nu_e$ and $\bar{\nu}_e$. Lastly, we consider the scattering opacities for all
species given by Eq.~(A6) in RJS.

The calculation of the terms above requires the knowledge of neutrino temperature, nucleon number density,
neutrino chemical potential, and net electron fraction, as well as terms such as the Pauli blocking factor. We evolve
the nucleon density, while the temperature is computed as in Appendix~\ref{app:FTH}. In the following we
summarize our treatment to obtain the other ingredients.  

%%%%%%%%%%%%%%%%%%%%%
%%%   Apendix D1  %%%
%%%%%%%%%%%%%%%%%%%%%
%
\subsection{Chemical Potentials}
\label{subapp:ACP}

For the chemical potential of neutrinos, which are used in computing the blocking factors, we
adopt the expressions in~\cite{RF1996},
\begin{equation}
  \label{mu_nue}
\mu_{\nu_e} = \mu^{\text{ceq}}_{\nu_e}\,[1 -\exp(-\tau_{\nu_e})] + \mu^{0}_{\nu_e}\,\exp(-\tau_{\nu_e})\,,
\end{equation}
where $\tau_{\nu_e}$ is the optical depth, which is estimated as  the minimum of the line integrals of total opacity along the three Cartesian directions, and $ \mu^{\text{ceq}}_{\nu_e} = \mu_e + \mu_p - \mu_n - Q$ is the chemical
potential of electron neutrinos in chemical equilibrium. Here $Q = 1.2935 \rm~MeV \approx
2\times 10^{-6} \, \rm erg$ is the rest-mass energy difference between a neutron and a proton, and
$\mu^{0}_{\nu_e}$ is the chemical potential in low-density, transparent matter which, by simplicity, we set to zero.
For $\mu_e$, $\mu_p$, and $\mu_n$, we use Eq.~(11.2.4) in~\cite{ShapiroBook83}.

%%%%%%%%%%%%%%%%%%%%%
%%%   Apendix D2  %%%
%%%%%%%%%%%%%%%%%%%%%
%
\subsection{Pauli Blocking Factors}
\label{subapp:OandE}

We need to compute the Pauli blocking factor that appears in the absorption and scattering
opacities, and emissivity in various neutrino processes. They are obtained as integrals over
momentum of the Fermi distribution function for the relevant nucleons and electrons involved in
the reactions. We adopt expressions for $Y_{\rm NN}$, $Y_{np}$, $Y_{n}$, $Y_{p}$, and $Y_e$
obtained in~\cite{Bruenn85} for equilibrium mixtures. These quantities are defined as follows:
\begin{itemize}
\item $Y_{\rm NN}$ defines the phase space fractions of free neutrons (when $N\equiv n$) or
  protons (when $N\equiv p$), respectively, due to the Pauli blocking effects in nucleon
  scattering reaction (see~Eq.~(\ref{eq:scattering})).
\item $Y_{np}$ defines the phase space fraction of free nucleons due to the absorption
  by neutrons (the inverse reaction of Eq.~(\ref{eq:charged_e-}))
\item $Y_{n}$, $Y_{p}$, and $Y_e$ define the number fraction of neutron, proton, and electron.
  For completely dissociated matter the nucleon fractions are $Y_n = 1 - Y_e$, and $Y_p = Y_e$.
\end{itemize}

%%%%%%%%%%%%%%%%%%%%
%%%   Apendix E  %%%
%%%%%%%%%%%%%%%%%%%%
%
\section{Shear Viscosity Comparison}
\label{subapp:nu-vis}

\begin{figure}
    \centering
    \hspace{-5mm}\includegraphics[scale=0.37]{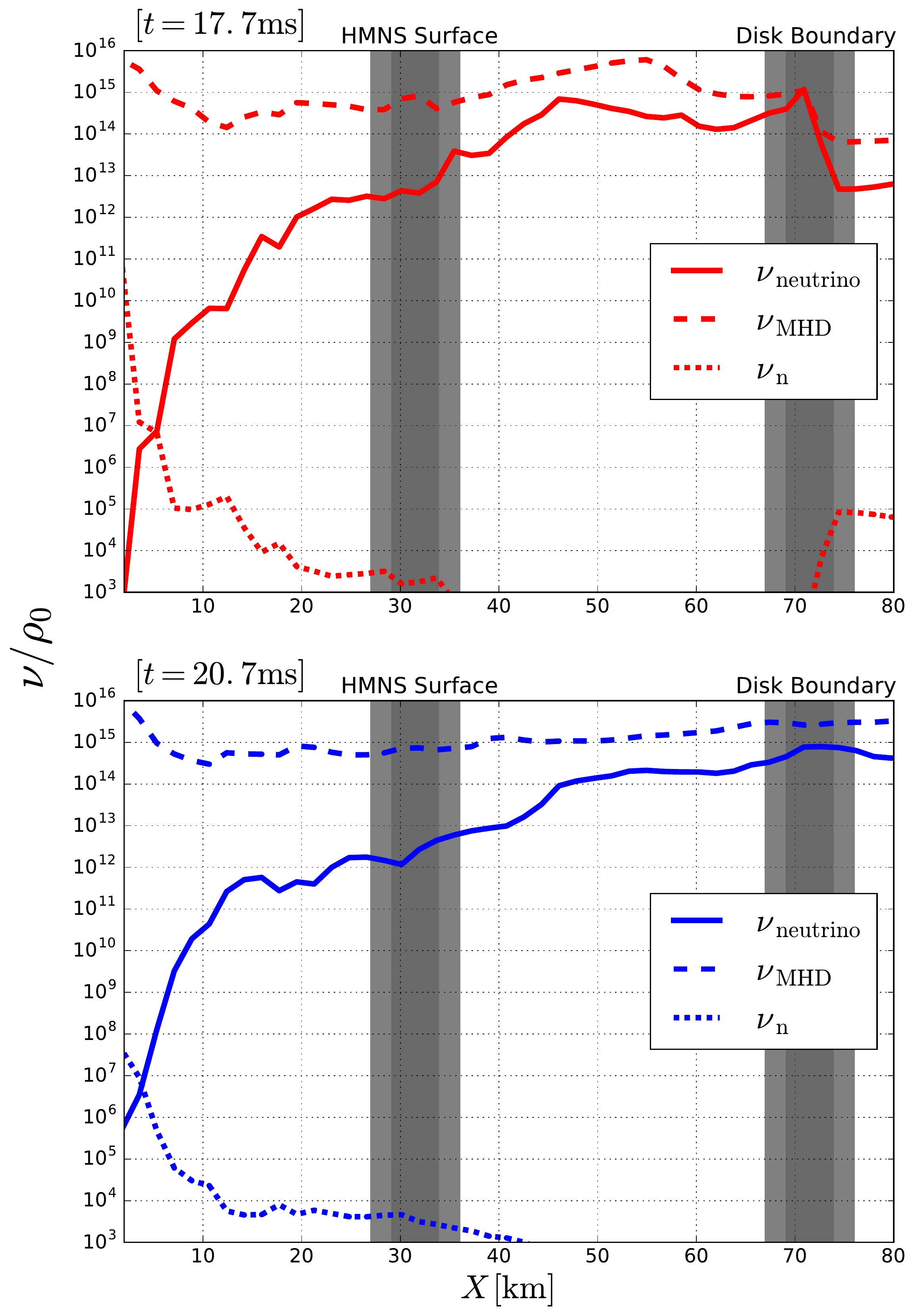}
    \caption {Effective shear viscosity $\nu/\rho_0$ due to neutrinos (solid curves), MHD turbulence
      (dashed curves), and nuclear interactions (dotted curves) along the x-axis Mag+Rad-Full during the HMNS (top
      panel) and BH + disk (bottom panel) phases.
      The gray vertical regions mark the approximate HMNS surface and disk
      outer boundary, respectively.}
    \label{fig:vis_vs_x}
\end{figure}

As suggested by~\cite{Guilet15}, the diffusion of neutrinos trapped inside a HMNS can induce momentum transport, acting like an effective shear viscosity on length-scales longer than the neutrino
mean free path. In addition, if the neutrino mean free path is longer than the MRI wavelength,
neutrinos induce a drag on the velocity field due to the Doppler effect between the fluid motion
and the background radiation field~\cite{Agol98, Jedamzik98}. According to~\cite{Guilet17},
the effective viscosity and drag from neutrinos may then have a significant impact on the MRI inside
a HMNS if magnetic fields are weaker than $10^{14}\,\rm G$. However, high-resolution simulations of NSNS
mergers have shown that seed magnetic fields can be boosted from $\sim 10^{11}\,\rm G$ to $\sim 10^{16}\,\rm G$
within the first $\sim 5\rm\,ms$ following merger (see e.g.~\cite{kkssw14, Palenzuela:2021gdo,Aguilera-Miret20}).
Therefore, the viscous effect on MRI due to neutrinos may be neglected. To assess the effect of neutrino momentum transport 
in our simulations, we compute the effective neutrino viscosity using the approximate analytical expression as in~\cite{KeiJanMue96} 
\begin{equation}\label{eq:nu_neutrino}
\begin{aligned}
  \nu_{\rm neutrino} &= 3.8 \times 10^{22} \\ &\times \left(\frac{T}{1 ~\rm MeV} \right)^2
  \left(\frac{\rho_0}{2.8 \times 10^{14}\,\rm g\, cm^{-3}}\right)^{-1} \rm g \,cm^{-1} \, s^{-1}\,,
\end{aligned}
\end{equation}
which considers six species of non-degenerate neutrinos in local thermodynamic equilibrium and takes
neutrino scattering as the dominant source of opacity. For the effective magnetic turbulent viscosity induced by MHD
turbulence, we adopt the shear viscosity of an ``$\alpha$-disk" model~\cite{Artymowicz94, MacFadyen_2008}
\begin{equation}
  \nu_{\rm MHD} = \frac{2}{3}\,\frac{P}{\rho_0}\,\alpha_{\rm SS}\,\Omega^{-1}\,,
\end{equation}
where $\alpha_{\rm SS}$ is the Shakura-Sunyaev stress parameter (see Sec.~\ref{sec:3-6}) and $\Omega$ is
the characteristic angular frequency of the HMNS. Fig.~\ref{fig:vis_vs_x} displays $\nu_{\rm neutrino}$
and $\nu_{\rm MHD}$ as functions of radius along the x-axis Mag+Rad-Full by solid and dashed curves, respectively.
Note that $\nu_{\rm MHD}$ is computed using an averaged value of $\alpha_{\rm SS}\sim 0.1$. The gray vertical
regions span the approximate radii of the HMNS surface and disk outer boundary. Top and bottom panels show
their behaviors at $t \sim 18~\rm ms$ (or $\sim 0.5\,\tau_{\rm HMNS}$;
see Table~\ref{table:Tab2}) and $t \sim 21~\rm ms$ (just after BH formation), respectively.
We find that $\nu_{\rm MHD}$ is a factor of $\gtrsim 10$ larger than $\nu_{\rm neutrino}$ inside the bulk of
the HMNS. Therefore, we conclude that neutrino effective viscosity due to diffusion can be ignored whenever the
magnetic field strength is $\gtrsim 10^{14} \rm G$, as is the case here.
We evaluate the effect of viscosity on the linear growth of the MRI using the viscous Elsasser
number $E_\nu \equiv v^2_A/(\nu_{\rm neutrino} \Omega)$~\cite{Pessah_2008, Longaretti10, Guilet15}, which
needs to be smaller than 1 for the effective neutrino viscosity to significantly affect the MRI growth. Using the
average value of Alfv\'en speed and angular frequency in our cases, we find that $E_\nu > 10^4$ inside the
HMNS and $E_\nu > 10^2$ in the disk. We also evaluate the critical field strength below which neutrino viscous
effects become important~\cite{Guilet15}, scaled to typical parameters inside our HMNS:
\begin{equation}
\begin{aligned}
  B_{\rm crit} & =  3.1 \times 10^{13} \left(\frac{\rho_0}{2.8\times 10^{14}\, ~\rm g \,cm^{-3}} \right)\\
  & \times \left( \frac{\nu_{\rm neutrino}}{3.8\times 10^{22}\, ~\rm g \, cm^{-1} \,s^{-1}} \right)^{1/2} \left( \frac{\Omega}{10^{4} ~\rm s^{-1}}
  \right)^{1/2} ~\rm G\,.
\end{aligned}
\end{equation}
We find $B_{\rm crit}$ is at least a factor of 100 smaller than the typical
field strength in the HMNS remnant~(see Sec.\ref{sec:Evolution}). Similarly,
the growth rate of MRI is hardly affected by the neutrino drag. As the neutrino drag can only be induced
if the wavelength of the fastest growing MRI mode is shorter than the neutrino mean free path, there exists
an upper limit of the magnetic field strength which cannot exceed $\sim 10^{14}~\rm G$ in the HMNS remnant
because $\lambda_{\rm MRI}$ is proportional to the magnetic field strength~\cite{Guilet15, Guilet17}.
Note that from~\cite{Guilet17} neutrino drag is only important inside the HMNS but not in the disk, where
the drag damping rate is smaller than the angular frequency. Therefore, we conclude that the effective viscosity
due to MHD turbulence dominates over that induced by neutrino radiation in our simulations. 

We also consider the microscopic shear viscosity of nucleon matter in the  cores of the HMNS,
which, for non-superfluid matter, is dominated by neutron-neutron and neutron-proton scattering mediated by strong
interactions~\cite{Shternin08}. Note that the temperatures of interest in our models are mostly greater than the critical
temperature $T_{\rm} \sim 10^{8-10} \rm K$ below which dripped neutrons are likely to form Cooper pairs and undergo phase
transitions to superfluid~\cite{Gezerlis2014, Bertoni2015}. The shear viscosity due to in-vacuum nucleon-nucleon scattering
in the non-superfluid liquid core was first estimated by~\cite{FloIto76, FloIto79} and an approximate and widely
used fitting formula is given by~\cite{Cutler87}
\begin{equation}
   \begin{aligned}\label{eq:vis_shear}
  \nu_n & =  8.2 \times 10^{14} \\ \times  &\left(\frac{T}{1 \, \rm MeV}\right)^2 \left(\frac{\rho_0}
     {2.8 \times 10^{14} \, \rm g\, cm^{-3}}\right)^{9/4} \, \rm g \, cm^{-1} \, s^{-1}\,.
\end{aligned} 
\end{equation}
Another possible contribution to the shear viscosity comes from the collision of electrons and muons.
According to~\cite{FloIto79}, the shear viscosity is dominated by neutrons in non-superfluid matter. Therefore,
we ignore the effects from electrons and muons. 

In Fig.~\ref{fig:vis_vs_x}, the dotted curves represent the viscosity due to the nucleon processes
above using Eq.~(\ref{eq:vis_shear}). Near the core of the HMNS, the microscopic shear viscosity
is comparable to $\nu_{\rm n}$. Nevertheless, at radius greater than 5 km, it quickly diminishes
and becomes negligibly small compared to $\nu_{\rm MHD}$ and $\nu_{\rm n}$. Therefore, in our
simulations, the shear viscosity due to nucleon scattering is minuscule and can be ignored
as well. We again note that, the bulk viscosity, which has larger magnitude, the nucleon
shear viscosity is most important during the inspiral and does not redistribute angular momentum in the remnant. It is the shear viscosity, which helps drive the HMNS to
collapse and drives disk accretion onto the remnant BH. To conclude, the viscous effects due to
neutrino diffusion, drag, and nucleon-nucleon scattering are small compared to the MHD turbulence
in our systems. 

%%%%%%%%%%%%%%%%%%%%%
%%%  Appendix F   %%%
%%%%%%%%%%%%%%%%%%%%%
%
\section{Code Test: Heated Oppenheimer-Snyder Collapse}
\label{App:OS}
%
%%%%%%%%%%%%%%%%%%%%%%
%%%  Appendix F1   %%%
%%%%%%%%%%%%%%%%%%%%%%
%
\subsection{General Description}
\label{subapp:OS-gen}

Here we summarize results from one important test of our radiative transport scheme in a
strong gravitational field. Other tests, including those involving shocks, have been previously performed in~\cite{grrmhd} and reproduced here. The collapse of a general relativistic,
homogeneous, and pressureless sphere is given by the analytic Oppenheimer-Snyder (OS)
solution~\cite{PhysRev.56.455}. We consider a thermal radiation perturbation to OS collapse in ~\cite{Shapiro89},
where the radiation is treated in the relativistic, thermal diffusion approximation
for an optically thick radiation field that remains dynamically unimportant to the spacetime and
collapsing dust sphere, and where LTE is maintained throughout the collapse. The solution for the interior radiation field and outgoing flux as a function of radius and time were determined analytically
in this case (see also Exercise 5.20 in~\cite{bsbook10}). In~\cite{Shapiro96}, we solved the
full Boltzmann radiative transfer equation coupled to the radiation moment equations numerically to obtain the exact interior radiation
intensity and flux for both optically thick and thin cases. The two methods showed close agreement
in the optically thick limit. In~\cite{grrmhd}, we again performed numerical simulations of ``Heated OS collapse", now
with the radiative transfer module in our {\tt Illinois GRMHD} code, constrained to the optically thick limit.
Again the results closely matched the analytic solution. Recently, we carried out a similar treatment using a
two-moment radiation transport scheme in the reference-metric formalism, again for
an optically thick radiation field~\cite{bs20} in spherical symmetry. To test the reliability of our
M1 implementation here, we again simulate heated OS collapse. For the first time, we not only treat
the radiation field in the optically thick interior but also the outgoing field in the vacuum exterior,
which is optically thin. 
%
%%%%%%%%%%%%%%%%%%%%%%%%
%%%    OS Collapse   %%%
%%%%%%%%%%%%%%%%%%%%%%%%
%
\begin{figure}
  \hspace*{-4mm}
  \includegraphics[scale = 0.175]{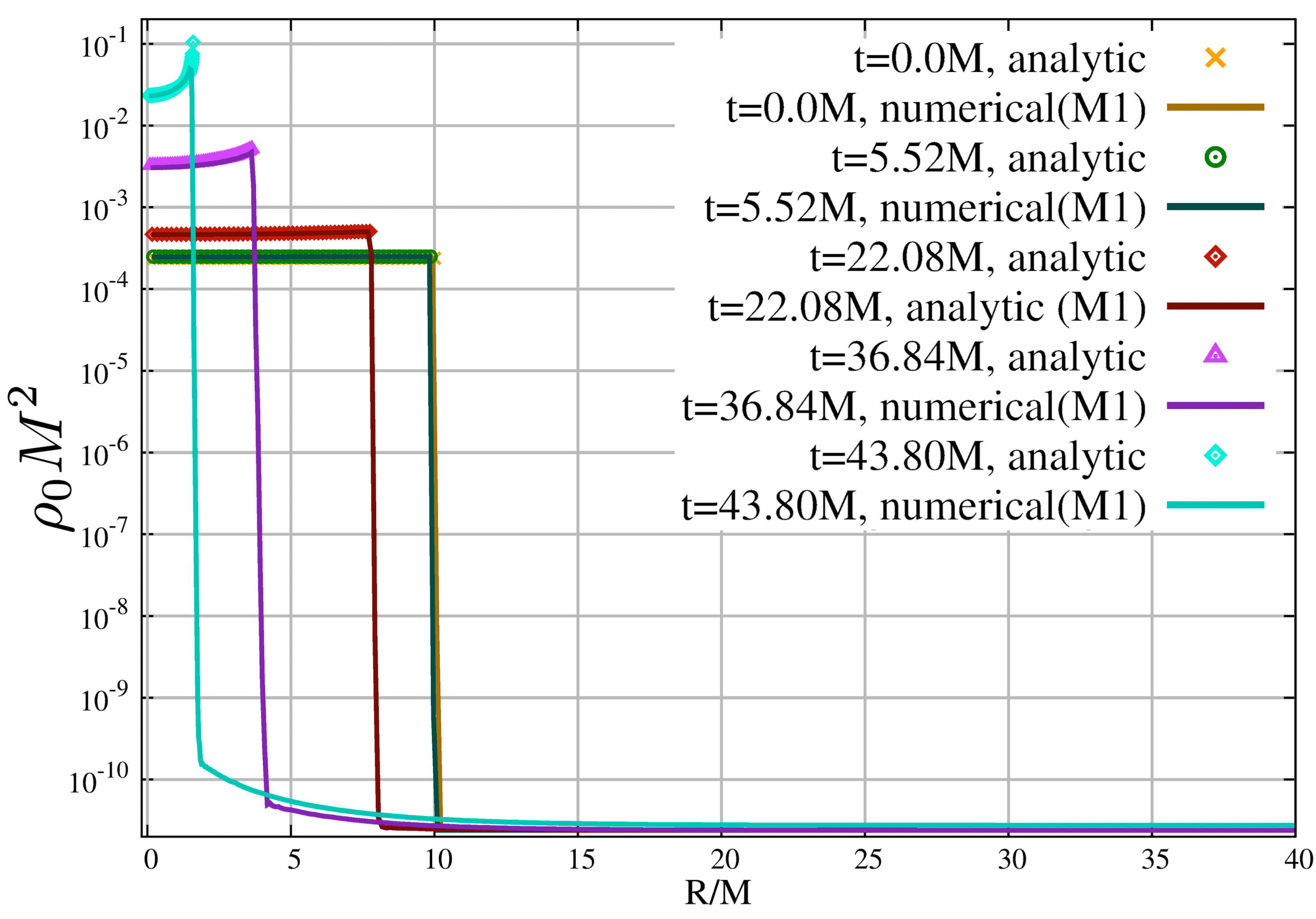}
  \hspace*{-4mm}
  \includegraphics[scale = 0.175]{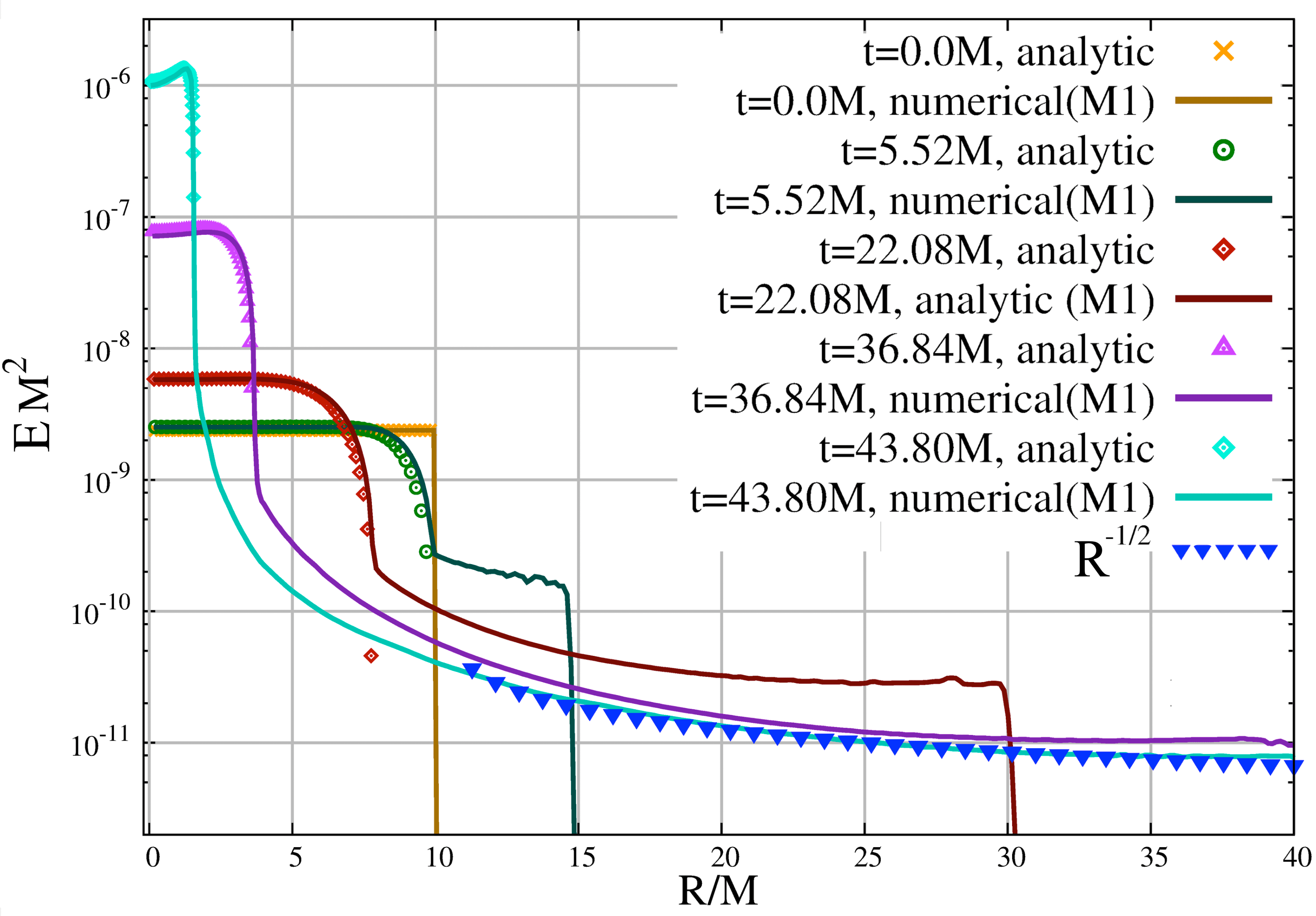}
  \hspace*{-4mm}
  \includegraphics[scale = 0.168]{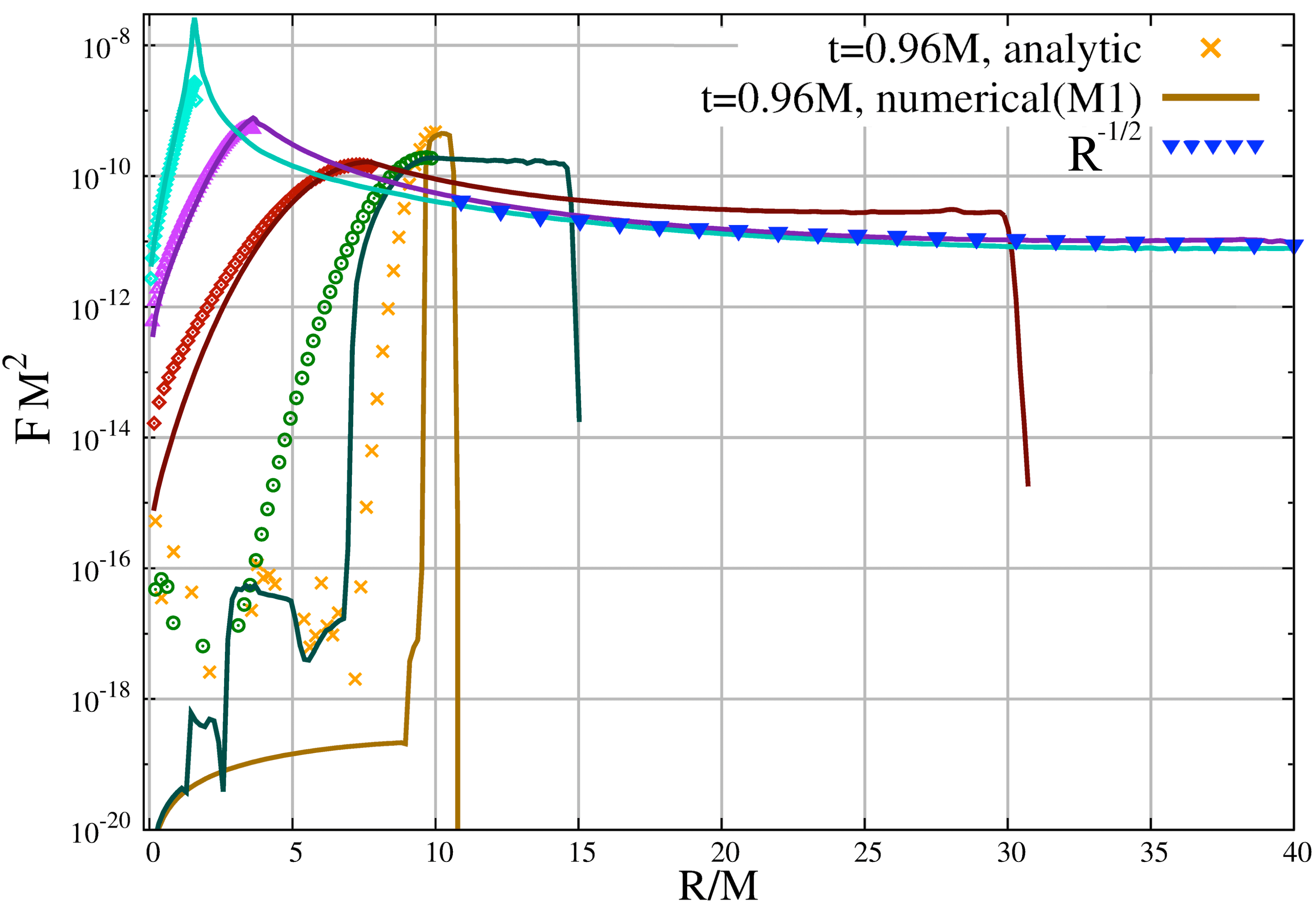}
    \caption{Numerical (dotted curves) and analytical (solid curves)
      solutions at various times for the rest-mass density (top panel), radiation energy
      density $E$ (middle panel), and radiation flux $F$ (bottom panel)
      vs. areal radius $R$ for a heated OS collapse with initial areal radius $R_0 = 10$M. The blue triangular
      dotted curve in the stellar exterior at the latest time represents the curve $\sim R^{-1/2}$.}
    \label{fig:OS_Erad_ana}
\end{figure}

In our test, we set the initial areal radius $R_0 = 10 M$,
from which the initial rest mass density $\rho_{0, \rm ini} = 2.39 \times 10^{-4} M^{-2}$. We set
$P_{0, \rm ini} = 10^{-6} \rho_{0, \rm ini}$ to make matter pressure dynamically unimportant, and
$E_{0, \rm ini} = 10^{-5} \rho_{0, \rm ini}$ to make radiation pressure unimportant, to recover heated OS collapse.
We also set $v^i_{\rm ini} = F^i_{\rm ini} = 0$. The temperature of
the sphere is initially set by the LTE condition $E = a\,T^4$. We choose the grey absorption
opacity by setting the initial optical depth $\tau^a = R\,\kappa^a\,\rho_0 = 25$, and set the scattering opacity
$\kappa^s$ to zero. During collapse, the interior remains optically thick as the density increases with
$R^{-3}$. We construct the initial metric based on the analytical OS solution, which matches a closed-Friedmann
interior to the Schwarzschild exterior~\cite{Misner:1973prb}. To be consistent with previous work, we adopt 
the hyperbolic driver conditions for the lapse and shift functions (see Eqs.~(17) and (18)
in~\cite{grrmhd}) with the same choice of freely specifiable constants $a_1, a_2, a_3, b_1,$ and $b_2$.
Our grid consists of four nested refinement boxes centered in the sphere,
with adjacent levels differing in size and resolution by factors of two. The outer boundary is
at $x_{\rm max} = 60M$ and $dx = 0.12M$ on the coarsest level. We note that the sphere is initially  resolved by
$\approx 2167$ grid points across the diameter on the four levels. We endow the initial exterior regions with
a low-density atmosphere with $\rho_{0,\rm atm} = 10^{-12} \rho_{0,\rm ini}$. To track both the interior and
exterior evolution, we lift the ``zero-temperature approximation" ($E = 0$)
imposed by~\cite{Shapiro89, bs20} at the stellar boundary and instead impose an outgoing radial radiation
boundary condition in the near vacuum at the outer edge of our grid.

The profiles of rest mass density, radiation energy density, and radiation flux as functions of radius
at select moments of our heated OS collapse model are displayed in Fig.~\ref{fig:OS_Erad_ana}. The dotted
and solid curves represent the numerical and analytical solutions, respectively.
In the following sections, we describe our results and assess their reliability in regions with different
optical thickness. We first compare our results in the optically thick region (stellar interior) with previous
thermal OS collapse simulations using the Eddington closure schemes (with Eddington factor $1/3$), then we locate
the photon trapping radius at the transition region and explain the behavior of radiation fields across it.
Finally we describe the free propagation of the radiation field in the optically thin exterior. 

%%%%%%%%%%%%%%%%%%%%%%
%%%   Appendix F2  %%%
%%%%%%%%%%%%%%%%%%%%%%
%
\subsection{Optically thick region: diffusion approximation}
\label{app:subsec}

In Fig.~\ref{fig:OS_Erad_ana}, we show the comparisons of our numerical results computed using 50
Lagrangian fluid tracers for the rest-mass density $\rho_0$ (top panel), radiation energy density
$E$ (middle panel), and radiation flux scalar magnitude $F \equiv F^\mu F_\mu$ (bottom panel) with
the analytic solutions summarized by Eqs.~(D11)-(D13) in ~\cite{grrmhd}. Our numerical solutions lie on top of the analytical ones in the stellar
interior for the rest-mass density and radiation energy density (top and middle panels
in Fig.~\ref{fig:OS_Erad_ana}). The radiation flux matches
the analytical solution after the initial transient phase (bottom panel in
Fig.~\ref{fig:OS_Erad_ana}). Our numerical values also match the results in~\cite{bs20} in the stellar
interior. Notice that we also recover the expected difference in $F$ between the numerical and analytical results
during the early phase of collapse, as reported and explained in~\cite{bs20} (which refers it as the ``initial
transient" phase). The difference originates from the computation of a (vanishingly) small quantity by
subtracting two nearly equal quantities during the recovery of primitive variables in the region with a
spatially non-constant lapse. Nevertheless, at late stages, when the gauge and radiation characteristics
have reached the stellar center, the numerical result of $F$ becomes close to the analytical solution
from the diffusion equation, as expected. Therefore, we conclude that the code reproduces the evolution
of heated OS collapse consistent with previous works in the stellar interior (optically thick regions).
Also, for this limit, our primitive solver matches the analytical expressions
for the primitive variables (see Eq.(65)-(67) in ~\cite{grrmhd}) using Eddington closure.
% 
%%%%%%%%%%%%%%%%%%%%%%%%%%%
%%%   Trapping surface  %%%
%%%%%%%%%%%%%%%%%%%%%%%%%%%
%
\begin{figure}
\includegraphics[scale=0.39]{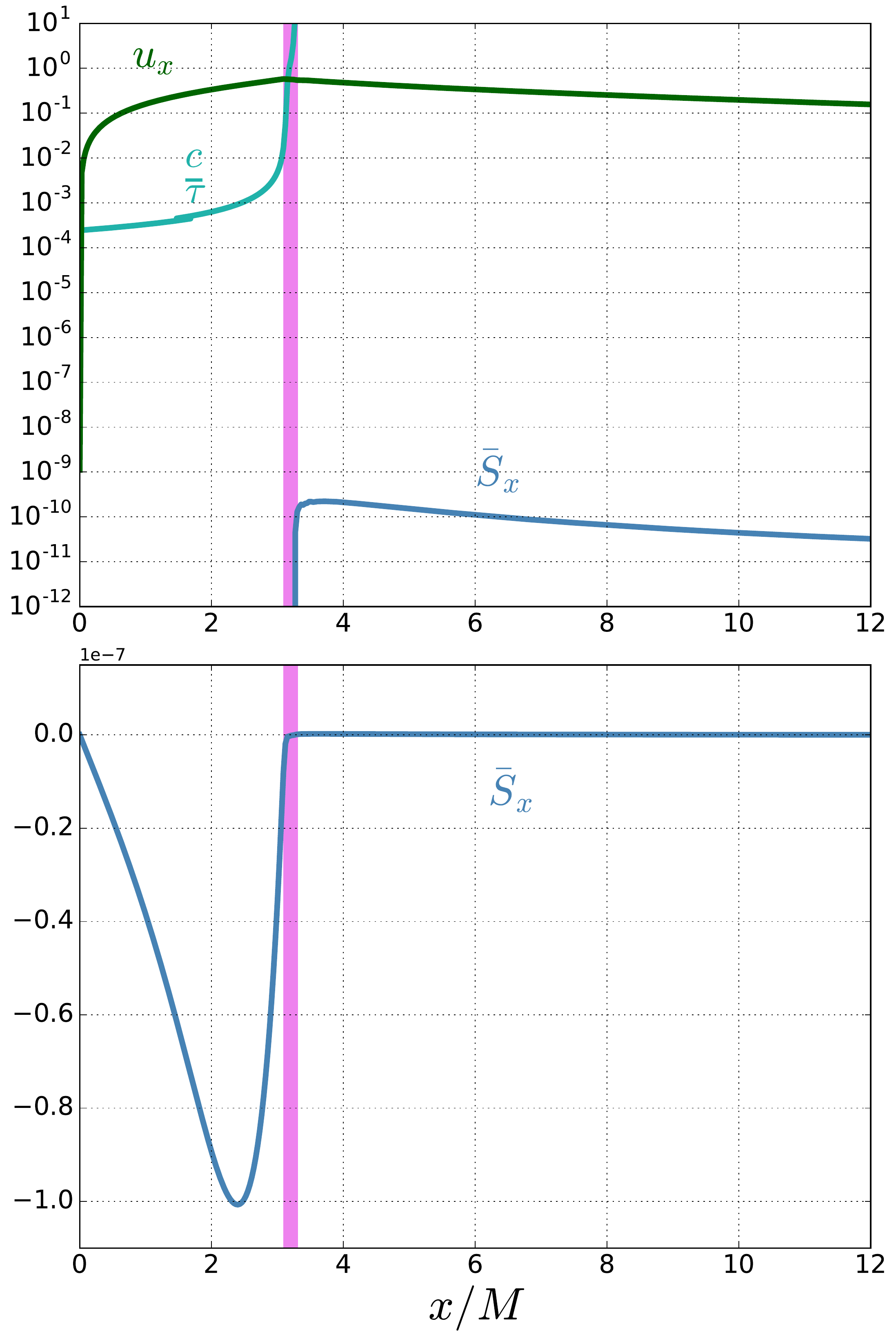}
\caption{Top panel: Photon-trapping radius during OS-collapse (magenta) where the
  radiation momentum density $\bar{S}_x$ flips direction (blue). This radius can be also identified by equating the
  radius at which the gas velocity (green) equals the approximate radiation diffusion speed (cyan).
Bottom panel: The change of sign (direction) of $\bar{S}_x$.
  Results are shown at $t = 36.84M$, when the areal surface of the star is at $R \approx 3.7M$.}
\label{fig:photon_trap}
\end{figure}

%%%%%%%%%%%%%%%%%%%%%%
%%%   Appendix F3  %%%
%%%%%%%%%%%%%%%%%%%%%%
%
\subsection{Photon-trapping Radius}
\label{app:subsec2}
According to~\cite{Beg79}, if the matter accretion rate is greater than $\dot{M}_{\rm E} \equiv L_{\rm E}/c^2$,
where $L_{\rm E}$ is the Eddington limit, there exists a surface at radius $r_{\rm trap}$ within which the
radiation is trapped and eventually falls into the BH with the gas. This photon trapping radius $r_{\rm trap}$
is defined by setting the infall speed of the gas equal to the approximate outward diffusion speed of the
radiation ($\sim c/\tau$). At this radius, the radiation momentum density changes its direction, from inward
inside $r_{\rm trap}$ to outward outside.

To approximate $r_{\rm trap}$ in our OS collapse model, we compute the optical depth by integrating
$\kappa^a \rho_0$ along the x-axis. The top panel of Fig.~\ref{fig:photon_trap} shows the stellar dust velocity,
the diffusion speed of the radiation, $c/\tau$, and the radiation
momentum density along the x-axis at $t = 36.84 M$ (corresponding to the maroon curve
in Fig.~\ref{fig:OS_Erad_ana}). From the plot, the intersection between the diffusion speed and the gas velocity
is at around $x = 3.2 \,\rm M$, which agrees with the location at which the radiation momentum density $\bar{S}^x$
turns over. As shown in the bottom panel, $\bar{S}^x$ becomes negative at radius smaller than
$ 3.2 \,\rm M$, indicating the radiation field is trapped and falls inward following the collapse of the star.
We define the position $x = 3.2M$, labeled by the magenta vertical line in Fig.~\ref{fig:photon_trap}, as the
radius of photon trapping of the model. It is expected that the radius would move inward with time and eventually
freeze at $2M$ at BH formation. The consistency between trapping radius and flux direction turnover provides
another useful check on our code. 
%
%%%%%%%%%%%%%%%%%%%%%%%%
%%%    scalar flux   %%%
%%%%%%%%%%%%%%%%%%%%%%%%
%
\begin{figure}[]
  \includegraphics[scale=0.35]{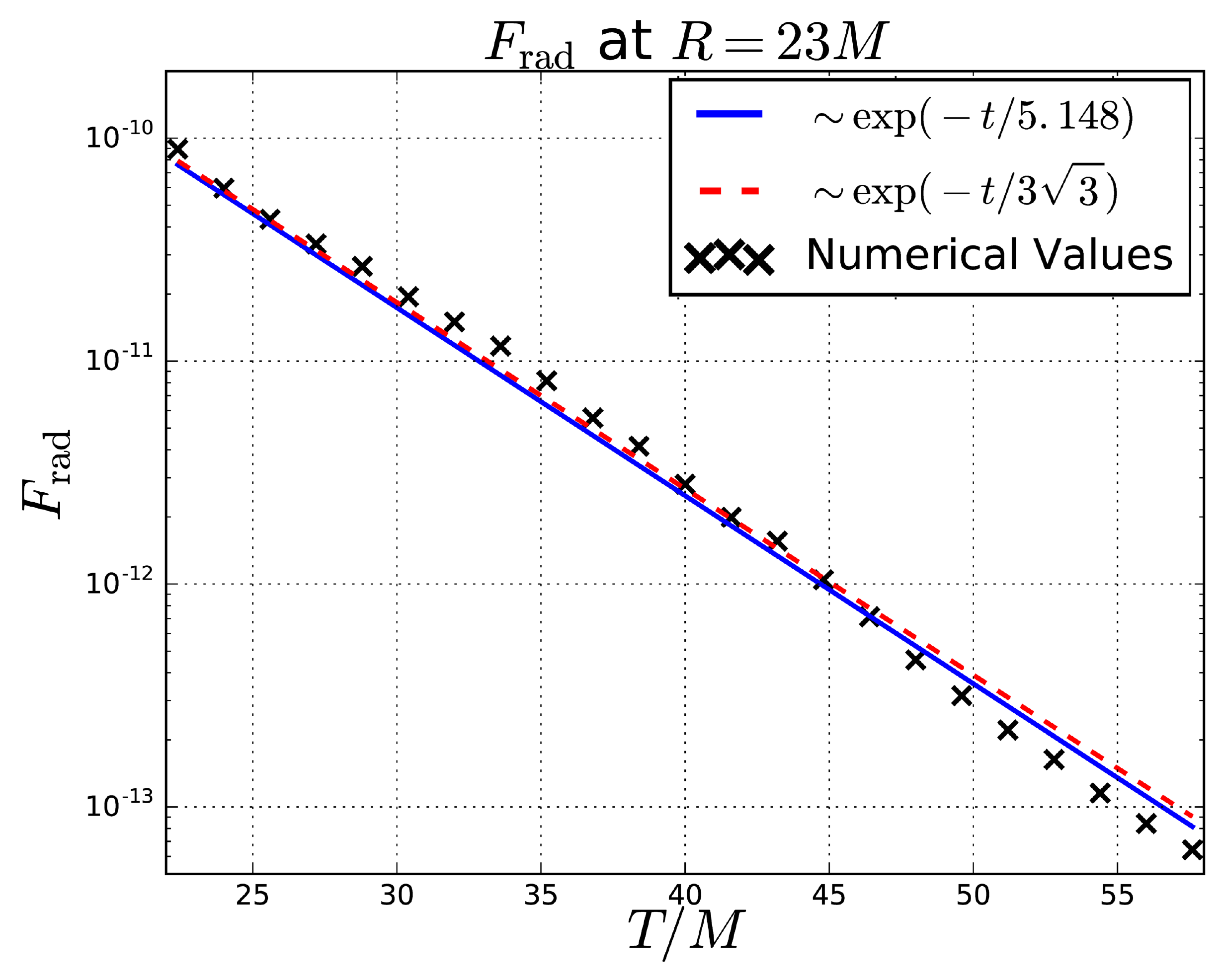}
  \caption{The asymptotic scalar flux (black crosses),  its best-fit line (blue)
    and the analytic curve Eq.~(\ref{eq:Lrad_AE}) (red) at $R=23M$ at the late stage of collapse for
    the stellar model with the initial areal radius $R_{\rm ini} = 4M$.}
    \label{fig:L_decay}
\end{figure}

%%%%%%%%%%%%%%%%%%%%%%
%%%   Appendix F4  %%%
%%%%%%%%%%%%%%%%%%%%%%
%
\subsection{Decay of Radiation Flux at Large Radius}
\label{app:subsec3}

With the M1 closure scheme of radiation, we are able to study radiation transport in the optically thin
exterior region of the system. In the exterior, free propagation of radiation at the speed of
light is expected. As shown in the middle and bottom panels of Fig.~\ref{fig:OS_Erad_ana},
the distance between the initial stellar surface at $R_0 = 10M$ and the radius marking the outermost
non-zero $E$ and $F$ profiles, approximates the time of evolution, indicating free propagation at the speed
of light. At large radius, the magnitude of energy density equals that of the radiation flux $E = F$. This is confirmed in the two panels as
well, where both $E$ and $F$ slowly fall together for radius $R > 30 M$ roughly as $R^{-1/2}$, as indicated in Fig.~\ref{fig:OS_Erad_ana}.
Note that as the star is collapsing, the radiation source is not steady, therefore the flux does not
decay as $R^{-2}$ as it would for a constant source.

One of the most important quantitative results is the exponential attenuation of the luminosity measured
by a distant observer in the late stage of collapse, after the stellar surface has reached its gravitational
radius at $R = 2M$~\cite{Misner:1973prb, Podurets65, Ames68}:
\begin{equation}
  \label{eq:Lrad_AE}
    L_{\rm rad} \sim {\rm exp} \left(-\frac{t}{3\sqrt{3} M}\right)\,.
\end{equation}
The light is dominated by photons deposited in the unstable photon circular orbit at $R = 3 M$. They leak out at an
exponential decay rate. To check our code, we perform a collapse model with the initial areal radius $R_{\rm ini} = 4M$
and follow the collapse up to $t = 57 M$. This is approximately four times the collapse time, which is defined the time
at which the stellar surface passes through an areal radius of $R = 2M$ for the first time. We fit our data
for the flux $F$  at $R = 23M$ from $t = 22M$ to $t = 56M$ with the exponential expression
\begin{equation}
  F \rightarrow ~{\rm exp} (-t/t_1)\,,
\end{equation}
where $t_1$ is constant. We find that $t_1/M = 5.15^{+0.18}_{-0.17}$. Comparing with Eq.~(\ref{eq:Lrad_AE}),
the fitted decay rate of the asymptotic flux data has less than $1 \, \%$ deviation from the theoretic result,
i.e. $t_1/M = 3\sqrt{3} \approx 5.20$. The corresponding plot is shown in Fig.~\ref{fig:L_decay}. We conclude
that the exponential decay of the asymptotic flux at large radius is correctly identified by our M1 code,
showing its reliability in optically thin regions as well. 
  
%%%%%%%%%%%%%%%%%%%%%%%
%%%   References    %%%
%%%%%%%%%%%%%%%%%%%%%%%
\bibliography{reference}
\end{document}